\documentclass[12pt]{article}

\usepackage[square,numbers]{natbib}

\usepackage{makeidx}

\newcommand{\bm}[1]{\mbox{\boldmath $#1$}}

\title{\large{\textbf{THEORETICAL BASIS FOR A SOLUTION TO}}\\
\large{\textbf{THE COSMOLOGICAL CONSTANT PROBLEM\footnote{Cite as: arXiv:gr-qc/0603005v18}}}}

\author{J.C. Lindner\footnote{Email address: research@jclindner.ca}\\
Department of Physics, Universit\'{e} de Montr\'{e}al\\
Montr\'{e}al, QC, Canada}

\date{\normalsize{3 March 2006 (updated 3 June 2026)}}

\makeindex

\begin{document}

\maketitle

\begin{abstract}
Following a short discussion of some unresolved issues in the standard model of cosmology (considered to be a generic $\Lambda$CDM model with flat geometry and an early period of inflation), an update on the current state of research regarding the problem of negative energy is provided. Arguments are then given to the effect that traditional assumptions concerning the behavior of negative-energy matter give rise to various inconsistencies, including a violation of the requirement of relational definition of physical attributes. An alternative set of axioms is proposed that would govern the behavior of negative-energy matter if it is to be considered a viable element of physical theories upon which cosmological models are build. A generalization of the core mathematical framework of general relativity theory is derived from the proposed axioms, which enables the formulation of quantitative predictions concerning the interaction of positive- and negative-energy objects. Based on those developments, solutions are proposed to the problem of the discrepancy between theoretical and experimental values of the average density of vacuum energy and to several other related issues in theoretical cosmology, including the problem of the nature of dark matter and dark energy and that of the origin of thermodynamic time asymmetry.
\end{abstract}

%\pacs{04.20.Cv, 95.35.+d, 95.36.+x, 98.80.-k}

\newpage

\section*{Note to the reader}

This is the abridged version of a report which was previously released in extended form \cite{Lindner-4} by the present author. It is aimed at providing a short, self-contained account of the most significant results which are discussed in the extended version. It covers a good portion of the material which is contained in the larger version, except for what regards discrete symmetries and black hole thermodynamics, quantum-mechanical aspects of gravitation theory and cosmology, and the problem of the interpretation of quantum theory itself. For reading convenience, adapted versions of both the extended and the abridged version of this report have been published in print under the title \textit{Negative-Energy Matter and the Direction of Time}.

The present update provides many significant improvements and corrections to most sections, but more particularly concerning the problem of the origin of repulsive gravitational forces between opposite-energy objects and the analogy of voids in an expanding matter distribution from section \ref{sec:3}, as well as regarding the formulation of the generalized gravitational field equations and the question of energy conservation from section \ref{sec:5}. Even if you have been in contact with the author's earlier publications, it is highly recommended to read those sections again in order to fully appreciate the value of the many profound changes they introduce to our understanding of the matters discussed.

\newpage
\tableofcontents
\newpage

%\newpage
%\listoffigures
%\newpage
%\listoftables
%\newpage

\section{Introduction\label{sec:1}}

The main unresolved issue of the cosmological models currently considered to best fit observational data concerns the cosmological term\index{cosmological term} $-\Lambda g_{\mu\nu}$ that enters the left-hand side of the gravitational field equations of general relativity theory. This term was first introduced by Einstein\index{Einstein, Albert} \cite{Einstein-1} in order to balance the mutual gravitational attraction of matter to enable a static universe. It was later reintroduced into physics to describe the consequences on the expansion of the universe of a vacuum with average energy density $\rho_v=c^4\Lambda/8\pi G$. This constant energy density, generally assumed to be positive, would be the source of a negative pressure\index{negative pressure} $p_v$ whose value is provided by an equation of state\index{equation of state} of the form $p_v=-\rho_v$.

Given that a positive cosmological constant\index{cosmological constant} $\Lambda$ contributes positively to the right-hand side of the second Friedmann equation\index{second Friedmann equation} for the scale factor\index{scale factor} $a(t)$
\begin{displaymath}
\frac{\ddot{a}}{a}=-\frac{4\pi G}{3}\left(\rho+\frac{3p}{c^2}\right)+\frac{\Lambda c^2}{3}
\end{displaymath}
due to its negative pressure, while the positive energy density $\rho$ and the positive pressure $p$ of ordinary matter and radiation contribute negatively to the variation $\ddot{a}$ of the rate of expansion, then it follows that this negative pressure would enable to explain the acceleration of universal expansion first observed in 1998 by two independent groups \cite{Perlmutter-1} \cite{Riess-1}. The density of vacuum energy $\rho_v$ would then provide the missing energy that is needed to account for the fact that we measure, by various means, a total energy density $\Omega_0\sim 1$ (as a fraction of the critical density), but an energy density of normally gravitating matter (visible and dark) that amounts to only $\Omega_{0,m}\sim 0.3$. Regardless of one's personal opinion concerning this issue, it must be recognized that the hypothesis of a non-vanishing density of vacuum energy is the only explanation of the acceleration of the rate of expansion of space that is not based on speculative theoretical constructs and that relies on well tested aspects of elementary particle physics and relativity theory.

The problem which then arises is that, the energy density of the vacuum as a fraction of the critical density, must be adjusted to this non-zero value $\Omega_{0,\Lambda}\sim 0.7$ which happens to be of the same order of magnitude as that of the present, average energy density $\Omega_{0,m}$ of normal and dark matter, without any apparent justification in the underlying physics for this unlikely state of affairs (as those two densities scale differently in the course of universal expansion). This situation is even more problematic given that this value of vacuum energy density appears to be in disagreement with most theoretical predictions from the standard model of elementary particles or its extensions, which produce a value 120 orders of magnitude larger \cite{Weinberg-1}. Prior to the reintroduction of the cosmological term, as a means to rectify original cold-dark-matter (CDM) models\index{cold-dark-matter model}, which were no longer in agreement with experimental data, it was still possible to argue that the density of vacuum energy was zero due to some as yet unknown symmetry principle. But with the advent of $\Lambda$CDM models and their small, yet non-vanishing cosmological term, such a solution may appear to no longer be possible.

In what follows, however, I will suggest that, upon the introduction of a revised concept of negative-energy matter into relativity theory, an understanding of the acceleration of universal expansion as originating from a non-zero value of average vacuum energy density is still possible. This solution is particularly attractive given that it provides the basis for a solution to other outstanding problems in the field of classical cosmology. In fact, from this perspective, it becomes possible to solve at once, in a simple and straightforward manner, both the problem of the discrepancy between the observed versus predicted values of vacuum energy density and the problem of flatness\index{flatness problem}, which is usually tackled using inflation\index{inflation!theory} theory. But I will also explain that some of the effects which are normally attributed to the presence of ordinary cold dark matter\index{cold dark matter} appear to arise from local variations in the density of vacuum energy attributable to the presence of inhomogeneities in the large-scale matter distribution, as well as from the presence of underdensities in the distribution of invisible negative-energy matter. I will conclude this report with a discussion on how those developments can be used to provide a satisfactory solution to the problem of the origin of the thermodynamic arrow of time\index{thermodynamic arrow of time}.

\section{The current situation\label{sec:2}}

Our vision of the cosmos has changed drastically in the previous decades, but it now seems that, experimentally, we have reached a stable point where new data only come to confirm previous observational findings, despite the fact that those empirical results may not fully agree with current theoretical expectations. It is consequently an ideal time to re-examine the basic theoretical assumptions underlying the standard model of cosmology\index{standard model of cosmology} in the light of this fixed experimental background. One of the most basic and often implicit assumption that is made is that energy must be positive. The mathematical expression of this hypothesis is called the weak energy condition\index{weak energy condition} and it states that for every timelike 4-vector $u^\alpha$ we must have
\begin{equation}\label{eq:01}
T_{\alpha\beta}u^\alpha u^\beta\geq 0
\end{equation}
where $T_{\alpha\beta}$ is the stress-energy tensor. There is another more restrictive condition on the values of energy which is called the strong energy condition\index{strong energy condition} and which states that for every timelike 4-vector $u^\alpha$
\begin{equation}\label{eq:02}
(T_{\alpha\beta}-\frac{1}{2}T g_{\alpha\beta})u^\alpha u^\beta\geq 0
\end{equation}
where $T=T_\alpha{}^\alpha$ is the contraction of the stress-energy tensor\index{stress-energy tensor}. If this stronger condition is obeyed in all situations, then gravity must always be an attractive interaction.

However, those equations are classical equations and even though they may be valid when we are considering the expectation values of energy of quantum systems, they do not reflect the subtleties related to the definition of energy at the quantum level. So, what do we mean exactly by positive energy? To answer that question, we must first examine what \textit{negative} energy might be. Negative energy made its first appearance in physical theory when Paul Dirac\index{Dirac, Paul} and others tried to integrate special relativity into quantum mechanics. To obtain Lorentz-invariant equations for the wave function it seemed that one had to sacrifice the positivity of energy. While trying to make sense of those negative-energy\index{negative energy!solutions} solutions, Dirac was led to introduce antiparticles \cite{Dirac-1}.

Of course, antiparticles do not have negative energy (at least in the sense we usually understand) otherwise we would run into a number of problems ranging from violation of the conservation of energy to the possibility of producing perpetual motion\index{perpetual motion} machines. It does not appear however that negative energy alone is to blame for those inconsistencies. What really constitutes a problem is negative energy applied to antiparticles. Nevertheless when faced with the prospect of having to introduce negative energy states into physics, Dirac saw the apparently insurmountable difficulties that this would entail and rather than recognizing the validity of his deduction he turned it into an argument for the existence of a new class of positive-energy particles identical to normal particles, but with reversed electrical charge. This became one of the greatest theoretical predictions of the history of physics.

The relation between antiparticles and negative energy in Dirac's theory arises from the hypothesis that antiparticles\index{antiparticles!holes} consists of holes in a completely filled negative-energy matter distribution\index{filled negative-energy matter distribution} (which is not to be confused with a uniform distribution of negative vacuum energy\index{negative vacuum energy!uniform distribution}). Thus, it was argued, that the absence, in this negative-energy sea, of a negative-energy electron with a normal, negative electric charge (which may arise when such an electron is excited to a higher, positive energy state), is equivalent to the presence of a positive-energy electron with a reversed (positive) charge. Due to the exclusion principle\index{exclusion principle}, a positive-energy electron wouldn't be able to make a transition to the already occupied, lower, negative energy states, although it could fall into one of the holes and radiate energy in the process. It was further assumed that there are no observable effects from the presence of this filled distribution of negative-energy matter. Given that such a model cannot apply to bosons, which are not ruled by the exclusion principle, it is hard to see how Dirac's solution makes things any better than simply accepting the existence of negative-energy particles and this is why this theory was soon abandoned.

Most people today consider that the theory of relativistic quantum fields\index{relativistic quantum fields} that came to replace Dirac's model has eliminated the `problem' of negative energy states\index{negative energy states!problem of} at the source and that it simply doesn't matter anymore if some antiquated equations describing a single particle allow for the existence of negative energy states, because those states are not `physical'. But upon closer examination I realized that if we are not bothered with negative energy states in quantum field theory\index{quantum field theory} it is because we simply choose to ignore those solutions in the first place and then integrate that choice into the formalism. Basically, this amounts to say that the negative energies predicted by the single-particle relativistic equations\index{single-particle relativistic equations} are simply transition energies or energy differences between two positive energy states and there is no reason why those variations couldn't be negative as well as positive. But I'm getting ahead of myself here, so let's go back to the question of what is actually meant by negative energy.

The next step in our understanding of antiparticles was accomplished by Richard Feynman\index{Feynman, Richard} \cite{Feynman-1} (with a little help from John Wheeler\index{Wheeler, John} and following Ernst St\"{u}ckelberg\index{St\"{u}ckelberg, Ernst}). What's interesting with Feynman's original approach is that it made clear the fact that we really \textit{choose} to exclude negative-energy particles, but it also helped us understand what exactly it is that we mean by negative energy. Feynman recognized that it was a mistake to exclude the negative energy states because among other things ``\textit{it leads to an incomplete set of wave functions} [and] \textit{it is not possible to represent an arbitrary function as an expansion in functions of an incomplete set}''. It also appeared that those energy states were required from a physical viewpoint because there were well-defined predictions of transition probabilities\index{transition probabilities} to those states and if they are excluded there would be contradictions with observations (if the theory is indeed right). This is actually the same argument that motivated Dirac\index{Dirac, Paul} to introduce antiparticles. Feynman's solution also involves antiparticles, but this time the negative energy states\index{negative energy states} are not swept under the rug, they are rather considered to be the states of particles propagating backward in time. Feynman shows that an electron propagating backward in time with negative energy, but with unchanged electrical charge, would be equivalent from our overall, unidirectional-time point of view\index{unidirectional-time viewpoint} to an electron with positive energy and reversed charge.

What is essential to understand here is this dependence of the definition of energy on the direction of propagation in time. It appears that simply saying that a particle has negative energy doesn't make sense. We must in addition always specify the direction of propagation of this energy with respect to time. What's more, if we are to find any physical distinction, particularly with respect to gravitation, between ordinary matter and matter with negative energy, then we must consider either negative energy propagating \textit{forward} in time or \textit{positive} energy propagating backward in time, because it can be expected that antiparticles behave identically to particles in any gravitational field, given that they have positive energy when considered from the viewpoint of the conventional, positive direction of time (for a review of the arguments against the idea that antiparticles could behave in unusual ways in a gravitational field see Ref. \cite{Nieto-1}).

Given those considerations it appears that an antiparticle is really just an ordinary particle that reverses its energy to go backward in time, as when a particle reverses its momentum to move backward in position space. But a particle with real negative energy propagating forward in time could have completely different properties. We may use the term `negative action'\index{negative action} to differentiate such particles from the positive-action ones we are familiar with, but I will continue to use the term `negative energy' in place of negative action when the context clearly indicates that I mean negative energy propagating forward in time.

To return to Feynman's approach to quantum electrodynamics\index{quantum electrodynamics!Feynman's approach}, it appears that what prevents negative action from being present in the theory is merely a choice of boundary conditions\index{boundary conditions}. There are several possible choices for the propagation kernel\index{propagation kernel} or propagator\index{propagator} (giving the probability amplitude\index{probability amplitude} of a transition from point 1 to point 2 in spacetime) which all constitute valid solutions of the basic equations. For example (see Ref. \cite{Feynman-4}, p. 74), in a time-stationary field if the wave functions $\phi_n$ are known for all states of the system, the kernel $K_+^A$ may be defined by
\begin{eqnarray*}
K_+^A(2,1) & = & \sum_{E_n>0}e^{-iE_n(t_2-t_1)}\phi_n(\bm{x}_2)\tilde{\phi}_n(\bm{x}_1) \\
 & & \mbox{ for }\;t_2>t_1 \\
 & = & -\sum_{E_n<0}e^{-iE_n(t_2-t_1)}\phi_n(\bm{x}_2)\tilde{\phi}_n(\bm{x}_1) \\
 & & \mbox{ for }\;t_2<t_1
\end{eqnarray*}
another solution of the equations is
\begin{eqnarray*}
K_0^A(2,1) & = & \sum_{E_n>0}e^{-iE_n(t_2-t_1)}\phi_n(\bm{x}_2)\tilde{\phi}_n(\bm{x}_1) \\
 & & +\sum_{E_n<0}e^{-iE_n(t_2-t_1)}\phi_n(\bm{x}_2)\tilde{\phi}_n(\bm{x}_1) \\
 & & \mbox{ for }\;t_2>t_1 \\
 & = & 0\;\mbox{ for }\;t_2<t_1.
\end{eqnarray*}
Although the kernel $K_0^A$ is also a satisfactory mathematical solution of the equations it is not accepted as a meaningful proposition because it requires the idea of an electron in a real negative energy state. Only those kernels propagating positive frequencies\index{positive frequencies} (or energies) forward in time and negative energies backward in time are usually considered physical (this is usually done through the selection of a particular contour of integration\index{contour of integration} for the relativistic propagator) and that is why quantum field theory\index{quantum field theory} is assumed not to involve true negative energies or, more accurately, negative actions\index{negative action}. Of course, there is nothing wrong with those assumptions, because they are validated by experiments. We never see negative-energy particles propagating forward in time and such particles do not appear to influence the outcome of experiments involving ordinary matter and antimatter, however precise those experiments actually are.

The only problem with the modern approach to quantum field theory is that the formalism is usually introduced in a way that encourages us to believe that after all, antiparticles\index{antiparticles!backward-in-time propagating particles} are not really propagating backward in time with negative energy and that a positron is simply another particle identical to the electron, but with a positive charge. From this viewpoint charge is the decisive aspect and the spacetime relationship between matter and antimatter uncovered by Feynman shouldn't be considered as anything more than an analogy. But it must be clear that we can hold on to such a viewpoint only at the expense of losing the best explanation we have for the existence of antimatter. If we retain the most simple and effective viewpoint under which antiparticles are ordinary particles propagating backward in time, then we must accept that there exists in nature matter whose energy is definitely negative. That may open the door to further insight.

Now, I just said that negative energy does not explicitly enter quantum field theory\index{quantum field theory} (with the exception of the negative energy of particles propagating backward in time), but this doesn't mean that truly negative energies do not appear in this theory at the phenomenological level. In fact, it is well-known that a particular prediction of quantum field theory, that would not apply classically, is that the local energy density may not always be positive definite \cite{Epstein-1}. The most easily accessible experimental setting which can be used to observe some manifestation of a state where energy density takes on negative values is the one where two parallel mirrors\index{parallel mirrors in vacuum} are placed a very small distance $L$ from each other in the vacuum. This experiment was first described by Hendrik Casimir\index{Casimir, Hendrik} \cite{Casimir-1} who calculated that there would be a very small, but detectable force equal to
\begin{displaymath}
F=\frac{\hbar c\pi^2 A}{240 L^4}\;\;(A\gg L^2)
\end{displaymath}
pulling the reflecting plates together (where $A$ is the area of a plate), due to the fact that some positive-energy quantum modes are absent from the vacuum between the mirrors. This force was eventually observed in the laboratory \cite{Lamoreaux-1} \cite{Onofrio-1} and the results confirm theoretical predictions. It must be clear, however, that we are not directly measuring a negative energy density with this setup, only the indirect effects of an absence of positive energy from the vacuum, which is assumed to imply that the density of energy is negative in the small volume between the mirrors.

The realization that quantum indeterminacy appears to allow negative energy densities\index{negative energy densities} has led many authors to propose a modified version of the weak energy condition (equation (\ref{eq:01})) that tries to take into account the fluctuations of energy which arise in the quantum realm. This is the \textit{averaged}, weak energy condition\index{averaged weak energy condition}
\begin{equation}\label{eq:03}
\int_{-\infty}^\infty \langle T_{\alpha\beta}u^\alpha u^\beta\rangle d\tau\geq 0
\end{equation}
where $u^\alpha$ is the tangent to a timelike geodesic and $\tau$ is the proper time of an observer following that geodesic. Here we consider only quantum expectation values of the stress-energy tensor\index{stress-energy tensor} averaged over the entire world-line of the observer, rather than idealized measurements at a point. Consequently, this inequality does seem to allow for the presence of large negative energies over relatively large regions, if there is compensation by a larger amount of positive energy somewhere on the observer's path. It is not entirely clear, however, if even this relaxed positive-energy condition is respected by the predictions of quantum field theory.

Nevertheless, it turns out that the theory places strong limits on the values of negative energy density which can actually be observed. Ford, Roman and Pfenning have found inequalities \cite{Ford-1} \cite{Ford-2} \cite{Pfenning-1} \cite{Pfenning-2} which constrain the duration and magnitude of negative energy densities for various fields and spacetime configurations. They show that for quantized, free, massless scalar fields\index{massless scalar fields} in four-dimensional Minkowski space, the renormalized energy density\index{renormalized energy density}, written in covariant, form obeys the following equation
\begin{equation}\label{eq:04}
\rho=\frac{\tau_0}{\pi}\int_{-\infty}^\infty\frac{\langle
T_{\alpha\beta}u^\alpha u^\beta\rangle}{\tau^2+\tau_0^2}\,d\tau\geq
-\frac{3}{32\pi^2\tau_0^4}
\end{equation}
for a static observer who samples the energy density by time-averaging it against a Lorentzian function with characteristic width $\tau_0$. Basically, this means that the more negative energy there is in a given interval, the shorter this interval must be. This can be seen to drastically limit the consequences of the negative energy densities which are allowed by quantum field theory, in particular for what regards the ability to use those states of matter to achieve causality-violating paradoxes\index{causality-violating paradoxes} or violations of the second law of thermodynamics\index{thermodynamics!violations of second law} (in section \ref{sec:3} I will challenge the idea that negative-action matter could actually be used to produce such inconsistencies even in the absence of those limitations). But again, one must admit that it is not possible to simply rule out the existence of negative energy states even though their observation appears to be severely restricted in the context of a theory that does not explicitly involve such states.

Unlike quantum field theory, classical physics doesn't predict the existence of negative energy densities (although there is no a priori reason why it would forbid them). But it turns out that, even in a classical framework, some form of negative energy must sometimes be taken into account. This is the case with bound systems\index{bound systems}, for which the total energy is smaller than the energy at rest of their constituent subsystems. As those bound systems have lower energy after they are formed, they must emit energy during the process of their formation and consequently they can be considered as physically different from the sum of their parts. This is possible only if we assume that the energy of the attractive field maintaining them together contributes negatively to their total energy.

We need not consider only elementary particles here. Systems as large as the Earth-Moon system\index{Earth-Moon system} can be shown to have asymptotically-defined total masses\index{asymptotically-defined total mass} smaller than those of their constituent planets and experiments confirm those predictions. In this case, the gravitational field responsible for binding the two planets together must have negative energy and it is this negative contribution that diminishes the total mass. However, we can only deduce that the interaction field\index{interaction field!negative energy} has negative energy, but we cannot measure that energy directly. The energy of the attractive electromagnetic field binding the proton and the electron together in the hydrogen atom also has negative energy, but this energy cannot be observed \textit{independently} from that of the rest of the system, even if its contribution to the total energy of the system is well defined. One simply cannot isolate the interaction field from its sources and the same argument is valid for larger systems.

One may wonder, though, if this binding energy could get large enough (negative enough) that it would make the total energy of the bound system itself negative. Once again, however, theory comes to the rescue to put constraints on the values that observable total energy may take\footnote{
It was shown \cite{Brill-1} \cite{Brill-2} \cite{Deser-1} \cite{Brill-3} \cite{Schoen-1} \cite{Schoen-2} \cite{Schoen-3} \cite{Schoen-4} \cite{Schoen-5}, concerning the gravitational interaction in particular, that the energy of matter (everything except gravitation) plus that of gravitation is always positive when the dominant energy condition\index{dominant energy condition|nn} is assumed to be valid, which actually amounts to assume that the energy of the component particles is itself positive.}.
 If we compress a positive-energy object too tightly, it simply collapses into a black hole before its surface area is allowed to become so small and its energy density so large that the magnitude of its negative gravitational potential energy\index{gravitational potential energy!negative} would be larger than the positive energy of the matter. Thus, positive-energy matter cannot turn into negative-energy matter through an increase of negative gravitational potential energy. As with the other cases I discussed, it happens that as soon as we find it, negative energy disappears from our view. But one must admit that, this time, it cannot be considered an unusual and exotic phenomenon with limited consequences, because it appears to be as pervasive and as commonplace as bound systems themselves.

Before concluding this section, I would like to point out the existence of an astronomical phenomenon, not directly related to the existence of negative-energy matter, but which nevertheless involves some kind of gravitational repulsion\index{gravitational repulsion} analogous to that which would presumably be exerted by negative-energy objects on positive-energy matter. It is well-known, in effect, that while an overdense region in an otherwise uniform, expanding distribution of positive-energy matter produces a local deceleration of the rate of expansion of the surrounding matter, a spherical underdense region\index{spherical underdense region} with $\delta\rho/\rho<0$ must give rise to a local acceleration of the rate of expansion of the surrounding positive-energy matter, as James Peebles\index{Peebles, James} once showed \cite{Peebles-1}, using Birkhoff's theorem\index{Birkhoff's theorem}. It was emphasized by Tsvi Piran\index{Piran, Tsvi} \cite{Piran-1} that this would give rise to a situation similar to that we would normally expect to observe if such an underdense region was attributed both a negative gravitational mass\index{negative gravitational mass} and a positive inertial mass\index{inertial mass}, given that an underdense region would not only appear to repel surrounding positive masses, but would also attract other underdense regions.

Now, if this was the manner by which negative-mass or negative-energy matter interacted with itself and with positive-energy matter, then it would be impossible for the equivalence principle\index{equivalence principle} to apply in the way it is usually assumed to apply, and therefore it is usually assumed that this analogy between voids in an expanding matter distribution\index{voids in an expanding matter distribution} and negative mass\index{negative mass} objects cannot be valid in general. But what I will try to explain, in the following section, is that the hypothesis that negative-mass matter, if it exists, would have the properties we expect of voids in an expanding matter distribution, is in fact a basic consistency requirement, imposed by some of the most fundamental principles upon which all successful physical theories are based.

\section{The analysis\label{sec:3}}

As energy (more precisely the stress-energy tensor\index{stress-energy tensor}) replaces mass as the source of gravitational fields in relativity theory (where gravitational fields are actually represented by the curvature of spacetime) and as the only physically significant instance of negative energy is the one I have defined as negative action\index{negative action}, we can meaningfully discuss the problem of negative action by considering that of negative mass\index{negative mass}. The first discussion of negative mass in general relativity is Hermann Bondi's\index{Bondi, Hermann} 1957 paper \cite{Bondi-1} and for a long time it was the only paper about negative mass in modern gravitation theory. In it Bondi explains what is the currently held view on the subject of negative mass. He proposes four possible combinations of gravitational and inertial mass and exposes what is believed to be the behavior of matter endowed with such properties.

First there is the case where all mass is positive, which is ordinary matter. It is assumed that matter of this kind responds normally to non-gravitational forces, responds normally to gravitational forces, and produces attractive gravitational fields. Then there is the case where inertial mass\index{negative inertial mass} is negative and gravitational mass\index{gravitational mass} is positive. It is assumed that matter of this kind would respond perversely to all forces whether gravitational or non-gravitational and would produce attractive gravitational fields. There does not seem to be any justification for the existence of this type of matter and so I will not discuss this case any further.

Next is the case where inertial mass is positive and gravitational mass is negative. It is usually presumed that matter of this kind would respond normally to non-gravitational forces, would respond perversely to gravitational forces and would produce repulsive gravitational fields\index{repulsive gravitational field}. This would seem to imply that such masses would be submitted to mutual gravitational attraction, while they would repel ordinary matter and be repelled by it. This is the behavior one would expect from negative-mass matter if it behaved similarly to voids in an expanding positive-energy matter distribution\index{voids in an expanding matter distribution}. Finally, there is the case where both inertial and gravitational mass is negative\index{negative gravitational mass}. The conventional viewpoint, defended by Bondi, is that matter of this kind would respond perversely to non-gravitational forces, would respond normally to gravitational forces, and would produce repulsive gravitational fields.

Bondi then explains that in general relativity we are not left with as many choices of combinations, because the principle of equivalence\index{equivalence principle} requires that inertial mass\index{inertial mass} and passive gravitational mass\index{passive gravitational mass} be the same (in the Newtonian approximation passive and active gravitational masses\index{active gravitational mass} are also considered equal). I believe that he is right, not only because of the validity of general relativity, but simply because it appears contradictory to assume that what is positive is also at the same time negative. I believe that, even in a Newtonian framework, all mass (whether inertial or gravitational) should always be considered either positive or negative as a simple consistency requirement, even if it looks like we can separate its physical attributes into two categories. This is important, even if general relativity has completely superseded Newton's theory, because we can still obtain a lot of useful results by considering a Newtonian approximation\index{Newtonian approximation}.

Yet I also strongly disagree with Bondi. I believe that a negative-mass or negative-action object would actually have some of the properties that he incorrectly attributes to matter with both a positive inertial mass and a negative gravitational mass. I believe that a negative-mass object, the only possible type of negative-mass object, with both negative inertial mass and negative gravitational mass, would be repelled by normal positive masses (which it would also repel) while it would be attracted by other negative masses. I will explain what motivates this conclusion below, after I discuss the consequences of Bondi's problematic assumption concerning the behavior of negative masses.

While reading Bondi's paper you may notice how difficult it appears to be for him to justify the outcome of his own views on the subject of negative-mass matter. This is not surprising as one of the consequences of the existence of a negative-mass object (with both negative gravitational mass and negative inertial mass) that would obey Bondi's rules is a violation of the principle of inertia\index{principle of inertia!violation}. Bondi describes the motion of a pair of (very massive) particles, one is a normal particle with positive mass and the other is one of his negative-mass particles, which is assumed to produce repulsive gravitational fields, but to respond like ordinary matter to gravitational forces. As such, the negative-mass particle should be attracted to positive-mass particles, while it would also repel them gravitationally. Such a pair of particles, left alone in space and initially at rest in an inertial reference system\index{inertial reference system}, would spontaneously accelerate in one direction, the negative-mass particle chasing the positive-mass particle. That process would enable them to reach arbitrarily large opposite energies without any work being done on the whole system.

One can immediately notice that a strange, and I believe suspicious, aspect characterizes the phenomenon described here, beyond the fact that it seems highly unlikely that it could ever be observed. Indeed, why is it that it is necessarily the positive-mass particle that is chased and the negative-mass particle that pursues the other one? Shouldn't there be an equivalence between the viewpoint of the positive mass and that of the negative mass? How can we define attraction and repulsion in an absolute manner, such that one system always attracts other systems and another one always repels other systems, when the only property that distinguishes those systems is the sign of their mass? It does not just appear that there is some principle violated here, there is actually a violation of the notion that we can only define the properties of an interaction based on the effect it has on other systems and not with respect to some absolute notion of positivity and negativity\index{absolute positivity and negativity}. A mass cannot be said to be absolutely positive or to absolutely attract everything, because there would then be no reference attribute to which you could relate that arbitrary distinction.

This argument is so important that I will discuss it a little further. If the sign of mass\index{sign of mass} is to have any physical meaning then it must indicate that there can be a reversed or opposite value to a given mass and if there is a reversed value it can only be reversed relative to a non-reversed mass, that is, relative to a positive mass, for example, and to nothing else; mass cannot be reversed with respect to an absolute (non-relationally defined) point of reference, without any physical meaning. Consequently, if a repulsive gravitational field\index{repulsive gravitational field} results from taking a minus sign for the gravitational mass\index{gravitational mass} $m$ of its source, then it must be repulsive for positive masses and positive masses only. According to the conventional approach defended by Bondi, for a negative-mass system, the attractive or repulsive nature of the gravitational field of a positive-mass system would be the same as it is for a positive-mass system, given that it is defined in an absolute manner. As a consequence, he is forced to conclude that given the choice we have for active gravitational mass\index{active gravitational mass}, either all masses (positive and negative) will be attracted or all masses will be repelled.

But I believe that this just can't be true, because the sign of mass $m$ and more precisely of energy\index{sign of energy!relative physical attribute} $E$ (as it propagates in a given direction of time) is a purely relative physical attribute and does not relate to anything of an absolute nature. Failure to understand this means that we allow for the case where a positive mass attracts a negative mass (since all objects are attracted by it) while the negative mass repels the positive mass (since all objects are repelled by it) and under such conditions we should expect to observe the pair to move off with uniform acceleration to arbitrarily large velocities, in gross violation of the principle of inertia\index{principle of inertia!violation}. The mistake we do by following Bondi's approach is that we assume that the attractive or repulsive nature of a gravitational field is defined in an absolute manner, in the sense that it is not dependent merely on the sign of mass or energy of the object that is experiencing it. We shouldn't be surprised, then, that we end up with a theory that predicts the existence of phenomena so blatantly implausible that they defy common sense.

Clearly, if physical systems with negative mass\index{negative mass} or negative action\index{negative action} are assumed to exist, they cannot have the properties which Bondi assumes they possess. But why is it that the case which he rejected of a positive inertial mass combined with a negative gravitational mass\index{negative gravitational mass} appears to produce better agreement with the requirement that the sign of mass or energy be purely relative? Under this alternative proposal, from a purely phenomenological viewpoint there is an equivalence between positive- and negative-action matter, because particles of any one type are submitted to mutual gravitational attraction, while particles of opposite energy sign gravitationally repel each other. It then looks like repulsion and attraction \textit{could} be defined in a purely relative manner, but in fact they are still defined in an absolute manner and it is a coincidence if we obtain a model which appears, from a superficial point of view, to be invariant with respect to mass sign.

If we appear to obtain the desired results when keeping inertial mass positive while gravitational mass is reversed, it is simply because this is equivalent to assuming that the inertial mass of a negative-mass object is reversed a second time (from negative to positive) which would have the same effect as a reversal of the absolutely defined, equivalent gravitational field\index{equivalent gravitational field} attributable to acceleration relative to an inertial reference system\index{inertial reference system}. What I have come to understand is that while the inertial mass\index{negative inertial mass} of a negative-mass object is negative, just like its gravitational mass, the direction of the equivalent gravitational field due to its acceleration relative to an inertial reference system must by necessity be reversed compared to that which is experienced by a similarly accelerating positive-mass object, if this gravitational field is to itself be determined in a relational manner\footnote{
In a more extensive report \cite{Lindner-4}, published after the first version of the present one, I have explained more at length why it is that we can expect the equivalent gravitational field associated with acceleration to be reversed from the viewpoint of a negative-mass object, compared to what it is for a positive-mass object submitted to the same acceleration.}.
 As a consequence, a correctly described negative-mass object with negative gravitational mass and negative inertial mass would actually follow the rules traditionally assumed to be obeyed by a gravitating object with negative gravitational mass and positive inertial mass.

What I'm suggesting is that the gravitational field produced by a given source is not attractive or repulsive \textit{per se}, because this property depends on the signs of action of the interacting particles involved. The effects produced by a given gravitational field on a negative-mass object are actually different from those produced on a positive-mass object. There are in fact four possible situations which can arise when we limit ourselves to changing the signs of the interacting masses. First, the source of the field could have conventional positive mass density and the field be attractive because the test particle has normal positive mass. Next, the source of the field could have conventional negative mass density and the field be repulsive, again because the test particle has normal positive mass. Another possibility is that the source of the field could have conventional positive mass density and the field be \textit{repulsive} because the test particle has negative mass. Finally, the source of the field could have conventional negative mass density and the field be \textit{attractive}, still because the test particle has negative mass.

In such a context like-masses are gravitationally attracted to one another and masses of opposite signs gravitationally repel each other and this is all possible even though inertial mass is assumed to be reversed when gravitational mass is reversed. What's interesting is that this description remains valid even when positive mass is considered to be negative mass and negative mass is considered to be positive mass. We may say that the viewpoint under which what we normally call positive mass actually has positive mass is the natural viewpoint of what we normally consider to be a positive-action observer, while the viewpoint under which what we normally call positive mass actually has negative mass is the natural viewpoint of what we would normally consider to be a negative-action\index{negative action!observer} observer. I will show in the next section how these notions can be more precisely expressed from within the mathematical framework of a generalized gravitation theory\index{generalized gravitation theory}, based on conventional relativity theory.

However, we may want to retain a certain definition of inertial mass\index{inertial mass} that would clearly be related to the physically significant properties with which it is traditionally associated. This appears necessary once we realize that the energy or the mass could be zero in a given region of space, even in the presence of matter. For example, when two astronomical objects of sufficiently low mass densities with opposite mass signs are superposed, the total energy, as determined by the gravitational field they produce, would be zero. But because the energies of the two objects can be measured independently, the system does not have one unique vanishing inertial mass (it would take a relatively large amount of work to move it). We may define the \textit{absolute inertial mass}\index{absolute inertial mass} to be the measure of mass obtained by taking the absolute value of the mass of each independently evolving physical system. In the example of the two superposed objects with opposite masses, we would have zero total mass including inertial mass, but nonzero absolute inertial mass.

Now, it appears that a truly consistent notion of negative-mass or negative-energy matter, one that correctly takes into account both the principle of inertia\index{principle of inertia} and the principle of relativity\index{principle of relativity}, must have the \textit{apparent} consequence of enabling a distinction between gravity and acceleration. This tension between the concept of negative-energy matter that I propose and the equivalence principle\index{equivalence principle} can be easily pictured with the help of Einstein's gedanken, accelerated elevator experiment\index{accelerated elevator experiment}. What happens is that it seems that we would be able to tell when it is that we are simply accelerating far from any large mass and when it is that we are really standing still in the gravitational field of a planet. This is because near a planet or another large mass (of either type) positive- and negative-action test particles would accelerate in different directions (one upwards and the other downwards), while when the elevator is simply accelerating far from any large mass, both test particles would accelerate in the same direction, betraying the fact that the acceleration is `real'. Consequently, the principle that the effects of acceleration are totally equivalent to those of a gravitational field (the principle of equivalence), doesn't seem to be valid when we introduce a more consistent concept of negative-energy matter.

Faced with that prospect, one may be tempted to consider the view that it is better to sacrifice the very old and uncertain principle of inertia than anything we take for granted about the principle of equivalence, as the latter is the one principle upon which all of relativity theory and our modern concept of gravity is founded. But this viewpoint is hardly justifiable, given that the validity of the principle of equivalence is dependent on the validity of the principle of inertia. If a violation of the principle of inertia could happen, as would be the case in the presence of conventional negative-energy matter, then we cannot tell what would actually be the consequences of such a violation in a general-relativistic context. We don't even know if Bondi's prediction that the two particles should spontaneously accelerate in a given direction would hold under such conditions. The only argument we are left with, if we assume the validity of the conventional concept of a negative-mass or negative-energy system, is that such systems do not exist, so that we are not faced with the annoying and clearly unpredictable consequences of this proposal.

I believe that, in fact, the principle of equivalence\index{equivalence principle} and the principle of relativity\index{principle of relativity} on which it stands are not really threatened by my proposal. First, the equivalence principle always applies only in a limited portion of space. It is clear that we can tell that there is a real gravitational field when we consider a portion of space that is sufficiently large. If two elevators are suspended on two opposite sides of the Earth and we consider them together, it is obvious that even though observers in each elevator are free to believe that they are only accelerating far from any large mass, from the viewpoint of the ensemble of the two systems there is definitely a local force field directed towards the center of the planet. Even within a single elevator standing still on the surface of the Earth, freely falling (positive-mass) test particles would have a tendency to converge slightly towards one another, betraying the presence of a large mass nearby, attracting them towards its center.

In the end, one can say that the equivalence between the effects of acceleration and a gravitational field is still valid, only it applies separately for positive- and negative-energy particles (just as it applies separately for separate portions of space), because each of those two kinds of matter particle is to be attributed its own free-fall reference system\index{free-fall reference system} defined in relation to its mass sign. What's significant is that those two sets of free-fall reference systems and the two distinct measures of space and time intervals with which they are associated, can be related to one another by a simple, unique transformation, as all particles of one type share the same free-fall motion. But it is only when you consider two particles with different signs of action together, that you can tell the difference between the case of a uniform acceleration and that of a local gravitational field. Therefore, all particles with the same sign of energy still share the same local inertial reference system and this is all that is truly required for a general-relativistic gravitational field theory to apply that allows the sign of energy itself to be relativistically defined.

Now, it remains that one single reference system (or more exactly one set of reference systems) appears to be singled out by the combined behavior of the two types of test particles. This is the reference system relative to which both positive- and negative-action particles are not accelerating in the absence of local perturbations. It would then look like we can determine a state of absolute acceleration\index{absolute acceleration} relative to a metaphysical, non-accelerating reference system. This would seem to violate not only the principle of equivalence\index{equivalence principle}, but also the more general principle of relativity\index{principle of relativity} that motivates both the equivalence principle and my very introduction of unconventional negative-action matter (energy sign is a relative notion). However, this is not the case and in fact it should appear all the more natural, in the framework of general relativity theory, that some reference systems have unique properties.

As general relativity\index{general relativity theory} is a theory of gravitation, the local, inertial reference systems\index{local inertial reference systems} are defined by the effects of the surrounding matter. There is no doubt that there exists one very particular reference system in our universe, this is the reference system relative to which the majority of masses in the universe do not accelerate. One may call that system the global, inertial reference system\index{global inertial reference system}. It is while thinking about the necessity of a causal explanation to the existence of such a reference system that Einstein\index{Einstein, Albert} was led towards general relativity as a theory of gravitation. Once we recognize the validity of the general theory of relativity there should be no more mystery associated with the existence of the global, inertial reference system, as there may have been in Newton's age. That reference system is simply the outcome of the combined gravitational forces exerted by all masses in our universe (at least those which exert on a given object an influence that had the time to reach it since the Big Bang).

As it is, even far from any large mass, there remains the effect of the universe as a whole. For example, the reference system with respect to which we feel no rotation is the one which is not rotating relative to the average distribution of matter, that is, relative to the farthest galaxies. We are not surprised by the existence of such a privileged reference system\index{privileged reference system} and for the same reason we shouldn't look unfavorably at the possibility that there can exist one set of reference systems where at once positive- and negative-energy objects do not have accelerated motion when free from external non-gravitational forces. We are not faced with a reference system associated with absolute\index{absolute acceleration}, or non-relative acceleration, but only with a reference system in which the combined gravitational forces of all matter particles in the universe (with which an object actually interacts) imposes an absence of acceleration between positive- and negative-action objects.

So, we have an operational definition of the principle of equivalence\index{equivalence principle} that still works in a limited manner and that will enable me to generalize the existing mathematical framework of relativity theory\index{relativity theory!mathematical framework} so as to accommodate the presence of negative-energy matter. As well, I have provided arguments to the effect that the epistemological foundation of relativity theory, which basically consists in the principle of relativity\index{principle of relativity}, can only be satisfied in the context where negative-energy systems are governed by a theory according to which the sign of mass or energy does not have an absolute meaning. Before such a concept of negative-energy matter can become part of a generalized gravitation theory\index{generalized gravitation theory} that would provide an alternative framework to tackle the outstanding problems of cosmology, however, it is necessary to explain how the various difficulties it raises can be satisfactorily resolved.

There is one major issue, in particular, that must be faced before one can admit the possibility that there exists negative-action or negative-energy matter obeying the requirement of relational definition of physical attributes\index{requirement of relational definition of physical attributes}. The very basic difficulty we encounter upon the introduction of this kind of matter has to do with the energy\index{energy of interaction fields} of the fields responsible for the interaction between positive- and negative-action particles. It appears, in effect, that the energy of the field mediating such an interaction would not be well-defined, given that it is impossible to tell whether it is positive or negative (with respect to a given direction of time). Normally the energy of a field associated with a repulsive interaction between two particles, for example the energy of the electromagnetic field between two electrons, must be positive, while the energy of a field associated with attraction between two particles, for example the energy of the electromagnetic field between an electron and a positron, must be negative. As I mentioned before, we cannot isolate this energy from that of the source particles, but nevertheless, in all cases we deal with normally, this contribution to the total energy is well-defined.

The case of gravitation doesn't seem to require any special treatment. From the viewpoint of positive-energy observers, when two positive-action particles are attracted towards one another the contribution of the gravitational field to the energy of the system must be negative. When two negative-action particles are attracted towards one another or bound together in a single system, the contribution of the gravitational field to the energy of the system should be positive, as we can expect from the symmetry of this situation with the one involving two positive-action particles and as required for a relational definition of the sign of energy\index{sign of energy!relational definition} or action. But what happens in the case of the interaction of a positive-action particle with a negative-action particle?

The fact that what's positive energy for one observer could be considered negative energy for an observer of opposite energy sign doesn't change anything here. From the viewpoint of any observer, given that we are expecting a repulsive interaction, the energy of the gravitational field would have to be positive, if a positive-action particle is involved, but it would need to be negative if a negative-action particle is involved. It appears that the energy of the interaction field\index{energy of interaction fields} simply cannot be defined in a unique and non-ambiguous manner and there is a similar problem for all the other hypothetical interactions, attractive or repulsive, between positive- and negative-action particles. This is the most serious problem we must face upon the introduction of a more consistent concept of negative-energy matter in physical theory, but as I will explain, it can be turned into a clear advantage.

I believe that the difficulty, here, merely has to do with the fact we fail to recognize that there can be no direct interactions, mediated by interaction bosons\index{interaction bosons}, between positive- and negative-action particles. This suggestion may appear absurd at first; how could we have any indication whatsoever concerning the existence of negative-action matter if it doesn't interact with ordinary matter? It turns out, however, that even though this proposal amounts to proscribe most interactions, it still leaves a possibility for some sort of indirect gravitational interactions\index{indirect gravitational interaction} between positive- and negative-energy objects. It is true, however, that the other interactions are completely absent between opposite-action particles and this is justified by the impossibility to define either the energy sign of the fields which would mediate those interactions or the strength of any \textit{indirect} non-gravitational interaction\index{indirect non-gravitational interaction!strength} between those particles. This conclusion is very significant, from an astronomical viewpoint, as it actually means that a positive-energy observer cannot directly observe negative-energy matter.

To explain how it is possible to have a theory of the gravitational interaction between positive- and negative-energy matter, despite the absence of interaction between opposite-action particles, it will be necessary to consider the previously discussed analogy between the gravitational dynamics of positive and negative masses and that of voids in an expanding, uniform matter distribution\index{voids in an expanding matter distribution}. Now that I have explained that the only possible form of negative-mass or negative-energy matter would have properties traditionally attributed to matter with a negative gravitational mass\index{negative gravitational mass} and a positive inertial mass\index{inertial mass} (even though inertial mass is actually also negative), the significance of this analogy becomes clearer.

I previously explained, in effect, that it is well understood already that a void in an expanding, uniform distribution of positive-energy matter distribution would contribute to accelerate the local rate of expansion of the surrounding positive-energy particles, \textit{as if} the void was exerting a gravitational repulsion\index{gravitational repulsion} on those particles. But it must be clear that there is no (repulsive) gravitational field on the boundary of such a void in the matter distribution (this is implied by Birkhoff's theorem\index{Birkhoff's theorem} in the context where the density of matter energy is, in effect, null inside the void) and that it is merely because the rate of expansion of the positive-energy matter is not slowed down around the location of the void, due to the absence of gravitational attraction by the positive-energy matter that is missing, that the surrounding matter is allowed to expand more rapidly, as if it was submitted to some gravitational repulsion.

This conclusion would remain valid even if a uniform distribution of negative-energy matter was present throughout space that would make the density of energy negative (instead of null) inside a void in the positive-energy matter distribution, because, as I will explain below, what is implied by the absence of interaction between positive- and negative action particles is that no gravitational force of any kind can exist that would be exerted by a \textit{uniform} distribution of negative-energy matter on positive-energy matter. It is not possible, therefore, to conclude that when positive-energy matter is absent in a region of space, some gravitational repulsion\index{gravitational repulsion} may be exerted by the uniformly distributed negative-energy matter that is still present in this void, that would normally be compensated by the gravitational attraction of the positive-energy matter that is missing. Even in the presence of a homogeneous distribution of negative-energy matter it is still appropriate to assume that no repulsive gravitational field\index{repulsive gravitational field} can exist on the boundary of a void in the positive-energy matter distribution, that would produce a gravitational force on the surrounding positive-energy matter, and of course the same conclusion would be valid for negative-energy matter in the presence of a void in a uniform negative-energy matter distribution.

When I discussed the occurrence of negative energies in quantum field theory\index{quantum field theory}, however, I mentioned the existence of an experiment, first described by Hendrik Casimir\index{Casimir, Hendrik}, that enabled the measurement of a secondary effect of the absence of positive energy from the vacuum, attributable to the physical constraints exerted by the presence of two parallel mirrors\index{parallel mirrors in vacuum} placed a very small distance apart. I explained that such an absence of positive-energy from the vacuum implies that the energy density is actually negative in the small volume between the mirrors. But if we may, in effect, measure energy to be negative in a certain region of space where positive energy is missing from the vacuum, then there is no reason why we could not consider that negative energy states\index{negative energy states} in general are equivalent, in a certain way, to a local absence of positive energy from the vacuum, if from a phenomenological viewpoint there is no distinction between those two situations. In such a context, it may be appropriate to assume that the presence of negative-energy matter itself is equivalent to an absence of positive energy from the vacuum.

Now, in the context where we would also assume that positive-energy matter is equivalent to a local absence of \textit{negative} vacuum energy, it would follow that the interaction of positive-energy matter with the uniformly distributed, negative portion of vacuum energy\index{vacuum energy!negative portion} is actually an interaction of this negative portion of vacuum energy with itself and therefore there is no reason to assume that such a uniform distribution of negative vacuum energy\index{negative vacuum energy!uniform distribution} would exert no gravitational force on positive-energy matter, like a uniform distribution of negative \textit{matter} energy. But even if the uniform, negative portion of vacuum energy would seem to exert an equivalent (repulsive) gravitational force\index{repulsive gravitational force} on positive-energy particles, it would not follow that the negative vacuum energy that is present in a local void in the positive portion of vacuum energy would produce a repulsive gravitational force on the surrounding positive-energy matter, as one may expect to occur in the context where the gravitational force attributable to the positive portion of vacuum energy no longer compensates that which is attributable to the negative portion of vacuum energy, because, as I explained above, negative-energy matter particles themselves, as local voids in positive vacuum energy, cannot exert any gravitational force on positive-energy particles.

What I have come to understand is that, if negative-energy matter appears to exert a repulsive gravitational force\index{repulsive gravitational force} on surrounding positive-energy matter, this can only be a consequence of the fact that, the positive vacuum energy that goes missing\index{missing positive vacuum energy!negative-energy matter} locally, as a result of the presence of negative-energy matter, no longer contributes to balance the gravitational attraction exerted on this positive-energy matter by the surrounding positive portion of vacuum energy. When two maximum contributions with opposite signs exist for vacuum energy\index{vacuum energy!maximum positive and negative contributions}, it is necessary to assume, in effect, that in the absence of any positive or negative energy object, that is to say, in the absence of \textit{local} voids in the negative and positive portions of the uniform distribution of vacuum energy (respectively), no gravitational force would be exerted by vacuum energy on a positive-energy particle.

The equilibrium of gravitational forces\index{equilibrium of gravitational forces} that exists under such conditions, however, is not attributable merely to the presence of two opposite contributions of equal maximum magnitudes to the energy of the vacuum, but also to the absence of local inhomogeneities in one or another of those two opposite-energy distributions. When a void is present locally, in the positive portion of vacuum energy, this equilibrium is broken, not because the gravitational force exerted by the negative vacuum energy that is still present in the void is no longer compensated by that which would otherwise be exerted by the missing positive vacuum energy, but because the gravitational forces exerted on a positive-energy particle by the uniform portion of the surrounding \textit{positive} vacuum energy distribution no longer cancel out locally, while the gravitational forces exerted by the uniform portion of the surrounding negative vacuum energy distribution still cancel out locally, as if they were actually inexistent.

Thus, when positive vacuum energy is missing\index{missing positive vacuum energy} locally, it is as if the portion of the surrounding uniform distribution of positive vacuum energy above that which is present in the void and which would otherwise exert no gravitational force on either positive- or negative-energy particles, can now exert some kind of attractive gravitational force on positive-energy particles, even when the average density of vacuum energy remains null outside the void.

It is important, therefore, to differentiate between the equilibrium of gravitational forces that is attributable to the presence of two uniformly distributed, opposite contributions to the energy of the vacuum\index{vacuum energy!uniformly distributed opposite contributions}, which are allowed to cancel out on the cosmic scale (thereby possibly giving rise to a zero cosmological constant\index{cosmological constant!zero} that does not affect the rate of expansion of matter), and the local equilibrium of gravitational forces\index{equilibrium of gravitational forces} exerted on positive-energy particles, which arises from the uniformity of the \textit{spatial} distribution of positive vacuum energy and which may be affected by the absence of positive vacuum energy in a limited portion of space (the gravitational force exerted on positive-energy particles as a result of a local absence of negative vacuum energy being merely the outcome of an ordinary interaction between particles with the same positive sign of energy).

What we can actually expect, therefore, is that a local void in the uniform, positive portion of vacuum energy, which is equivalent to the presence of a local concentration of negative-energy matter, would gravitationally repel a positive-energy particle, not because the positive-energy particle is gravitationally repelled by the negative vacuum energy that is present in the void, but because the positive vacuum energy that is missing\index{missing positive vacuum energy} does not exert the gravitational attraction it would otherwise exert in the direction of the void, while the surrounding positive portion of vacuum energy still exerts an attractive gravitational force\index{attractive gravitational force!outward-directed} directed outward from the center of the void. This is possible, despite the fact that the uniform distribution of vacuum energy may have a null density outside the void, because when positive vacuum energy is missing locally, from a uniform distribution of vacuum energy that is already null, the gravitational attraction exerted on a positive-energy particle in the direction of the void is weaker than it would be even in the total absence of any energy, which means that it is actually pointing in a direction opposite that in which it would be pointing if it was attributable to the presence of positive vacuum energy in the void.

Given that there could be no gravitational force on positive-energy particles, arising from a void in vacuum energy\index{void in vacuum energy}, if equal amounts of positive and negative vacuum energies were missing at the same time in this void, while such a situation would only differ from that in which only positive vacuum energy is missing\index{missing positive vacuum energy} due to the fact that there would be no negative vacuum energy in the void, then it is necessary to assume that it is the surrounding uniform distribution of positive vacuum energy\index{positive vacuum energy!uniform distribution} that is exerting a gravitational force on a positive energy particle when the only energy that is missing is positive vacuum energy, because when negative vacuum energy is missing\index{missing negative vacuum energy!positive-energy matter} as well in the void, the absence of forces is not due to an absence of gravitational repulsion\index{gravitational repulsion} by the negative vacuum energy that is missing, but to the presence of an attractive gravitational force by the positive-energy matter which is located \textit{inside} the void and which exists as a result of this local absence of negative vacuum energy.

It must be clear that, despite what one may be tempted to argue, Birkhoff's\index{Birkhoff's theorem} theorem doesn't imply that it is impossible for the surrounding uniform distribution of positive vacuum energy\index{uniform distribution of vacuum energy} to be the cause of the gravitational force that pulls a positive-energy particle away from a region where positive vacuum energy is missing, because even though the spherical symmetry that characterizes the distribution of vacuum energy that exists in the presence of a single spherical void in the positive portion of vacuum energy\index{void in positive vacuum energy} does imply that the only solution to the gravitational field equations is the Schwarzschild solution\index{Schwarzschild solution}, as when a void in the uniform positive-energy \textit{matter} distribution is considered, in the present case the density of energy inside the spherical void is not null, but rather negative and under such conditions it is possible for a non-zero gravitational field to be present on the boundary of the void.

But even if, under certain circumstances, there may be an equivalence between a local imbalance in the sum of attractive gravitational forces\index{imbalance of attractive gravitational forces} attributable to the uniform distribution of positive vacuum energy and what would appear to be a gravitational repulsion\index{gravitational repulsion} exerted on a positive-energy particle by negative-energy matter, we are nevertheless always dealing with gravitational attraction. There is no question that it is the uncompensated gravitational attraction\index{uncompensated gravitational attraction} of surrounding positive vacuum energy that is responsible for the apparent gravitational repulsion which would be exerted on a positive-energy particle by a void in the otherwise uniform distribution of positive vacuum energy\index{uniform distribution of vacuum energy}. As I explained above, it is clearly as a consequence of the fact that positive vacuum energy is missing in the direction where the void is located, while the positive vacuum energy present in the opposite direction still exerts its gravitational pull, that there exists a net force directed away from the void.

Concerning the effects which I'm suggesting should be attributed to missing positive vacuum energy\index{missing positive vacuum energy}, we may ask to what extent such a void may actually be considered physically significant, in the sense of being merely an anomaly in an otherwise uniform distribution of vacuum energy\index{uniform distribution of vacuum energy}? Given that it must be the surrounding positive vacuum energy that exerts the \textit{outward}-directed attractive gravitational force\index{attractive gravitational force!outward-directed} that would be experienced as a gravitational repulsion by positive-energy particles, it follows that, as we consider voids of increasingly larger sizes, there may come a point when there would be no positive vacuum energy left outside the void to produce the uncompensated gravitational attraction\index{uncompensated gravitational attraction} that must exist to produce the equivalent repulsion.

This would happen in the presence of a uniform negative-energy matter distribution, given that the void in the positive-energy portion of the vacuum\index{void in positive vacuum energy!whole universe extent} which is equivalent to the presence of this matter would actually extend to the whole universe. This void would have been present in the vacuum from the very beginning of the universe's history and would not have developed through the production of a compensating overdensity of positive vacuum energy in its environment. In such a case it would no longer be possible to assume the existence of an uncompensated gravitational pull on positive-energy particles from the surrounding positive vacuum energy, because indeed there would be no vacuum energy with higher positive density outside the void to generate the attraction.

But if the indirect gravitational force\index{indirect gravitational force} exerted on positive-energy particles by voids in the positive-energy portion of the vacuum\index{void in positive vacuum energy} constitutes the only form of gravitational interaction between positive- and negative-energy matter, then it follows that a homogeneous distribution of negative-energy matter exerts no gravitational force at all on positive-energy matter, which means that positive-energy matter interacts with negative-energy matter only in the presence of inhomogeneities in any of the two matter distributions.

In an originally homogeneous matter distribution, however, a negative-energy matter overdensity can only form if a compensating underdensity also forms in its surroundings, which can be expected to produce an oppositely directed gravitational field, given that the energy that is the source of the gravitational fields produced by inhomogeneities in this matter distribution must be conserved when such a void forms. Therefore, one must conclude that, on the cosmological scale, the presence of an average density of negative-energy matter has absolutely no effect on the gravitational dynamics of positive-energy matter (and vice versa). This would mean, in particular, that the rate of universal expansion\index{rate of expansion!positive-energy observers} experienced by positive-energy observers is not influenced by the presence of negative-energy matter and similarly that the rate of expansion\index{rate of expansion!negative-energy observers} experienced by negative-energy observers is not affected by the presence of positive-energy matter.

If one can expect that an underdensity in a uniform distribution of negative-energy \textit{matter} would produce an attractive gravitational field, from the viewpoint of positive-energy particles, this is because, in the presence of such an inhomogeneity, the density of gravitationally attractive, positive vacuum energy would be larger than its average value locally, while there would be no gravitational force on positive-energy particles from the presence of a homogeneously distributed underdensity in positive vacuum energy which is equivalent to the presence of a homogeneous distribution of negative-energy matter. A negative-energy matter overdensity, on the other hand, can be expected to produce a repulsive gravitational field\index{repulsive gravitational field}, because under such conditions the density of gravitationally-attractive, positive vacuum energy would be smaller than its average value locally, thereby allowing the surrounding positive vacuum energy to exert an uncompensated gravitational attraction\index{uncompensated gravitational attraction} directed away from the center of the structure.

It is absolutely necessary to assume that a void\index{void in negative-energy matter distribution} in the uniform distribution of negative-energy matter, would produce an attractive gravitational force on positive-energy particles, because otherwise, as we consider voids of increasingly larger sizes in the negative-energy matter distribution, there would come a point when the void would occupy the whole observable universe, but such a void would still provide no contribution to the density of gravitationally-attractive, positive vacuum energy, while a totally uniform distribution of vacuum energy with a maximum positive density, as would exist in the absence of any negative-energy matter, must exert a gravitational force on positive-energy particles.

But it is also necessary to assume that a negative-energy matter overdensity would produce a repulsive gravitational field\index{repulsive gravitational field}, from the viewpoint of positive-energy particles, because if it was not the case, then the formation of a void\index{void in negative-energy matter distribution} in the negative-energy matter distribution would give rise to a violation of the conservation of energy\index{conservation of energy!violation}, given that this void must itself be the source of an attractive gravitational field, which means that if the local excess of negative-energy matter which must be produced in the surroundings of this void, as a result of its formation, was not giving rise to a repulsive gravitational field that did not exist originally, then no compensation would occur, as if positive energy could be created out of nothing.

As I mentioned above, however, the conclusion that such indirect interactions\index{indirect gravitational interaction} should exist between opposite-energy objects only applies to gravitation, because even if there must be an equivalence between non-gravitational charges\index{missing non-gravitational charges} missing from a neutral charge distribution\index{neutral charge distribution} and the presence of opposite-sign charges and even if it may be necessary to assume that the presence of matter is in fact equivalent to both missing energy and missing non-gravitational charges from zero-point vacuum fluctuations\index{zero-point vacuum fluctuations} (in order to avoid having to conclude that ordinary positive-action particles themselves, as voids in negative vacuum energy\index{void in negative vacuum energy}, have no charges), an additional difficulty would emerge if we were to assume that inhomogeneities in the distribution of negative-energy matter charges exert electromagnetic and other non-gravitational forces\index{non-gravitational forces} on charged positive-energy matter particles.

As I will explain in section \ref{sec:5}, opposite-energy objects do not share the same metric properties of spacetime\index{metric properties of spacetime} and this means that it is not possible to define the strength of an electric force\index{electric force strength} in a unique and consistent manner, when it is assumed that this force is attributable to an interaction between opposite-action particles, just like it is not possible to define the sign of energy of the field that would need to mediate such an interaction, while those differences would give rise to genuinely distinct effects (of a non-metric kind) from the viewpoints of opposite-energy observers. Therefore we have no choice but to recognize that such forces cannot exist, even in the presence of inhomogeneities in the distributions of positive- and negative-energy matter charges. It is merely the fact that gravitation is the interaction that determines the very metric properties of spacetime that allows the strength of the indirect gravitational forces\index{indirect gravitational force!opposite-energy objects} between opposite-energy objects which arise in the presence of inhomogeneities in the positive- and negative-energy matter distribution to be well-defined, unlike is the case for non-metric interactions\index{non-metric interactions}.

To return to the case of gravitation, it emerges that if an overdensity in the negative-energy matter distribution was to become large enough over a \textit{limited} region of space, then a spherical event horizon\index{event horizon} could form around it that would not merely prevent negative-energy matter from escaping the gravitational field of the object, but that would also prevent positive-energy matter from crossing its surface from outside. It would be as impossible for a positive-energy observer to explore the inside of such an object as it is impossible for the same observer to get out of an ordinary positive-energy black hole\index{black hole}. Such a negative-energy black hole\index{negative energy!black hole} would need to radiate negative thermal energy\index{negative thermal energy} for the exact same reason that a positive-energy black hole radiates positive thermal energy\index{thermal energy}. Indeed, given that the relation between the heat\index{heat} absorbed $dE$ and the variation of entropy\index{entropy variation!black hole} $dS$ which is provided by the first law of thermodynamics\index{thermodynamics!first law} $dS=dE/T$ (as applies to positive-energy black holes\index{black hole!thermodynamics} in the absence of changes to their angular momentum and electric charge) would also need to apply to those negative-energy black holes, then it follows that the temperature\index{temperature!negative} $T$ of the object would be negative, because its entropy would decrease in the course of such a process, while it would need to increase as the black hole absorbs negative-energy matter.

But given that, from the viewpoint of the alternative concept of negative-energy matter developed here, a positive-energy black hole would not be allowed to absorb negative-energy matter, just like a negative-energy black hole would be prevented from absorbing positive-energy matter (due to the overwhelming gravitational repulsion to which this matter would be submitted as it approaches the event horizon of the object), then it is not possible, as it would from a conventional viewpoint, to produce violations of the second law of thermodynamics\index{thermodynamics!violations of second law} by throwing negative-energy matter into a positive-energy black hole in small amounts. If negative energy states\index{negative energy states} are to be considered a true possibility, then the fact that the conventional concept of negative-energy matter would allow such violations of the second law of thermodynamics (given that negative-energy matter is usually assumed to be gravitationally attracted by positive-energy matter), while the alternative approach proposed in this report would forbid them, constitutes a strong motive to recognize that this latter proposal is more appropriate.

Now, I have argued above that once we recognize the validity of the general theory of relativity\index{general relativity theory} there should be no more mystery in relation to the existence of a unique global, inertial reference system\index{global inertial reference system} relative to which at once positive- and negative-energy objects do not accelerate, if it can be assumed that this reference system is determined by the gravitational forces attributable to the large-scale matter distribution\index{large-scale matter distribution}. But in the context of the above discussed developments, it emerges that if the matter distribution is homogeneous and isotropic on the largest scale, as observations appear to require, then only positive-energy matter can contribute to determine the global inertial reference system relative to which positive-energy objects have no acceleration in the absence of external forces, while only negative-energy matter can contribute to determine the global inertial reference system relative to which negative-energy objects have no acceleration, because a homogeneous distribution of negative-energy matter exerts no gravitational force on positive-energy objects and vice versa.

This conclusion is much more significant than it may perhaps appear to be, because if it had been the case that the uniform distribution of negative-energy matter was exerting a compensating gravitational force on positive-energy objects, then there would be no way for the phenomenon of inertia\index{inertia!phenomenon} to emerge, because it would be as if there was no matter on the largest scale, as all matter energies would add-up to zero (at least initially, in the first instants of the Big Bang, as I will explain in section \ref{sec:6.3}), which means that there would arise no inertial gravitational force\index{inertial gravitational force}, regardless of the state of acceleration of an object relative to the uniform large-scale matter distribution.

In the previous section I also discussed the case of negative energy as it arises in bound systems\index{bound systems}. I mentioned that such systems are physically different from the sum of their parts. For one thing, they may, in some cases, have lower energy after they have formed than there was energy in the isolated component subsystems from which they formed (even when we consider only the energy of mass\index{energy of mass}). To balance the energy budget, one must assume that the energy of the attractive field\index{energy of attractive field} of interaction is negative. This energy cannot be measured independently from the energy of the whole bound system, but it does contribute to it in a well-defined manner. Yet those very common physical systems were never found by any experiment to violate the principle of equivalence\index{equivalence principle} in any way.

What this means is \textit{not} that there is a problem with the hypothesis that negative-mass objects do not respond perversely to non-gravitational forces, but rather simply that one cannot consider that the interaction field with its negative energy contributes independently to the inertial properties of the system, which would result in the system having an \textit{absolute} inertial mass\index{absolute inertial mass} (as defined above) greater than its gravitational mass\index{gravitational mass}. Thus, when we are considering a single bound system of positive total energy, whose negative binding energy\index{negative binding energy} cannot be independently observed or measured, consistency dictates that we also cannot measure an independent contribution to the absolute inertial mass of the system by this field of interaction and of course this is also valid for negative-energy systems in the presence of positive energy attributable to a binding force field.

Before concluding this section, I would like to discuss another aspect of the problem of negative energy states\index{negative energy states!problem of}. I mentioned in the previous section that we can, and in fact we must, picture an antiparticle as really just an ordinary particle that reverses its energy to go backward in time, as when a particle reverses its momentum to move backward in position space. As such, antiparticles may be considered to merely be the same particles, observed from a different perspective. The distinction between a positive-action particle and a negative-action\index{negative action!particle} particle, however, is more significant, because a negative-action particle propagates a negative energy forward in time or a positive energy backward in time. One significant consequence of this distinction is that opposite-action particles cannot directly interact with one another through the exchange of interaction bosons\index{interaction bosons}, as I have explained earlier in this section.

But under such conditions one must conclude that a positive-energy particle cannot turn into a negative-energy particle on a continuous particle world-line\index{continuous particle world-line} without also changing its direction of propagation in time\index{direction of propagation in time} (as when a positron annihilates with an ordinary electron), while the same positive-energy particle cannot reverse its direction of propagation in time without reversing its sign of energy, because if this was allowed, then it would mean that from the conventional, forward-in-time viewpoint\index{forward-in-time viewpoint}, the particle that would have existed before the reversal could interact with the particle in the state it would occupy after the reversal, while those two particles would have opposite action signs and therefore are not even allowed to directly interact with one another.

This is another theoretically derived conclusion with many interesting, observable consequences. Indeed, if this constraint was not valid there could arise processes of annihilation\index{annihilation!opposite-action particles} of positive-action particles with negative-action particles that would leave no energy behind (as could have happened during the first instants of the Big Bang, when the densities of positive- and negative-action matter were required to be equally high, for reasons I will discuss in section \ref{sec:6.3}). But there could also arise processes in which pairs of opposite-action particles would be created out of nothing\index{creation out of nothing} in unlimited amounts. The principle of conservation of charge\index{conservation of charge} is of no help here, as among negative-action particles there are both positive- and negative-charge particles, that is, there are negative-action\index{negative action!antiparticles} antiparticles, just like there are positive-action antiparticles. An electron could always annihilate with the appropriate negative-action counterpart that has opposite electrical charge, if it was not for the constraint discussed above.

What emerges from those considerations is that it is not necessary to take into account any interaction vertex\index{interaction vertex!mixed action signs} with mixed action signs when calculating probability amplitudes\index{probability amplitude} for elementary particle transitions\index{particle transitions}. As a result no positive-action particle is allowed to decay into a negative-action particle by emitting a positive-action photon or another such interaction boson\index{interaction bosons}, because this would require the negative-action particle so produced to have interacted with this positive-action photon at the event in spacetime where it would have been produced. Thus, it seems that a certain constraint exists, from the viewpoint of the alternative approach advocated here, that would prevent electrically charged positive-energy particles from radiating energy while falling into the `lower' negative energy levels predicted to exist by quantum field theory\index{quantum field theory}. But this constraint would \textit{not} prevent a particle in a negative-energy state from reaching arbitrarily large negative energies, by absorbing negative-energy\index{negative energy!radiation} radiation.

This could be a problem from a conventional viewpoint, according to which the sign of energy\index{sign of energy!absolute meaning} has an absolute meaning, which allows an absolute direction on the energy scale\index{energy scale!absolute direction} to be singled out as being that of lower energies. But the difficulty no longer exists once we recognize the necessity to adopt a relational definition for the sign of energy\index{sign of energy!relational definition}. From this more consistent viewpoint, negative-energy observers would be allowed to consider the `lower', more negative energies, below the zero level, to actually be higher energies, while they would be allowed to consider the `higher', less negative energies, near the zero level, to actually be lower energies, in the sense that negative-energy particles would have a natural tendency to be drawn towards those lower energies as time goes on. This is due to the fact that the entropy\index{entropy!of matter} of matter is larger when negative energy is contained in a larger number of particles with smaller negative energies, for the same reason that entropy is larger when positive energy is contained in a larger number of particles with smaller positive energies.

Those conclusions are all made unavoidable by the essential requirement that there should be no absolute distinction between positive- and negative-energy particles, so that if positive-energy particles decay to less positive energy levels, then negative-energy particles should necessarily decay to less negative levels of energy, under similar conditions.

The outcome of the progress achieved in the latest portion of this section is that it becomes possible to assume that negative-energy\index{negative energy!states} states can be occupied by matter without giving rise to all sorts of instabilities and without interfering with the description of processes involving positive-action particles which is provided by ordinary quantum field theory\index{quantum field theory}. It would be like considering that there exists an additional theory, similar to the current one, but concerned only with negative-action systems. This relative independence of the processes described by the two equivalent theories guarantees that the near perfect agreement of the current theory with experimental data isn't jeopardized.

\section{The postulates\label{sec:4}}

Eleven basic principles or postulates help define the alternative concept of negative-energy matter developed in the previous section. They were used to derive the generalized gravitational field equations\index{generalized gravitational field equations} that will be introduced in the following section and which constitute the foundations of the solutions which will be proposed in section \ref{sec:6} to the cosmological constant\index{cosmological constant!problem} problem and other related issues in classical cosmology\index{classical cosmology}.

The first principle is the most fundamental and a recognition of its validity opens the way for a derivation of all the other results.
\begin{quote}
\textbf{Principle 1}: The distinction between a positive-action particle and a negative-action particle (propagating negative energy forward in time or positive energy backward in time) can only be defined by referring to the difference or the equality of the sign of action of one particle in comparison with that of another, so that this sign of action or energy has no absolute meaning.
\end{quote}
Another rule applies to the Newtonian concept of mass, but is necessary to derive the modified general relativistic gravitational field equations which will be discussed in the next section. It states that:
\begin{quote}
\textbf{Principle 2}: When mass is reversed from its conventional positive value (as a result of a reversal of the sign of action\index{sign of action!reversal}), both gravitational mass\index{gravitational mass} and inertial mass\index{inertial mass} are reversed and together become negative.
\end{quote}
This requirement is equivalent to assume that there is indeed only one physical property to which we may refer to as being that of mass, but it does not have the consequences one usually expect it would have, as it actually means that negative-mass objects do not respond perversely to applied forces.

The third principle is both theoretically and experimentally unavoidable.
\begin{quote}
\textbf{Principle 3}: There can be no direct interactions, either gravitational or non-gravitational, mediated by the exchange of interaction bosons\index{interaction bosons}, between positive- and negative-action particles.
\end{quote}
Compliance with this principle means that positive-energy observers cannot directly observe negative-energy matter (and vice versa).

Another important result emerges in the context where it must be assumed that there exist two opposite contributions of maximum magnitude\index{opposite contributions of maximum magnitude!vacuum energy} to the energy of zero-point vacuum fluctuations\index{zero-point vacuum fluctuations!energy}. This result simply states that:
\begin{quote}
\textbf{Principle 4}: The distribution of vacuum energy that surrounds a local void in an otherwise uniform distribution of positive vacuum energy\index{positive vacuum energy!uniform distribution} must give rise to uncompensated gravitational forces\index{uncompensated gravitational forces} which attract positive-action particles away from the center of mass of the void, as if the void itself was producing a repulsive gravitational force\index{repulsive gravitational force}.
\end{quote}

The following principle is probably the most decisive after principle 1, given that it is the result that allows the whole concept of negative-energy matter to have a significance despite the validity of principle 3 and the absence of direct interactions between positive- and negative-action particles. It states that:
\begin{quote}
\textbf{Principle 5}: The presence of negative matter energy in a given location is equivalent to the absence of an amount of energy of equal magnitude from the portion of zero-point vacuum fluctuations that provides a maximum positive contribution to the uniform distribution of vacuum energy\index{uniform distribution of vacuum energy!maximum positive contribution}.
\end{quote}
And by principle 1 it would also follow that the presence of positive matter energy is equivalent to the absence of an amount of energy of equal magnitude from the portion of zero-point vacuum fluctuations that provides a maximum negative contribution to the uniform distribution of vacuum energy\index{uniform distribution of vacuum energy!maximum negative contribution}.

But given that the void in the positive portion of vacuum energy\index{void in positive vacuum energy!homogeneous matter distribution} that is equivalent to a homogeneous distribution of negative-energy matter leaves no surrounding positive vacuum energy to produce an uncompensated gravitational attraction\index{uncompensated gravitational attraction} which, according to principle 4, would be equivalent to a gravitational repulsion\index{gravitational repulsion} arising from the void itself, then it is necessary to assume that:
\begin{quote}
\textbf{Principle 6}: The void of universal proportion\index{void in positive vacuum energy!universal proportion} in the uniform distribution of positive vacuum energy that is equivalent to a uniform distribution of negative-energy matter exerts no gravitational force on positive-energy matter and does not contribute to determine the curvature of space experienced by positive-energy observers.
\end{quote}
A similar limitation applies, which expresses the absence of gravitational forces on negative-energy matter arising from the void of universal proportion\index{void in negative vacuum energy!universal proportion} in the uniform distribution of negative vacuum energy that is equivalent to a uniform positive-energy matter distribution. As a result only the inhomogeneities present in the negative-energy matter distribution can affect the gravitational dynamics of positive-energy matter and vice versa.

An additional rule can be deduced from those stated above, in the context where it is recognized that the absence of direct interactions between positive- and negative-action particles does not prevent positive-energy matter, conceived as missing negative vacuum energy\index{missing negative vacuum energy}, from interacting with that very portion of vacuum energy.
\begin{quote}
\textbf{Principle 7}: Despite its negative sign the uniform, negative portion of vacuum energy\index{vacuum energy!negative portion} does exert a gravitational force on positive-energy matter.
\end{quote}
This deduction would also apply to the uniform, positive portion of vacuum energy\index{positive portion of vacuum energy} from the viewpoint of negative-energy matter.

Now, even though it is appropriate to assume that the presence of matter is in fact equivalent to both missing energy and missing non-gravitational charges\index{missing non-gravitational charges} from zero-point vacuum fluctuations\index{zero-point vacuum fluctuations} and while such a local absence of charges is equivalent to the presence of opposite-sign charges, it must be considered a consistency requirement to assume that:
\begin{quote}
\textbf{Principle 8}: Even in the presence of inhomogeneities in the distribution of negative-energy matter charges\index{negative-energy matter charge distribution!inhomogeneities}, no \textit{indirect} forces\index{indirect forces!non-gravitational interactions} can be exerted by negative-energy matter on positive-action particles which would result from a non-gravitational interaction between those particles and the portion of zero-point vacuum fluctuations\index{zero-point vacuum fluctuations!maximum positive energy contribution} that provides a maximum positive contribution to the uniform distribution of vacuum energy.
\end{quote}
And the same conclusion would apply for negative-action particles in the presence of inhomogeneities in the distribution of positive-energy matter charges.

The following principle must also apply, even when it is required that inertial mass\index{inertial mass} reverses when gravitational mass\index{gravitational mass} does.
\begin{quote}
\textbf{Principle 9}: The equivalence of the effects of acceleration and a gravitational field does not apply merely for matter in the same location, but only for matter with the same sign of mass or energy in the same location.
\end{quote}

Another rule must be obeyed in the context where a negative-mass object does not respond anomalously to non-gravitational forces despite the reversal of its inertial mass.
\begin{quote}
\textbf{Principle 10}: As the negative contribution of a field of interaction to the energy of a bound physical system\index{bound systems!negative contribution to energy} with overall positive energy cannot be independently and directly observed, only the diminished total energy of the bound system contributes to its (previously defined) absolute inertial mass\index{absolute inertial mass}.
\end{quote}
Again, this also applies in the case of a physical system with overall negative energy, concerning positive binding energy\index{positive binding energy}.

One last constraint is theoretically motivated, but is also required from a phenomenological viewpoint. It is the following:
\begin{quote}
\textbf{Principle 11}: A particle cannot reverse its direction of propagation in time\index{direction of propagation in time} on a continuous particle world-line\index{continuous particle world-line} without also reversing its energy and equivalently, a particle cannot reverse its energy on a continuous particle world-line without also reversing its direction of propagation in time.
\end{quote}
Here by `negative energy' I mean negative energy\index{negative energy!relative to direction of propagation in time} relative to the true (even though relationally defined) direction of propagation in time, as in the case of the positron as a negative-energy electron propagating its negative electric charge backward in time.

\section{The equations\label{sec:5}}

In this section I will lay out the mathematical framework of the generalized gravitation theory\index{generalized gravitation theory!mathematical framework} that emerges from the developments discussed in the preceding sections of the present report. The essential requirement that must be imposed on a formulation of the gravitational field equations\index{gravitational field equations} is that the gravitational field attributable to a local source shall not be attractive or repulsive depending merely on the sign of energy of this source. This can be satisfied by assuming that the gravitational field experienced by a negative-energy particle is different from the one experienced by a positive-energy particle in the same conditions. For this purpose it is necessary to assume that the sign of energy or action of the particle submitted to the gravitational field is an invariant property that may be chosen to be positive-definite, while it is the sign of energy of the source that constitutes the variable, observer-dependent physical property\index{observer-dependent physical property!sign of energy}.

In a Newtonian context this would mean that both the gravitational mass\index{gravitational mass} and the inertial mass\index{inertial mass} of the particle experiencing the field would be kept positive, while the equivalent gravitational field\index{equivalent gravitational field} due to acceleration far from any local source would be an invariant property. This is equivalent to assume that both the inertial mass and the equivalent gravitational field are reversed for a negative-energy particle with negative gravitational mass, as required if those attributes are to be defined in a relational manner. The crucial assumption here is that, while the gravitational fields attributable to local sources are observer-dependent physical properties\index{observer-dependent physical property!gravitational field}, the equivalent gravitational field associated with acceleration far from local sources is, for its part, an observer-independent property\index{observer-independent property!equivalent gravitational field}. Such a requirement can be satisfied when matter with a given energy sign only interacts with the matter distribution with the same sign of energy on the global scale.

Under such conditions, it is possible to write observer-dependent gravitational field equations\index{observer-dependent gravitational field equations} which allow the gravitational field to vary as a function of both the energy sign of the source and the energy sign of the particle submitted to it, so that only the difference or the identity between the energy sign of the source and that of the matter experiencing its gravitational field determines the repulsive or attractive nature of their interaction. In a general-relativistic context this means that opposite-energy observers may actually experience the metric properties of spacetime\index{metric properties of spacetime} in a different way. After I realized the necessity of such an approach, I became aware of the fact that at least one trivial instance of the general mathematical framework that can accommodate those requirements had already been proposed \cite{Petit-1} that simply amounted to allow for negative contributions to the stress-energy tensor\index{stress-energy tensor} of matter, while implicitly (but in a more general context, unsatisfactorily) trying to conform to the requirement of symmetry under an exchange of positive and negative energy signs.

But it soon became clear to me that the set of equations proposed could not satisfy all the constraints I have identified in the previous section as being essential to the elaboration of a consistent theory of gravitation compatible with the existence of negative-energy matter. One of the problems which affected all early formulations of such a bi-metric theory\index{bi-metric theory} is that, when vacuum energy is not null, they cannot really achieve the objective of providing a theory that conforms to the requirement of symmetry under exchange\index{requirement of exchange symmetry!positive- and negative-energy matter} of positive- and negative-energy matter. Also, the few predictions that they allowed to derive appeared to be incompatible with astronomical observations\index{astronomical observations}, in particular those which allow to determine the rate of expansion\index{rate of expansion} of space in the primordial universe. Anyhow, what was initially proposed is the following set of equations:
\begin{eqnarray}
R_{\mu\nu}-\frac{1}{2}g_{\mu\nu}R=\frac{8\pi G}{c^4} (T_{\mu\nu}-T^-_{\mu\nu}) \label{eq:05} \\
R^-_{\mu\nu}-\frac{1}{2}g_{\mu\nu}R^-=\frac{8\pi G}{c^4} (T^-_{\mu\nu}-T_{\mu\nu}) \label{eq:06}
\end{eqnarray}
Here and in what follows $G$ is Newton's constant, $c$ is the speed of light in a vacuum, and the Greek indexes $\mu$ and $\nu$ run over the four general coordinate system labels (assuming a metric with diagonal elements $+1$, $-1$, $-1$, $-1$ in an inertial coordinate system).

In those equations, the usual notation is used for the curvature tensors\index{curvature tensors} $R_{\mu\nu}$ and $R$ experienced by positive-energy observers and for the stress-energy tensor\index{stress-energy tensor} $T_{\mu\nu}$ of what we conventionally consider to be positive-energy matter, as measured by a positive-energy observer, while $-T^-_{\mu\nu}$ is the stress-energy tensor of what we usually consider to be negative-energy matter, as determined by a positive-energy observer. The curvature tensors\index{curvature tensors!negative-energy observers} experienced by negative-energy observers are for their part denoted $R^-_{\mu\nu}$ and $R^-$, while the stress-energy tensor of what we would conventionally consider to be negative-energy matter, as measured by a negative-energy observer, is here denoted $T^-_{\mu\nu}$ and of course $-T_{\mu\nu}$ is the stress-energy tensor of what we usually consider to be positive-energy matter, as determined by a negative-energy observer (who would measure a negative contribution by positive-energy matter to the density of matter energy).

The first of those two equations can be used to determine the geodesics\index{geodesics} followed by positive-energy particles, while the second equation determines the geodesics followed by negative-energy particles. Here, all stress-energy tensors would correspond with positive-definite energy densities\index{positive-definite energy densities} if it was not for the negative sign in front of the second stress-energy tensor on the right-hand side of each equation, which allows for observer-dependent negative contributions to the stress-energy tensor\index{stress-energy tensor!observer-dependent negative contributions} of matter. It is because there are two different measures for the gravitational field, associated with the two different ways by which the positive and negative contributions to the total energy of matter can be attributed, that there are two different equations for the gravitational field\index{gravitational field!two different equations}, instead of the single one that is usually considered. Otherwise, however, those equations are fairly conventional and were certainly the most simple and straightforward that one could derive for a bi-metric theory\index{bi-metric theory}.

To try to address the shortcomings I have identified above, I proposed, in the first version of the present report,
%% FOR BOOK VERSION: ...in the first version of the report \cite{Lindner-1} from which this book originates,
 equations which can be written in the following form:
\begin{eqnarray}
R^+_{\mu\nu}-\frac{1}{2}g_{\mu\nu}R^+=\frac{8\pi G}{c^4} (T^+_{\mu\nu}+\check{T}^-_{\mu\nu}-\hat{T}^-_{\mu\nu}) \label{eq:07} \\
R^-_{\mu\nu}-\frac{1}{2}g_{\mu\nu}R^-=\frac{8\pi G}{c^4} (T^-_{\mu\nu}+\check{T}^+_{\mu\nu}-\hat{T}^+_{\mu\nu}) \label{eq:08}
\end{eqnarray}
Here $R^+_{\mu\nu}$ and $R^+$ are simply the curvature tensors\index{curvature tensors} experienced by positive-energy observers, while $R^-_{\mu\nu}$ and $R^-$ are the curvature tensors\index{curvature tensors!negative-energy observers} experienced by negative-energy observers. Also, $T^+_{\mu\nu}$ is the stress-energy tensor\index{stress-energy tensor} of what we consider to be positive-energy matter, as measured by a positive-energy observer, while $\check{T}^-_{\mu\nu}$ is the stress-energy tensor associated with the (positive) measure of energy of negative-energy matter below its average cosmic density\index{average cosmic densities!negative-energy matter} (as determined by a positive-energy observer) and $-\hat{T}^-_{\mu\nu}$ is the stress-energy tensor associated with the (negative) measure of energy of negative-energy matter above its average cosmic density (as determined by a positive-energy observer). Similarly, $T^-_{\mu\nu}$ is the stress-energy tensor of what we would usually consider to be negative-energy matter, as measured by a negative-energy observer (and which actually provides a positive contribution to the energy of matter) while $\check{T}^+_{\mu\nu}$ is the stress-energy tensor associated with the (positive) measure of energy of positive-energy matter below its average cosmic density\index{average cosmic densities!positive-energy matter} (as determined by a negative-energy observer) and $-\hat{T}^+_{\mu\nu}$ is the stress-energy tensor associated with the (negative) measure of energy of positive-energy matter above its average cosmic density (as determined by a negative-energy observer).

This formulation of the generalized gravitational field equations\index{generalized gravitational field equations} allowed me to take into account the fact that there are two distinct categories of contributions to the total energy density experienced by positive-energy observers, one positive definite for all densities of positive-energy matter and one that can be either positive or negative depending on the value of energy density of negative-energy matter relative, not to the zero-energy ground state, but to the density of this negative-energy matter averaged over the entire volume of the observable universe.

Basically, what that means is that the energy measures of the second category of contributions experienced by a positive-energy observer are shifted from the conventional zero level of energy\index{zero-energy level} toward a lower (more negative) energy level below which energies are negative and above which energies are positive, up to a maximum value which is reached when no negative-energy matter is present at all in the considered location. This redefinition of the measures of energy\index{redefinition of energy measures} associated with what we consider to be negative-energy matter simply amounts to subtract the true, negative, average density of energy of this matter (add the absolute value of this density) from every measure of its energy density that contributes to determine the gravitational field experienced by what we consider to be positive-energy matter.

The refinement discussed here is justified by the fact that in the context where negative-energy matter is understood to consist of voids in the positive-energy portion of the vacuum, it must be assumed that the uniform portion of the distribution of negative-energy matter exerts no gravitational force on positive-energy matter (because no surrounding positive vacuum energy is present to produce an uncompensated gravitational attraction\index{uncompensated gravitational attraction}), which  requires considering the contributions of negative-energy matter to the stress-energy tensor\index{stress-energy tensor} experienced by positive-energy observers as being significant merely relative to the average density of negative matter energy (and therefore to actually be positive in the presence of underdensities in the otherwise uniform distribution of negative-energy matter), as I explained in section \ref{sec:3}.

The equations I had initially proposed also allowed to express the fact that a similar requirement exists for the contributions of positive-energy matter to the total stress-energy tensor experienced by negative-energy observers. But, still, I did not find the set of equations I had proposed completely satisfactory. I thought that the right solution should bring about a simplification of the gravitational field equations\index{gravitational field equations} (particularly those involving a non-zero cosmological constant\index{cosmological constant!non-zero}), while, visibly, the equations I had derived were not even as simple and elegant as the equations originally proposed by Einstein\index{Einstein, Albert}, despite the fact that, in their compact form, they were similar.

As I now understand, however, the equations I had proposed also fell short of meeting a certain mathematical requirement which I have come to appreciate as being essential to a consistent bi-metric theory\index{bi-metric theory} of gravitation of the kind I sought to develop. This became clear when a paper \cite{Hossenfelder-1} was published (whose existence came to my attention as a result of the fact that its author cited the original version of the present report)
%% FOR BOOK VERSION: ...the original version of the report on which this book is based)
 in which new equations were proposed, which introduced a further refinement to bi-metric theories, by not assuming that there is a unique predefined relationship between the metric properties of spacetime\index{metric properties of spacetime} experienced by positive-energy observers and those experienced by negative-energy observers in the same situation. As a consequence of this revised assumption, additional variables had to be considered that affected the contribution of negative-energy matter to the total stress-energy tensor\index{stress-energy tensor} experienced by positive-energy observers, or the contribution of what we consider to be positive-energy matter to the total stress-energy tensor experienced by negative-energy observers.

The equations proposed were the following, in which the additional factors are written in their explicit form, using my notation, and the quantities are now expressed in units where $c=1$ and $G=1/8\pi$:
\begin{eqnarray}
R^+_{\mu\nu}-\frac{1}{2}g_{\mu\nu}R^+=T^+_{\mu\nu}-\sqrt{\frac{g^{-+}}{g^{++}}}a_{\nu}^{\;\underline{\nu}}a_{\mu}^{\;\underline{\mu}} T^-_{\underline{\nu\mu}} \label{eq:09} \\
R^-_{\nu\mu}-\frac{1}{2}g_{\nu\mu}R^-=T^-_{\underline{\nu\mu}}-\sqrt{\frac{g^{+-}}{g^{--}}}a^{\mu}_{\;\underline{\mu}}a^{\nu}_{\;\underline{\nu}} T^+_{\mu\nu} \label{eq:10}
\end{eqnarray}
In this notation, tensors which refer to positive or negative stress-energies, as determined from the viewpoint of positive-energy observers, are given a plus or minus upper right index, respectively. Tensors which refer to measures of spacetime curvature or metric properties as observed by positive-energy observers are also given a plus upper right index, while tensors which refer to the same kind of measures as observed by negative-energy observers are given a minus upper right index. Also, when the distinct, ordinary or underlined Greek letter indexes used in Ref. \cite{Hossenfelder-1} are not explicitly present to show the nature of the tensor considered, I simply add another plus or minus index to the right of that which already characterizes this tensor to define it as an object associated with physical properties as they are experienced by positive- or negative-energy observers, respectively, and associated with their own specific metric. For all such tensors, therefore, the first plus or minus index refers to the matter or gravitational field that is observed while the second plus or minus index (to the right) refers to the matter that is observing.

In those equations, the decisive additional factors are the determinants of what the author calls the pull-overs\index{pull-overs}, which are the maps $g^-_{\nu\mu}$ and $g^+_{\underline{\mu\nu}}$ (originally denoted $h_{\nu\mu}$ and $g_{\underline{\mu\nu}}$), which we may also write as $\bm{g}^{-+}$ and $\bm{g}^{+-}$ in tensor form. Those determinants are written here as $g^{-+}=\det(g^-_{\nu\mu})$ and $g^{+-}=\det(g^+_{\underline{\mu\nu}})$, while $g^{++}=\det(g^+_{\nu\mu})$ is the determinant of the usual metric tensor\index{metric tensor} related to properties of positive-energy matter as observed by positive-energy observers and $g^{--}=\det(g^-_{\underline{\mu\nu}})$ is the determinant of the metric tensor related to properties of negative-energy matter as observed by negative-energy observers (the map $\bm{a}$ is simply used as a means to transform the metric $\bm{g}^{++}$ into the $\bm{g}^{-+}$ pull-over or the metric $\bm{g}^{--}$ into the $\bm{g}^{+-}$ pull-over). It is clear, therefore, that the pull-over $\bm{g}^{-+}$ is the map which allows to describe the metric properties\index{metric properties of spacetime} obeyed by negative-energy matter as they are observed by positive-energy observers, while the pull-over $\bm{g}^{+-}$ is the map which allows to describe the metric properties obeyed by positive-energy matter as they are observed by negative-energy observers (which justifies my notation).

To better illustrate the relationships involved we may rewrite those equations as:
\begin{eqnarray}
R^+_{\mu\nu}-\frac{1}{2}g_{\mu\nu}R^+=T^+_{\mu\nu}-\gamma^{-+}\sqrt{\frac{g^{--}}{g^{++}}}a_{\nu}^{\;\underline{\nu}}a_{\mu}^{\;\underline{\mu}} T^-_{\underline{\nu\mu}} \label{eq:11} \\
R^-_{\nu\mu}-\frac{1}{2}g_{\nu\mu}R^-=T^-_{\underline{\nu\mu}}-\gamma^{+-}\sqrt{\frac{g^{++}}{g^{--}}}a^{\mu}_{\;\underline{\mu}}a^{\nu}_{\;\underline{\nu}} T^+_{\mu\nu} \label{eq:12}
\end{eqnarray}
where $\gamma^{-+}$ is the absolute value of the determinant of the previously considered map of the metric properties of spacetime experienced by negative-energy matter as negative-energy observers measure them, to the metric properties of spacetime experienced by negative-energy matter as positive-energy observers measure them, and vice versa for $\gamma^{+-}$. We can then rewrite those gravitational field equations\index{gravitational field equations!compact tensor form} in compact tensor form by making use of those \textit{metric conversion factors}\index{metric conversion factors} as:
\begin{eqnarray}
\bm{G}^+=\bm{T}^{++}-\gamma^{-+}\bm{T}^{-+} \label{eq:13} \\
\bm{G}^-=\bm{T}^{--}-\gamma^{+-}\bm{T}^{+-} \label{eq:14}
\end{eqnarray}
where $\bm{G}^+$ is the Einstein tensor\index{Einstein tensor} $G^+_{\mu\nu}=R^+_{\mu\nu}-\frac{1}{2}g_{\mu\nu}R^+$ related to positive-energy observers, $\bm{G}^-$ is the similar Einstein tensor related to negative-energy observers, $\bm{T}^{++}$ is the stress-energy tensor\index{stress-energy tensor} of positive-energy matter as measured by positive-energy observers, $-\gamma^{-+}\bm{T}^{-+}$ is the stress-energy tensor of negative-energy matter as measured by positive-energy observers, $\bm{T}^{--}$ is the stress-energy tensor of negative-energy matter as measured by negative-energy observers and finally $-\gamma^{+-}\bm{T}^{+-}$ is the stress-energy tensor of positive-energy matter as measured by negative-energy observers.

As is apparent, however, the proposed equations were still of the conventional kind, in the sense that they did not allow to take into account the requirement that negative-energy matter be experienced as voids in the positive-energy portion of the vacuum\index{void in positive vacuum energy}, or the fact that a uniform distribution of negative-energy matter exerts no gravitational force on positive-energy particles (and vice versa for a uniform distribution of positive-energy matter from the viewpoint of negative-energy particles). The complexity of those equations and their lack of symmetry under exchange\index{requirement of exchange symmetry!positive- and negative-energy matter} of positive- and negative-energy matter can be made more apparent by explicitly adding a term for the observed positive value of (average) vacuum energy density\index{average vacuum energy density}:
\begin{eqnarray}
\bm{G}^+=\bm{T}^{++}+\bm{T}_{\Lambda}^{++}-\gamma^{-+}\bm{T}^{-+} \label{eq:15} \\
\bm{G}^-=\bm{T}^{--}-\bm{T}_{\Lambda}^{+-}-\gamma^{+-}\bm{T}^{+-} \label{eq:16}
\end{eqnarray}
This cosmological term\index{cosmological term} $\bm{T}_{\Lambda}^{++}=\Lambda\bm{g}^{++}$ would provide the measure of stress-energy associated with the positive density of vacuum energy $\rho_{\Lambda}^{++}=\Lambda$ measured on a global scale by a positive-energy observer (with $\Lambda$ as the positive cosmological constant\index{cosmological constant} experienced by such an observer), while an additional cosmological term $-\bm{T}_{\Lambda}^{+-}=-\Lambda\bm{g}^{--}$ would provide the value of stress-energy associated with the same vacuum energy, as measured by a negative-energy observer.

The density of vacuum energy measured by a negative-energy observer must be the opposite of that measured by a positive-energy observer if the sign of energy is to remain an observer-dependent physical property\index{observer-dependent physical property!sign of energy} (which justifies the presence of a minus sign in front of the $\bm{T}_{\Lambda}^{+-}$ tensor that enters the gravitational field equations\index{gravitational field equations} for negative-energy observers). But given that we are indeed dealing with vacuum energy, it would seem inappropriate to assign to this tensor the same metric conversion factor\index{metric conversion factors} $\gamma^{+-}$ as apply to measures of positive-energy matter density performed by negative-energy observers, even if the sum of all positive and negative contributions to the energy of the vacuum is a positive number, because, in principle, the magnitude of all such contributions is the same for positive- and negative-energy observers on the cosmological scale. Anyhow, it is apparent that once all relevant contributions to the stress-energy tensors\index{stress-energy tensor} are considered, the symmetry of the original equations is lost, as their form becomes dependent on the actual sign of the average density of vacuum energy.

In order that such a formulation of bi-metric theory\index{bi-metric theory} be allowed to at least meet the requirements concerning the contribution of negative-energy matter to the measure of stress-energy experienced by a positive-energy observer which I had already identified and which were not taken into account by the preceding authors, it is necessary, first of all, to replace the usual stress-energy tensors associated with the measures of energy of negative- and positive-energy matter made by observers of opposite energy sign with the following \textit{irregular stress-energy tensors}\index{irregular stress-energy tensors}, which provide the observed densities of energy of negative- and positive-energy matter relative to their average cosmic densities\index{average cosmic densities!positive- and negative-energy matter}:
\begin{eqnarray}
-\gamma^{-+}\bm{\widetilde{T}}^{-+}=-\gamma^{-+}(\bm{T}^{-+}-\bm{\bar{T}}^{-+}) \label{eq:17} \\
-\gamma^{+-}\bm{\widetilde{T}}^{+-}=-\gamma^{+-}(\bm{T}^{+-}-\bm{\bar{T}}^{+-}) \label{eq:18}
\end{eqnarray}
where $-\gamma^{-+}\bm{T}^{-+}$ and $-\gamma^{+-}\bm{T}^{+-}$ are the usual measures of stress-energy of negative- and positive-energy matter experienced by observers of opposite energy signs (relative to the conventional zero level of energy\index{zero-energy level}) and $-\gamma^{-+}\bm{\bar{T}}^{-+}$ and $-\gamma^{+-}\bm{\bar{T}}^{+-}$ are the \textit{average} values of stress-energy\index{average values of stress-energy} of negative- and positive-energy matter determined by observers with an opposite energy sign, based on their own measures of spatial volume.

In such a context, it appears that negative-energy matter would contribute negatively to the total measure of stress-energy experienced by a positive-energy observer only when the magnitude of its local energy density is larger than the magnitude of its average energy density. Otherwise negative-energy matter would actually contribute positively to the total measure of stress-energy experienced by a positive-energy observer, up to a maximum level which is fixed by the average density of negative-energy matter that is measured by such a positive-energy observer. It must be noted, however, that even though positive contributions to the energy density measured by positive-energy observers may occur which would be attributable to the presence of underdensities in the negative-energy matter distribution, we must nevertheless apply the metric conversion factor\index{metric conversion factors} $\gamma^{-+}$ to such energy measures, because they still relate to measurements regarding the density of negative-energy matter, which are subject to the same mapping relationships as apply to other (truly negative) measures of negative-energy matter density determined by a positive-energy observer.

A more appropriate set of gravitational field equations\index{gravitational field equations!more appropriate set} would, therefore, take into account the shifted origin of the measures of stress-energy\index{stress-energy!shifted origin} related to positive- and negative-energy matter as they are experienced by observers of opposite energy signs:
\begin{eqnarray}
\bm{G}^+=\bm{T}^{++}+\bm{T}_{\Lambda}^{++}-\gamma^{-+}\bm{\widetilde{T}}^{-+} \label{eq:19} \\
\bm{G}^-=\bm{T}^{--}-\bm{T}_{\Lambda}^{+-}-\gamma^{+-}\bm{\widetilde{T}}^{+-} \label{eq:20}
\end{eqnarray}
But clearly, for what regards simplicity, we appear to be no better off than with the previous set of equations. Something is still missing from those equations. At this point, it appears necessary to take a bold step forward and simply guess what the final form of the equations should be that would generalize the set of equations (\ref{eq:19}) and (\ref{eq:20}) I have just proposed, which would otherwise constitute the most accurate description of the gravitational dynamics of positive- and negative-energy matter. As I have been able to understand, the crucial step in this process consists in reconsidering the meaning of the cosmological terms\index{cosmological term}.

What must be understood, basically, is that if the results of the above analysis are right, then all energy is vacuum energy, either present or missing. An additional insight is then necessary, which consists in recognizing that the magnitude of the positive and negative values of vacuum energy density relative to which are measured the missing energies which are equivalent to the presence of negative- and positive-energy matter (respectively) is actually determined by the natural scale of quantum-gravitational phenomena\index{quantum-gravitational phenomena!natural scale}. To be more specific, if the presence of negative-energy matter is to be considered as equivalent to the presence of a void in the positive portion of vacuum energy\index{void in positive vacuum energy}, then locally we should observe a value of fluctuating vacuum energy density\index{vacuum energy density!maximum value} that would be decreased from the maximum value associated with the Planck scale\index{Planck scale} in just the same measure as that of the magnitude of energy of the matter that is present.

Let me thus introduce the generalized gravitational field equations\index{generalized gravitational field equations} which allow to fulfill all the requirements I have identified as being essential aspects of a classical theory of gravitation\index{classical gravitation theory} that solves the problem of negative energy states\index{negative energy states!problem of}. The formula, in all its beauty and simplicity, is the following:
\begin{equation}\label{eq:21}
\bm{G}^{\pm}=\bm{V}^{\pm}
\end{equation}
where $\bm{G}^{\pm}$ is the Einstein tensor\index{Einstein tensor} associated with the metric properties experienced by what we would usually consider to be positive- and negative-energy observers and $\bm{V}^{\pm}$ is the \textit{vacuum stress-energy tensor}\index{vacuum stress-energy tensor} associated with the measures of vacuum energy effected by those same positive- and negative-energy observers. The similarity with the compact form of Einstein's own equation is very clear, but it is also somewhat misleading, as the right-hand side of the equation proposed here is a more general object than the stress-energy tensor\index{stress-energy tensor} of matter which appeared in the original theory. I will now define it with various levels of precision and generality.

If we first consider the significance of the equation for a positive-energy observer, we would obtain the following equation:
\begin{equation}\label{eq:22}
\bm{G}^+=\gamma^{-+}\bm{V}^{++}-\bm{V}^{-+}
\end{equation}
in which $\bm{G}^+$ is, again, the Einstein tensor associated with the gravitational field experienced by positive-energy observers, but now the vacuum stress-energy tensor\index{vacuum stress-energy tensor!positive- and negative-energy portions} is decomposed into its positive- and negative-energy portions $\gamma^{-+}\bm{V}^{++}$ and $-\bm{V}^{-+}$ as they would be measured by such positive-energy observers, either directly or based on the curvature of space they produce. This is the most basic form of the proposed generalized gravitational field equations\index{generalized gravitational field equations!most basic form} for a positive-energy observer.

In accordance with what was explained above we would then obtain the next level of decomposition of the equations, in which the two opposite contributions to the energy of vacuum fluctuations\index{energy of vacuum fluctuations!opposite contributions} determined by positive-energy observers are given their explicit form:
\begin{equation}\label{eq:23}
\bm{G}^+=(\gamma^{-+}\bm{V}_P^{++}-\gamma^{-+}\bm{T}^{-+})-(\bm{V}_P^{-+}-\bm{T}^{++})
\end{equation}
where $\gamma^{-+}\bm{V}_P^{++}$ and $-\bm{V}_P^{-+}$ are the \textit{natural vacuum-stress-energy tensors}\index{natural vacuum-stress-energy tensors} associated with the maximum, positive and negative contributions to the energy density of zero-point vacuum fluctuations\index{zero-point vacuum fluctuations!maximum contributions} set by the Planck scale\index{Planck scale} (as determined by positive-energy observers) and from which are subtracted the missing vacuum energies\index{missing vacuum energies} $\gamma^{-+}\bm{T}^{-+}$ and $\bm{T}^{++}$ which are equivalent to the presence of negative- and positive-energy matter, respectively. What justifies the attribution of the previously introduced metric conversion factor\index{metric conversion factors} $\gamma^{-+}$ to the positive measure of vacuum stress-energy\index{vacuum stress-energy} in equation (\ref{eq:22}) and therefore, also, to the maximum \textit{positive} contribution to the energy of zero-point vacuum fluctuations\index{zero-point vacuum fluctuations!maximum positive energy contribution} in equation (\ref{eq:23}) is precisely the fact that this is the portion of vacuum energy relative to which the negative measure of matter energy $-\gamma^{-+}\bm{T}^{-+}$ is determined and which we can therefore expect to be directly experienced (other than through the gravitational interaction) only by this negative-energy matter, even though it does exert an observer-dependent gravitational force\index{observer-dependent gravitational force} on positive-energy matter as well.

Given that the previously introduced metric conversion factors are made necessary as a result of the absence of predetermined relationships between the metric properties of spacetime\index{metric properties of spacetime} experienced by negative-energy matter and those experienced by positive-energy matter, it is natural to assume that if the density of negative-energy matter itself cannot be directly observed by a positive-energy observer, then the positive measure of vacuum energy density relative to which this matter energy is defined cannot be directly determined either, because if this was not true, then by directly measuring the density of energy contained in this positive portion of vacuum energy\index{positive portion of vacuum energy}, a positive-energy observer could determine the density of negative-energy matter which is experienced by negative-energy observers. What must be understood is that the fact that this portion of vacuum energy density is positive should not be assumed to invalidate the conclusion that it cannot be directly experienced by positive-energy observers \textit{other than through the gravitational interaction}.

The preceding equation can then be rewritten in the following form, by making use of the previously defined, irregular stress-energy tensor\index{irregular stress-energy tensors} $-\gamma^{-+}\bm{\widetilde{T}}^{-+}$ (equation (\ref{eq:17})) that provides the actual measure of stress-energy of negative-energy matter experienced by positive-energy observers, which are only affected by local \textit{variations} in the density of negative-energy matter:
\begin{equation}\label{eq:24}
\bm{G}^+=\bm{T}^{++}-\gamma^{-+}\bm{\widetilde{T}}^{-+}+(\gamma^{-+}\bm{V}_P^{++}-\bm{V}_P^{-+})
\end{equation}
This allows one to isolate a term, in the generalized gravitational field equations\index{generalized gravitational field equations}, that can be associated with pure vacuum energy\index{pure vacuum energy} and that would be provided by the following tensor:
\begin{equation}\label{eq:25}
\bm{T}_{V}^+=\gamma^{-+}\bm{V}_P^{++}-\bm{V}_P^{-+}
\end{equation}
where the positive index attributed to this \textit{vacuum-energy term}\index{vacuum-energy term} (associated with the energy that is present in the vacuum independently from the contribution of ordinary matter) now merely denotes the purely conventional energy sign of the observer experiencing it, without referring to an actual energy sign of the vacuum fluctuations themselves, which could in principle be either positive or negative (without affecting the form of the equations) and which is determined solely by the metric conversion factor\index{metric conversion factors} provided by the previously discussed map of the metric properties of spacetime\index{metric properties of spacetime} experienced by negative-energy observers onto those experienced by positive-energy observers.

But as I will explain in section \ref{sec:6.2}, it must be assumed that positive-energy matter would remain unaffected by the average portion of a locally varying component of negative vacuum energy\index{negative vacuum energy!locally varying component}. Therefore, only a redefined measure of vacuum-energy would actually contribute to determine the gravitational field experienced by a positive-energy observer, which is given by the following \textit{irregular vacuum-energy term}\index{irregular vacuum-energy term}:
\begin{equation}\label{eq:26}
\bm{\widetilde{T}}_{V}^+=\bm{T}_{V}^{+}-(-\bm{\bar{T}}_{VM}^{-+})
\end{equation}
where $-\bm{\bar{T}}_{VM}^{-+}$ is the average value of stress-energy arising from all locally variable negative contributions to the energy of zero-point vacuum fluctuations\index{zero-point vacuum fluctuations!locally variable negative energy contributions}
 (which can be associated with the presence of dark matter),
 as measured by a positive-energy observer on a global scale, and which would not include the contribution provided by a negative cosmological constant\index{cosmological constant!negative} (as the uniform portion of a negative density of vacuum energy).

It is now possible to write the generalized gravitational field equations\index{generalized gravitational field equations!most explicit form} associated with positive-energy observers in their most explicit form as:
\begin{equation}\label{eq:27}
\bm{G}^+=\bm{T}^{++}-\gamma^{-+}\bm{\widetilde{T}}^{-+}+\bm{\widetilde{T}}_{V}^+
\end{equation}
The formal equivalence of this equation with the previously derived equation (\ref{eq:19}), at which I had arrived on the basis of considerations of a physical nature, is quite clear. But while one may be tempted to deduce from this that the irregular vacuum-energy term $\bm{\widetilde{T}}_{V}^+$ is equivalent to the cosmological term\index{cosmological term} $\bm{T}_{\Lambda}^{++}$ which is present in the original version of the gravitational field equations\index{gravitational field equations!original version}, this would not be entirely appropriate, because contrarily to the cosmological term (associated with the cosmological constant\index{cosmological constant!uniform and invariant energy contribution} $\Lambda$), which must by necessity provide a uniform and invariant contribution, the vacuum-energy term\index{vacuum-energy term!variable in space and time} can vary in space and incidentally also with time, given that it is determined by the locally variable, metric conversion factor $\gamma^{-+}$. Thus, only the contribution associated with the uniformly distributed portion of vacuum energy\index{vacuum energy!uniformly distributed portion} contained in the irregular vacuum-energy term\index{irregular vacuum-energy term!average value} at one particular time can be expected to be equivalent to the original cosmological term associated with the cosmological constant.

Now, if we consider the above equation in a cosmological context, then the irregular stress-energy tensor\index{irregular stress-energy tensors!zero average value} $-\gamma^{-+}\bm{\widetilde{T}}^{-+}$ presumably reduces to zero on average (as the overdensities of negative-energy matter cancel out the underdensities present in the same matter distribution), so that the relevant equations, for positive-energy observers, take the following form:
\begin{equation}\label{eq:28}
\bm{G}^+=\bm{T}^{++}+\bm{\widetilde{T}}_{V}^+
\end{equation}
which is similar to their conventional form, except for the fact that the cosmological term $\bm{T}_{\Lambda}^{++}$ is here replaced by the irregular vacuum-energy term\index{irregular vacuum-energy term} $\bm{\widetilde{T}}_{V}^+$ that may vary with position. But given that \textit{local} variations would presumably cancel out for the variable component of negative vacuum energy\index{negative vacuum energy!locally varying component} as well, on a very large scale, and given the (relative) success of current cosmological models for predicting the relevant features of our universe's history, then this outcome appears to be appropriate from an observational viewpoint.

We may then also write the following set of equations, which would provide the various levels of decomposition of the general equation (\ref{eq:21}) which apply from the viewpoint of negative-energy observers:
\begin{eqnarray}
\bm{G}^- & = & \gamma^{+-}\bm{V}^{--}-\bm{V}^{+-} \label{eq:29} \\
\bm{G}^- & = & (\gamma^{+-}\bm{V}_P^{--}-\gamma^{+-}\bm{T}^{+-})-(\bm{V}_P^{+-}-\bm{T}^{--}) \label{eq:30} \\
\bm{G}^- & = & \bm{T}^{--}-\gamma^{+-}\bm{\widetilde{T}}^{+-}+\bm{\widetilde{T}}_{V}^- \label{eq:31}
\end{eqnarray}
where $\gamma^{+-}\bm{V}_P^{--}$ and $-\bm{V}_P^{+-}$ are the natural vacuum-stress-energy tensors\index{natural vacuum-stress-energy tensors} associated with the maximum, negative and positive contributions to the energy density of zero-point vacuum fluctuations\index{zero-point vacuum fluctuations!maximum contributions} set by the Planck scale\index{Planck scale} (as determined by negative-energy observers) and $\bm{\widetilde{T}}_{V}^-=\bm{T}_{V}^{-}-(-\bm{\bar{T}}_{VM}^{+-})$ is the irregular vacuum-energy term associated with a negative-energy observer, with $\bm{T}_{V}^-=\gamma^{+-}\bm{V}_P^{--}-\bm{V}_P^{+-}$ as the locally variable (positive or negative) value of vacuum energy density which would be observed by such an observer in the absence of any ordinary positive- or negative-energy matter and from which is subtracted the average cosmological value of stress-energy $-\bm{\bar{T}}_{VM}^{+-}$ arising from all locally variable negative contributions to the energy of zero-point vacuum fluctuations\index{zero-point vacuum fluctuations!locally variable negative energy contributions} which are experienced by a negative-energy observer (who measures a negative contribution from what would be positive vacuum energy to a positive-energy observer) and which would not include the contribution provided by what would appear to such an observer as a negative cosmological constant\index{cosmological constant!negative} (as the uniform portion of a negative density of vacuum energy which would appear positive to a conventional positive-energy observer).

The last equation, as well the other two, are now manifestly symmetric with the corresponding equations associated with positive-energy observers under a reversal of the sign of energy, as I have argued should be required. But the most remarkable feature of those equations and the related equations for the gravitational field experienced by a positive-energy observer is that they are obtained from a very simple expression (the first of the three equations) according to which the gravitational field experienced by an observer with a given energy sign is determined merely by the appropriate measures of (positive and negative) vacuum energy densities. This equation alone allows to embody the essence of the emerging framework.

What's interesting, also, with this particular approach is the fact that it naturally accommodates the modified measure of negative stress-energy provided by the irregular stress-energy tensor\index{irregular stress-energy tensors} $-\gamma^{-+}\bm{\widetilde{T}}^{-+}$ which enters the decomposed gravitational field equations\index{gravitational field equations!decomposed} associated with a positive-energy observer, given that the presence of negative-energy matter is here explicitly equivalent to an absence of positive energy from the vacuum.

This compliance of the proposed gravitational field equations\index{gravitational field equations} may perhaps appear to be of secondary concern, given how negligible the average density of positive-energy matter really is in comparison with the density variations encountered under most circumstances when we are dealing with astronomical objects of interest, like stars or even galaxies. But, if it was not for the modified measure of negative stress-energy provided by the second term of equation (\ref{eq:27}), or the corresponding term from equation (\ref{eq:31}), complex hypotheses would have to be introduced concerning the variation in time of the ratio of the average cosmic densities of positive- and negative-energy matter in order to try to maintain the agreement of the proposed models with astronomical observations\index{astronomical observations} regarding the rate of expansion\index{rate of expansion} of ordinary, positive-energy matter, which is already predicted with (relatively) good accuracy by conventional cosmological models\index{conventional cosmological models}, when no negative-energy matter is assumed to be present initially. This particularity, therefore, constitutes a decisive advantage of the framework developed here for integrating negative-energy matter into classical gravitation theory\index{classical gravitation theory}.

Now, before discussing the consequences of this model for cosmology, I would like to add a few comments concerning how it is possible for the principle of conservation of energy\index{principle of conservation of energy} to be obeyed despite the fact that, according to the equations introduced above, one must assume that energy can be exchanged between positive- and negative-energy objects. A problem may appear to arise, in effect, due to the fact that opposite-energy objects repel one another, which may allow them to collide, in which case any loss of kinetic energy by one of the two objects would be gained by the other object. But given that the energy that is lost can be positive, while the energy that is gained would then be negative, it seems that a total variation of kinetic energy twice as large as those individual changes would happen, thereby apparently allowing a violation of the law of conservation of energy\index{conservation of energy!violation}. The same problem would seem to affect momentum conservation, in the context where the momentum direction of a negative-action particle must be opposite its direction of propagation in space\index{direction of propagation in space}.

However, from the viewpoint of the generalized gravitational field equations\index{generalized gravitational field equations} introduced above, any variation in the energy of matter arising from such an interaction would have to come from a variation in the energy of the gravitational field\index{energy of gravitational field} attributable to changes in the energy of the vacuum. What really happens, when a positive-energy object loses some of its kinetic energy as a result of a collision with a negative-energy object, therefore, is that the gravitational force exerted on this negative-energy object by the \textit{negative} vacuum energy that surrounds the void in zero-point fluctuations\index{zero-point vacuum fluctuations} that is equivalent to the presence of the positive-energy object allows the magnitude of its energy to grow larger, which means that, following the exchange, there is more gravitational interaction between all negative-energy matter in the universe and the negative-energy object whose presence is equivalent to this absence of positive vacuum energy, but given that the energy of the field associated with this additional gravitational attraction is positive, then there must be a positive change in gravitational energy that exactly balances the negative change in positive matter energy.

It must be clear, though, that the interaction of the negative-energy object is with the distribution of negative vacuum energy and that the described change in positive gravitational potential energy\index{gravitational potential energy!variation} is a result of the gravitational force exerted by the void in the negative portion of vacuum energy\index{void in negative vacuum energy} on that \textit{distinct} negative-energy object. Thus, it is not the loss of negative gravitational potential energy attributable to the interaction of the positive-energy object with all the matter with the same energy sign in the universe that compensates its own loss of positive energy, because if that was the case then it would mean that no interaction would be required to trigger those changes, which could then occur without any identifiable cause (for both positive- and negative-energy matter).

What allows me to conclude that a local increase in the amount of negative matter energy will give rise to a variation of gravitational potential energy that is attributable only to the increased amount of interaction of this negative-energy matter with the ensemble of \textit{negative-energy} matter in the universe is the fact that this additional negative matter energy only interacts with the ensemble of matter in the universe that shares the same sign of energy and cannot directly interact with positive-energy matter, as I previously explained. If that was not the case, then one would need to conclude that the variation of gravitational potential energy discussed above is compensated by an opposite variation that would arise from the increased amount of interaction of this negative-energy matter with the ensemble of \textit{positive-energy} matter in the universe, which means that the variation of positive matter energy itself would remain uncompensated.

Once this is understood, it becomes possible to conclude that following any indirect, gravitationally-repulsive interaction\index{indirect gravitationally-repulsive interaction} between opposite-energy objects, there is actually a variation in the total energy of matter\index{energy of matter!variation}, but that this change is necessarily accompanied by an opposite variation in gravitational potential energy\index{gravitational potential energy!variation} that is attributable to those local variations in the energy of matter. In a general-relativistic context one can expect that a similar conclusion would apply for those \textit{momentum} variations which take place as a result of gravitationally-repulsive interactions. What this means, from a practical viewpoint, for positive-energy observers, is that kinetic energy\index{kinetic energy!positive-definite quantity} is exchanged between opposite-energy objects as if it was a positive-definite quantity.

\section{The consequences\label{sec:6}}

It is a well-known fact that theoretical cosmology is currently facing a major crisis. One of the difficulties is the uncertainty that surrounds the nature of dark energy\index{dark energy} and dark matter\index{dark matter}. In the introduction to the current report I explained that a problem emerges when we are trying to explain the origin of the acceleration of cosmic expansion\index{cosmic expansion!acceleration} as being an outcome of the presence of energy in the vacuum, due to the fact that such an explanation seems implausible in the context where the measured value of average vacuum energy density\index{average vacuum energy density} which would need to exist under such conditions is much smaller than the value we would expect to observe if this parameter is in effect not perfectly null. But there is also an additional problem, which is usually considered to not be directly related to the cosmological constant\index{cosmological constant!problem} problem and which arises from the fact that despite all the efforts which were dedicated to identifying the nature of the cold dark matter\index{cold dark matter} particles which are thought to give rise to the missing-mass effect\index{missing-mass effect} around visible structures in the large-scale matter distribution\index{large-scale matter distribution}, no viable candidate has been found that could explain this phenomenon.

Another major difficulty of the current standard model of cosmology\index{standard model of cosmology}, which is usually believed to not be directly related to the cosmological constant problem, is that of explaining how it is possible for the present rate of expansion of matter on the cosmological scale to be set to the critical value associated with a flat space\index{flat space}, despite the fact that this observation appears to require an extremely precise adjustment of parameters in the initial Big Bang state. This flatness problem\index{flatness problem} was once thought to have been solved by inflation\index{inflation!theory} theory, until it became clear that this theory offers so much predictive freedom that it is nearly unfalsifiable. Given the difficulty to actually confirm the validity of inflation theory, I think that it is still appropriate to seek an alternative solution to the flatness problem. But if an alternative solution to that problem can be implemented that is theoretically and observationally viable, then it becomes necessary to recognize that the phenomenon of inflationary expansion\index{inflationary expansion} cannot provide an adequate solution to the horizon problem\index{horizon problem} either.

The horizon problem has to do with the fact that it is not possible to explain the uniformity of the very-large-scale distribution of matter energy in the early universe as being a consequence of smoothing processes\index{smoothing processes} that would obey the principle of local causality\index{principle of local causality}. I only became interested with the horizon problem after I began to work on another decisive issue, which is not always recognized as a problem for cosmology, but which also originates from the absence of a viable explanation for the smoothness of the primordial matter distribution\index{primordial matter distribution!smoothness}. This is the problem of the origin of the thermodynamic arrow of time\index{thermodynamic arrow of time!origin problem}, which is probably the most serious and insoluble difficulty currently facing cosmology. The developments which have been introduced in the preceding sections of this report will allow me to approach the horizon problem from a different viewpoint and will culminate in the elaboration of a plausible explanation for how it can be that a fully time-symmetric fundamental theory\index{time-symmetric fundamental theory} can conspire to enforce boundary conditions which give rise to irreversible evolution and the second law of thermodynamics\index{thermodynamics!second law}.

\subsection{Dark energy\label{sec:6.1}}

If we consider, first, the problem of dark energy\index{dark energy}, what the generalized mathematical framework for classical gravitation theory\index{classical gravitation theory!generalized mathematical framework} developed in the preceding section implies is that there is actually a whole new class of contributions to the energy of the vacuum from this portion of zero-point vacuum fluctuations\index{zero-point vacuum fluctuations} which directly interacts, other than gravitationally, with negative-energy matter and this contribution to the average density of vacuum energy\index{average vacuum energy density} can be expected to compensate the sum of all currently considered contributions, as a consequence of the requirement of symmetry under exchange of positive and negative energy matter. This can be expected to occur regardless of the details of the Grand Unified Theory\index{Grand Unified Theory} chosen to describe elementary particles and their interactions and independently from how the symmetry of such a theory is spontaneously broken at lower energies. Yet it appears that it is still possible, in such a context, for the cosmological constant to take on non-zero values.

In section \ref{sec:5} I proposed, in effect, that the measure of vacuum energy density $\bm{T}_{V}^+$ which exists in the absence of any matter and whose uniform portion would be associated with the cosmological constant\index{cosmological constant} measured by a positive-energy observer be defined as the sum of the natural vacuum-stress-energy tensors\index{natural vacuum-stress-energy tensors} $\gamma^{-+}\bm{V}_P^{++}$ and $-\bm{V}_P^{-+}$, which provide the maximum positive and negative values of energy density contributed by those portions of zero-point vacuum fluctuations that directly interact only with negative- and positive-energy matter (respectively), but which both exert a gravitational influence on positive-energy matter:
\begin{equation}\label{eq:32}
\bm{T}_{V}^+=\gamma^{-+}\bm{V}_P^{++}-\bm{V}_P^{-+}
\end{equation}
From that particular viewpoint the cosmological term\index{cosmological term} $\bm{T}^{++}_{\Lambda}=\Lambda\bm{g}^{++}$ that enters the original form of the gravitational field equations\index{gravitational field equations} associated with a positive-energy observer (with $\Lambda$ as the cosmological constant) must be understood to consist of the uniform portion of the locally variable vacuum-energy term\index{vacuum-energy term} $\bm{T}_{V}^+$, measured at a given epoch of cosmic time\index{cosmic time}.

Given that the $\gamma^{-+}$ metric conversion factor\index{metric conversion factors} is the mathematical object that allows to map the metric properties of spacetime\index{metric properties of spacetime} experienced by negative-energy observers onto those experienced by positive-energy observers, if the portion of zero-point vacuum fluctuations\index{zero-point vacuum fluctuations} that provides a maximum positive contribution to the density of vacuum energy is directly experienced only by negative-energy observers, then from the viewpoint of positive-energy observers the measure of energy density involved must be submitted to the same metric conversion factor as applies to measures of negative-energy matter density effected by those same observers.

The important point is that this would also be valid on the cosmic scale, where the scale factor\index{scale factor!observer-dependent parameter} must be considered an observer-dependent parameter, because only the average density of positive matter energy can influence the variation of the rate of expansion\index{rate of expansion!variation} measured by positive-energy observers, so that even if the initial expansion rate\index{rate of expansion!initial}, in the first instants of the Big Bang, was the same for positive- and negative-energy observers, it could have come to differ later on, thereby allowing the scale factors determined by opposite-energy observers to diverge. This means that, even on a global scale, the metric properties of spacetime could differ for opposite-energy observers and this is what would allow the value of the cosmological constant\index{cosmological constant}, which is dependent on the $\gamma^{-+}$ conversion factor to differ from zero.

What's implied by the appearance of the metric conversion factors\index{metric conversion factors} in the proposed definitions of the density of vacuum energy, therefore, is that the magnitudes of the maximum, positive and negative contributions $\gamma^{-+}\bm{V}_P^{++}$ and $-\bm{V}_P^{-+}$ to the energy density of the vacuum determined by a positive-energy observer can be made to differ, as a consequence of the fact that opposite-energy observers do not necessarily share the same metric properties of spacetime\index{metric properties of spacetime}, even on the global scale, where it can be expected that matter is homogeneously distributed. The rule is that when the scale factor is measured as being proportionately smaller by a positive-energy observer, the density of the maximum positive contribution to the energy of the vacuum\index{vacuum energy!maximum positive contribution}, which cannot be directly measured by such an observer, would be increased from the viewpoint of this observer, in comparison with the density of the maximum negative contributions to the energy of the vacuum\index{vacuum energy!maximum negative contribution}, which can be directly measured by the same observer, so that according to equation (\ref{eq:32}) above, the average density of vacuum energy\index{average vacuum energy density} would be positive and our positive-energy observer would measure a positive cosmological constant $\Lambda$.

I'm therefore allowed to predict that if the scale factors\index{scale factor} experienced by opposite-energy observers were exactly the same in the initial state\index{initial state!maximum matter energy densities} of maximum positive and negative matter energy densities, then space must have expanded at a smaller rate, from the viewpoint of positive-energy observers, during a certain portion of the universe's history, in comparison with the rate at which it expanded from the viewpoint of negative-energy observers. Yet given that a positive value of average vacuum energy density would contribute to accelerate the rate of expansion\index{rate of expansion!acceleration} of space determined by positive-energy observers, due to the contribution of its negative pressure\index{negative pressure}, while it would contribute to decelerate the expansion rate\index{rate of expansion!deceleration} determined by a negative-energy observer, again as a result of its negative pressure, then it can be expected that any difference that would develop between the measure of the scale factor determined by positive-energy observers and that which is determined by negative-energy observers, would eventually be reduced.

It may perhaps appear doubtless, therefore, that the universe could have evolved in such a way that the scale factor experienced by negative-energy observers could have become so much larger, in comparison with the scale factor experienced by positive-energy observers, that the average value of vacuum energy density\index{average vacuum energy density} which results from this divergence could have grown into a positive value that is much larger than the density of positive-energy matter experienced by positive-energy observers, whose magnitude would already be larger than the density of negative-energy matter experienced by negative-energy observers. But even when we assume that the initial value of the cosmological constant\index{cosmological constant!initial value}, in the very first instants of the Big Bang, must have been null (as I will suggest in section \ref{sec:6.3}) it is still possible for the magnitude of the average density of vacuum energy to grow larger at later times due to the fact that the average density of matter energy may not only change as a result of expansion, but also as a consequence of the early annihilation of matter with antimatter\index{early matter-antimatter annihilation}.

Thus, it would appear necessary to assume that a comparatively smaller density of negative-energy matter was left over by the early annihilation of matter with antimatter that took place in the first instants of the Big Bang, due to a less substantial violation of time reversal symmetry\index{time reversal symmetry!violation} by this negative-energy matter, compared to that which affected positive-energy matter. As a result, the deceleration of the rate of expansion\index{rate of expansion!deceleration} experienced later on, during the matter-dominated era\index{matter-dominated era}, by positive-energy observers, must have been substantially larger than that experienced by negative-energy observers, which allowed the scale factor\index{scale factor} determined by positive-energy observers to become significantly smaller than that which is determined by negative-energy observers, which, in turn, allowed the cosmological constant\index{cosmological constant!growth} to grow to larger positive values early on, despite the accelerating effect it exerted on the expansion rate\index{rate of expansion!acceleration} experienced by positive-energy observers and the decelerating\index{rate of expansion!deceleration} effect it exerted on that which was experienced by negative-energy observers\footnote{
%%BOOK VERSION:
%%In the extended version of the report \cite{Lindner-4} on which this book is based I have provided a viable solution to the problem of the origin of the asymmetry between matter and antimatter\index{matter-antimatter asymmetry!problem of origin|nn} which is based on the hypothesis that there can exist baryonic dark matter\index{baryonic dark matter|nn} particles carrying reversed bidirectional charges\index{reversed-bidirectional-charge particles|nn} (charges opposite those carried by ordinary matter particles forward or backward in time) and while this solution requires that there exist equal numbers of positive- and negative-bidirectional-charge particles with a given sign of action, it allows the total number of negative-action particles of any kind to be smaller than that of positive-action particles, as required by observations, including those regarding the variation of the cosmological constant\index{cosmological constant!variation|nn} which are discussed here.}.
In the extended version of this report \cite{Lindner-4} I have provided a viable solution to the problem of the origin of the asymmetry between matter and antimatter\index{matter-antimatter asymmetry!problem of origin|nn} which is based on the hypothesis that there can exist baryonic dark matter\index{baryonic dark matter|nn} particles carrying reversed bidirectional charges\index{reversed-bidirectional-charge particles|nn} (charges opposite those carried by ordinary matter particles forward or backward in time) and while this solution requires that there exist equal numbers of positive- and negative-bidirectional-charge particles with a given sign of action, it allows the total number of negative-action particles of any kind to be smaller than that of positive-action particles, as required by observations, including those regarding the variation of the cosmological constant\index{cosmological constant!variation|nn} which are discussed here.}.

Eventually the average density of vacuum energy began to decrease under its own influence, while the rate of expansion\index{rate of expansion!deceleration} experienced by positive-energy observers kept decelerating, but at an ever slower rate, a situation which persisted until vacuum energy\index{vacuum energy!dominance} became dominant over positive matter energy, at which point the expansion rate actually began to accelerate from the viewpoint of positive-energy observers. But given that we do observe the present average density of positive matter energy (both visible and dark) to be somewhat smaller than the current average density of vacuum energy\index{average vacuum energy density}, then it can be expected that the average, positive density of vacuum energy and the magnitude of the cosmological constant\index{cosmological constant!decreasing value} are currently being reduced to smaller values as a result of their own influence on the expansion rates\index{expansion rates!positive- and negative-energy observers} determined by positive- and negative-energy observers.

Those deductions would appear to agree with astronomical observations, which indicate that the dominance of the average density of vacuum energy over that of matter energy occurred only recently on the cosmic time scale\index{cosmic time scale}, given that the expansion of space is observed to be decelerating immediately before it began accelerating, in the most recent period of the universe's history. But even if the positive, average density of vacuum energy\index{average vacuum energy density} did not become dominant over that of positive matter energy until recent times, it must have had a significant influence on the early rates of expansion\index{early rates of expansion} of positive- and negative-energy matter determined by observers with the same sign of energy.

I believe that this is what allows the Hubble tension\index{Hubble tension} (concerning the disagreement between low-redshift estimates of the current value of the Hubble constant\index{Hubble constant} $H_0$ \cite{Burns-1} \cite{Riess-2} and estimates of the same parameter deduced from observations of the cosmic microwave background\index{cosmic microwave background} \cite{Ade-1}) to be eased, ultimately, because a substantially larger early value for the positive, average density of vacuum energy, similar in magnitude to the early density of baryonic positive-energy matter\index{baryonic positive-energy matter!early density}, would allow the largest perturbations in the temperature of cosmic microwave background radiation to occur on a smaller scale, from the viewpoint of a positive-energy observer (because it would take less time for the average matter density to become small enough that the decoupling of matter from radiation\index{decoupling of matter from radiation} is allowed to occur, as required for the cosmic microwave background to be released), while this appears necessary in order to raise the current value of the Hubble constant derived from the standard model of cosmology\index{standard model of cosmology} (based on measurements of those CMB temperature fluctuations) up to the larger value that is determined by the direct method.

The validity of the approach advocated here is not compromised by the fact that it once seemed that empirical data was perhaps favorable to the hypothesis that the cosmological constant\index{cosmological constant} has not changed much during the recent history of the universe, because, given the current smallness of the observed average value of vacuum energy density (compared to the natural scale of quantum-gravitational phenomena\index{quantum-gravitational phenomena!natural scale}), it is natural to expect that the current rate at which the cosmological constant\index{cosmological constant!current reduction rate} is being reduced, which is determined by the very magnitude of this average density of vacuum energy, would have remained too small to be detected, but more precise measurements may allow to reveal this variation, which is one clear prediction of the approach proposed here, that may allow to confirm its validity\footnote{
After I wrote the latest versions of this document it was found that some puzzling astronomical observations of baryon acoustic oscillations\index{baryon acoustic oscillations} \cite{AbdulKarim-1} could be resolved if it is assumed that the (average) density of dark energy\index{dark energy!average density} has, in effect, been slowly decreasing since the epoch when the cosmic microwave background\index{cosmic microwave background} was released.}.

To avoid confusion, however, it must be understood that a positive cosmological constant contributes to accelerate the rate of expansion\index{rate of expansion!acceleration} of space measured by positive-energy observers (which we may call the \textit{specific expansion rate}\index{specific expansion rates} of positive-energy matter) and not merely to accelerate the rate of expansion of positive-energy matter, because the same metric conversion factor\index{metric conversion factors} that is involved in determining the net value of average vacuum energy density\index{average vacuum energy density} also affects the measure of average negative-energy matter density determined by positive-energy observers, thereby allowing this matter density to remain unchanged compared with the average density of positive-energy matter determined by the same positive-energy observers (which we may call the \textit{specific density}\index{specific matter densities} of positive-energy matter).

Now, one may ask why it is that the scale factors\index{scale factor} and the specific rates of expansion experienced by observers with opposite energy signs were so similar in the very first instants of the Big Bang that they only began to differ significantly during the matter-dominated era\index{matter-dominated era}, as a result of the early annihilation of matter and antimatter\index{early matter-antimatter annihilation}, which appears to have been more complete for negative-action matter? Despite the fact that a divergence of the specific rates of expansion\index{specific expansion rates!divergence} would produce a larger magnitude of average vacuum energy density, which would tend to reduce this divergence, there is no doubt that the average value of vacuum energy density could have been much larger (into positive or negative territory) initially, even before the early annihilation of baryons and antibaryons\index{early baryon-antibaryon annihilation} had an effect on the specific rates of expansion of positive- and negative-energy matter, in which case its present magnitude would still be much larger than the measured value. What I will explain in section \ref{sec:6.3}, however, is that additional constraints actually exist which allow to predict that the initial value of average vacuum energy density must have been perfectly null.

\subsection{Dark matter\label{sec:6.2}}

What I would like to explain, first of all, is that additional, attractive gravitational forces, similar to those we normally attribute to ordinary cold dark matter\index{cold dark matter}, but which are not taken into account by current cosmological models, can be expected to naturally arise in the context of the generalized gravitation theory\index{generalized gravitation theory} introduced in the preceding sections of this report. I previously mentioned that gravitational forces, indistinguishable from those usually attributed to positive-energy dark matter, would arise in the presence of an underdensity in an otherwise uniform distribution of invisible negative-energy matter, because, from the viewpoint of positive-energy particles, this local absence of negative-energy matter is equivalent to the presence of additional, gravitationally-attractive, positive-energy matter. Such a phenomenon could, in principle, occur around visible, positive-energy matter overdensities, given that those overdensities would repel negative-energy matter and create underdensities in this negative-energy matter distribution that could potentially enhance the gravitational attraction of the positive-energy objects.

The magnitude of those additional gravitational forces, however, is limited by the finite value of the average density of negative-energy matter over which the underdensities are measured, which, at the present epoch at least, is much smaller than the average density of positive-energy matter in a typical galaxy. It is necessary, therefore, to recognize that the presence of negative-energy matter underdensities cannot contribute significantly to the observed missing-mass effect\index{missing-mass effect} around positive-energy objects at the present epoch. But one can expect that the presence of negative-energy matter underdensities did accelerate the process of structure formation\index{structure formation!process} in the positive-energy matter distribution at the epoch, in the remote past, when the average matter density was still relatively large and the negative-energy matter distribution homogeneous enough on the scale of the structures considered, because under such conditions it is possible for this additional gravitational attraction to be concentrated around positive-energy overdensities.

This is certainly a positive development, given that, despite all the progress which was achieved in the last decades, the currently favored theory of structure formation\index{structure formation!theory}, involving only positive-energy cold dark matter\index{cold dark matter!positive energy}, is still inadequate to model the formation of large-scale structures\index{large-scale structures} in the early universe, as witness the fact that observations (see in particular Ref. \cite{Labbe-1}) keep revealing the presence of well-developed galaxies and clusters\index{galaxies and clusters!unexpectedly large masses} of galaxies with masses much larger than expected, at increasingly larger redshifts, corresponding to an epoch when there shouldn't be any such structures according to current models.

But this doesn't mean that negative-energy matter overdensities were present as well, at the same epoch, on a similar scale, which could have produced stellar- or galactic-size underdensities in the positive-energy matter distribution, that would have similarly accelerated the growth of those negative-energy structures, because from an observational viewpoint it appears necessary to assume that no significant amount of baryonic negative-energy matter\index{baryonic negative-energy matter} survived the early period of matter-antimatter annihilation\index{early matter-antimatter annihilation}, while it can also be expected that dark matter overdensities have a tendency to form and to grow where baryonic matter overdensities with the same sign of energy are located, for reasons I will soon discuss.

Now, if one recognizes that the presence of negative-energy matter underdensities would never allow to explain a significant portion of the missing-mass effects\index{missing-mass effect} which are observed around visible positive-energy structures at the present epoch, then one must admit that there definitely exist additional contributions of unknown origin to positive matter energy in our universe. Faced with the undeniable evidence that a certain form of dark matter\index{dark matter} must exist, the normal reaction is to seek to identify a weakly-interacting massive particle\index{weakly-interacting massive particles} that might constitute a viable candidate for this dark matter. The problem is that all attempts at detecting and identifying a weakly-interacting particle that does not interact with ordinary matter \textit{only} through the gravitational interaction have failed.

What I have come to understand, however, is that a phenomenon, distinct from that which can be attributed to the presence of underdensities in the negative-energy matter distribution, but which can also be expected to arise only in the context of a theory of gravitation governed by the generalized gravitational field equations\index{generalized gravitational field equations} introduced in section \ref{sec:5}, may allow to explain most the missing-mass effects observed around visible structures in the positive-energy matter distribution, without requiring the existence of hypothetical cold dark matter\index{cold dark matter} particles.

In the context where the curvature of space\index{curvature of space!observer dependent property} arising from the presence of inhomogeneities in the matter distribution is an observer dependent physical property it must, in effect, be possible for the density of vacuum energy\index{vacuum energy density!local variations} to vary locally, in both space and time, given that the metric conversion factors\index{metric conversion factors} which determine the value of vacuum energy density must themselves be allowed to vary as a function of the difference between the metric properties of spacetime\index{metric properties of spacetime} experienced by positive-energy observers and those experienced by negative-energy observers. Thus, in the presence of a positive-energy matter overdensity, one can expect to observe a positive increase in vacuum energy density, above that which is attributable to ordinary matter itself (as voids in negative vacuum energy\index{void in negative vacuum energy}), because the measures of spatial volume around that overdensity would appear to be comparatively smaller for a positive-energy observer, given that from such a viewpoint space is contracted and time dilated by the presence of positive-energy matter, while space is dilated and time contracted under similar conditions, from the viewpoint of a negative-energy observer.

This would be due to the fact that when the volume of space is measured to be smaller by a positive-energy observer and larger by a negative-energy observer, locally, then the maximum positive contribution to the density of vacuum energy\index{vacuum energy!maximum positive contribution} $\gamma^{-+}\bm{V}_P^{++}$, which is experienced by a positive-energy observer only through the gravitational force it exerts on positive-energy matter, becomes larger, locally, than the maximum negative contribution\index{vacuum energy!maximum negative contribution} $-\bm{V}_P^{-+}$ which is directly experienced by the same positive-energy observer and any such difference must add up to that which arises from the observer-dependent measures of cosmological scale factor\index{scale factor}, as if we were dealing with an independent measure of stress-energy, similar to that of ordinary matter and independent from the usually considered cosmological term\index{cosmological term} $\bm{T}^{++}_{\Lambda}=\Lambda\bm{g}^{++}$ associated with the cosmological constant\index{cosmological constant}.

This phenomenon can be referred to as \textit{vacuum dark matter}\index{vacuum dark matter}, because it would have consequences similar to those we normally attribute to the presence of ordinary dark matter\index{dark matter} (such as weakly-interacting massive particles\index{weakly-interacting massive particles}), given that it would contribute to significantly increase the mass of any astronomical object present on a sufficiently large scale, without raising the density of baryonic matter\index{baryonic matter!density}\footnote{
%%BOOK VERSION:
%%In the extended version of the report \cite{Lindner-4} on which is based this book I have explained that it can be expected that there should also exist an additional component of baryonic matter\index{baryonic matter!dark|nn} that would by necessity be dark and whose average density should be exactly the same as that of ordinary baryonic matter, which would have for consequence to double the density of baryonic positive-energy matter\index{baryonic positive-energy matter|nn} that survived the early annihilation of baryons and antibaryons\index{early baryon-antibaryon annihilation|nn}, but given that this baryonic matter is dark, its presence would not affect current estimates of the relative abundance of light elements\index{relative abundance of light elements|nn} derived from Big Bang\index{Big Bang!nucleosynthesis|nn} nucleosynthesis, even though the additional effect it exerted on the early rate of growth of the \textit{average} density of vacuum energy\index{average vacuum energy density|nn} is essential to resolve the Hubble tension\index{Hubble tension|nn}.}.
In a previously cited report \cite{Lindner-4} I have explained that it can be expected that there should also exist an additional component of baryonic matter\index{baryonic matter!dark|nn} that would by necessity be dark and whose average density should be exactly the same as that of ordinary baryonic matter, which would have for consequence to double the density of baryonic positive-energy matter\index{baryonic positive-energy matter|nn} that survived the early annihilation of baryons and antibaryons\index{early baryon-antibaryon annihilation|nn}, but given that this baryonic matter is dark, its presence would not affect current estimates of the relative abundance of light elements\index{relative abundance of light elements|nn} derived from Big Bang\index{Big Bang!nucleosynthesis|nn} nucleosynthesis, even though the additional effect it exerted on the early rate of growth of the \textit{average} density of vacuum energy\index{average vacuum energy density|nn} is essential to resolve the Hubble tension\index{Hubble tension|nn}.}.

It would, therefore, appear that it is the fact that the local variations of vacuum energy density\index{vacuum energy density!local variations} which are involved here are correlated, under most circumstances, with the presence of local inhomogeneities in the distribution of baryonic matter\index{baryonic matter}, due to the fact that such inhomogeneities are usually required to trigger the development of local variations in the density of vacuum energy, that allows them to provide the long sought explanation of the missing-mass effect\index{missing-mass effect} as being a particular aspect of the phenomenon of dark energy\index{dark energy}. The crucial point, here, is that the positive vacuum energy which is produced, locally, by the curvature of space\index{curvature of space} attributable to the presence of a positive-energy matter overdensity must itself contribute to produce additional space curvature, similar to that which is produced by ordinary positive-energy matter, which, in turn, produces additional positive vacuum energy.

It must be clear, however, that the portion of dark matter\index{dark matter} attributable to local variations of vacuum energy density already existed in a macroscopically homogeneous form on the cosmic scale, before it accumulated around baryonic matter\index{baryonic matter!overdensities} overdensities and therefore it only grows around inhomogeneities as a result of such an accumulation, without violating the principle of conservation of energy\index{principle of conservation of energy}. What's significant as well is that, given that we are dealing here with \textit{local} variations in the distribution of vacuum energy, then it follows that, from a gravitational viewpoint, vacuum dark matter\index{vacuum dark matter} would be equivalent to the presence of ordinary matter, despite the fact that, as a particular form of vacuum energy\index{vacuum energy!dark}, it must be dark.

This allows one to expect that the gravitational attraction of positive vacuum-dark-matter energy would not be compensated by a gravitational repulsion\index{gravitational repulsion} attributable to the presence of negative vacuum-dark-matter energy\index{vacuum-dark-matter energy!negative}, on the cosmological scale, at least as long as this negative-energy dark matter is uniformly distributed on a global scale. But it also allows one to expect that positive vacuum-dark-matter energy\index{vacuum-dark-matter energy!positive}, unlike a positive cosmological constant\index{cosmological constant!positive}, should not exert a negative pressure\index{negative pressure} that would repel positive-energy matter, either locally or globally, because the average density of positive vacuum-dark-matter energy does not remain constant in an expanding universe.

The one crucial aspect that differentiates vacuum dark matter\index{vacuum dark matter} from ordinary matter is the fact that, if we are to conceive of ordinary negative-energy matter as missing positive vacuum energy\index{missing positive vacuum energy}, then we have no choice but to assume that what is missing from zero-point vacuum fluctuations\index{zero-point vacuum fluctuations} under such conditions is not just positive energy, but also positive or negative non-gravitational charges, as the missing virtual particles\index{missing virtual particles} which are equivalent to the presence of matter also carry charges. Those charges, which are missing in the electrically neutral vacuum\index{electrically neutral vacuum}, are equivalent to the presence of charges of opposite signs, which appear to be carried by the matter particles themselves.

The presence of vacuum dark matter, however, does not result from a local absence of virtual particles from zero-point vacuum fluctuations, that would need to be correlated with an absence of charge, but arises from the differing measures of spacetime metric properties\index{metric properties of spacetime} experienced by opposite-energy observers, which alter the relative densities of the maximum positive and negative portions of vacuum energy\index{vacuum energy!maximum positive and negative portions} without affecting the electrical (or non-gravitational) neutrality of the vacuum. Thus, while ordinary positive-energy matter consists of both missing negative vacuum energy\index{missing negative vacuum energy} and missing positive \textit{or} negative vacuum charge\index{missing positive or negative vacuum charge}, positive-energy vacuum dark matter is only equivalent to an excess of positive over negative vacuum \textit{energy} and must remain electrically neutral under all conditions (it never carries any non-gravitational charges). This is the only aspect that differentiate vacuum dark matter from ordinary matter and which explains that it is, in effect, dark.

What makes it possible for a maximum, invariant amount of energy to have already been present as positive vacuum-dark-matter energy\index{vacuum-dark-matter energy!positive} and independently, as negative vacuum-dark-matter energy\index{vacuum-dark-matter energy!negative}, on the largest scale, is the fact that, even though the distribution of matter and radiation energy in the early universe was very homogeneous from a macroscopic viewpoint (a hypothesis that is theoretically and observationally unavoidable, as I will explain in section \ref{sec:6.4}), positive- and negative-energy matter particles couldn't occupy the exact same positions in the initial state\index{initial state!maximum matter energy densities} of maximum positive and negative matter energy densities and this means that, on the quantum-gravitational scale\index{quantum-gravitational scale}, there existed variations of considerable magnitude in the density of both positive and negative matter energies, to which were associated equally large amounts of vacuum-dark-matter energy\index{vacuum-dark-matter energy}.

As the universe expanded and the average density of matter decreased, along with the average kinetic energy\index{average kinetic energy!matter and radiation} of matter and radiation particles, this macroscopically homogeneous distribution of vacuum dark matter\index{vacuum dark matter} spread into the available space, along with the rest of matter. But when the small-amplitude inhomogeneities which were present on a macroscopic scale in the matter distribution (including that of vacuum dark matter) began to grow, later on, as a result of gravitational instability\index{gravitational instability}, the density of vacuum dark matter was allowed to vary on a macroscopic scale, as the positive-energy portion of it became more concentrated in positive-energy matter overdensities and more rarefied in positive-energy matter underdensities, even though the matter itself already existed in diffuse form.

From an observational perspective, it would appear possible to confirm that dark matter\index{dark matter} is, for the most part, an outcome of local variations in the density of vacuum energy\index{vacuum energy density!local variations}, because currently available data indicates \cite{McGaugh-1} that there is a strong correlation, in general, between the gravitational acceleration\index{gravitational acceleration} attributable to the total amount of matter inside an orbit (say around the center of a galaxy) and the gravitational acceleration attributable to the baryonic matter\index{baryonic matter} inside that orbit\footnote{
This correlation would probably be even stronger if we were taking into account the presence of baryonic dark matter\index{baryonic dark matter|nn} particles with reversed bidirectional charges\index{reversed-bidirectional-charge particles|nn}, whose average density can be expected to equal that of visible baryonic matter\index{baryonic matter|nn} particles, as I explained in the previously cited report \cite{Lindner-4} I have published on the problem of time directionality\index{time directionality|nn}.}.
 But if the presence of dark matter must be considered to be an effect of the curvature of space\index{curvature of space} (attributable to the matter that is present in a region of space) on the local measures of vacuum energy density\index{vacuum energy density!local measures}, then it is natural to expect that the more gravitational acceleration that there is as a consequence of the presence of baryonic matter, the more distinct the metric properties of spacetime\index{metric properties of spacetime} experienced by opposite-energy observers must be that gave rise to the accumulation of vacuum dark matter around that particular location.

The conclusion that there must exist a relationship between the amplitude of the gravitational field attributable to visible positive-energy matter overdensities and the amplitude of the missing-mass effect\index{missing-mass effect} would also imply that, even within galaxies and clusters, dark matter should be more concentrated around the visible elements of the structure. While this result is certainly unexpected, it does agree with some relatively recent observations \cite{Meneghetti-1}, which indicate that there is a greater than expected concentration of gravitational lensing\index{gravitational lensing!concentration} around individual galaxies within clusters.

It is important to point out, however, that vacuum dark matter\index{vacuum dark matter} would exert its own gravitational field, which would actually allow it to clump, just like conventional dark matter\index{dark matter!conventional}, despite the fact that it really is vacuum energy. This means that the observations which indicate that large overdensities of visible matter can sometimes become separated from their dark matter component (as a result of collisions between galaxy clusters\index{galaxy cluster!collisions} or in the course of galaxy mergers\index{galaxy mergers}) can be easily explained, unlike would be the case if the currently unexplained correlations I have just mentioned were the result of a more profound modification of the laws that govern the gravitational dynamics of astronomical objects, such as envisaged in the context of the theory known under the MOND acronym (which stands for modified Newtonian dynamics\index{modified Newtonian dynamics}).

When considering the effects arising from the presence of vacuum dark matter one must keep in mind that, it is possible for the average density of negative vacuum-dark-matter energy\index{vacuum-dark-matter energy!negative} to still be very similar to that of positive vacuum-dark-matter energy\index{vacuum-dark-matter energy!positive}, because the observationally confirmed absence of negative-energy matter overdensities on the scale of individual stars and galaxies is due to the fact that only a negligible amount of baryonic negative-energy matter\index{baryonic negative-energy matter} survived the early annihilation of matter with antimatter\index{early matter-antimatter annihilation} that took place during the radiation-dominated era\index{radiation-dominated era}. But vacuum dark matter is not submitted to mutual annihilation, given that it has no charge, and therefore its average density must have remained unchanged following the early annihilation of baryons with antibaryons\index{early baryon-antibaryon annihilation}.

But the presence of baryonic matter\index{baryonic matter} is necessary for initiating the formation of structures\index{structure formation} in a distribution of vacuum dark matter\index{vacuum dark matter} on smaller scales, because only such matter can reduce its average kinetic energy\index{average kinetic energy!matter} through the emission of radiation. Collapsing astronomical structures\index{collapsing astronomical structures!emission of radiation} can only begin to form through gravitational instability\index{gravitational instability} when they are allowed to release energy in such a way, as otherwise their internal pressure\index{internal pressure!astronomical structures} remains too large to allow them to stabilize. Given that vacuum dark matter is merely as cold as baryonic matter, due to the fact that zero-point vacuum fluctuations\index{zero-point vacuum fluctuations} involve the same particles fluctuating in and out of existence in their virtual form, then it follows that, on a scale where ordinary baryonic matter is allowed to collapse into stable structures only through the emission of radiation, the density of vacuum dark matter itself can only begin to grow if ordinary baryonic matter overdensities with the same sign of energy are already present, because without the additional gravitational attraction produced by such an overdensity, vacuum dark matter would rather tend to disperse.

That is not to say, however, that there can be no overdensities in the distribution of negative vacuum-dark-matter energy\index{vacuum-dark-matter energy!negative}. In fact, it appears necessary to assume that vacuum-dark-matter overdensities with very large negative masses\index{negative mass} would be present inside the largest voids in the distribution of positive-energy galaxies\index{positive-energy galaxy distribution!largest voids}, which do exert a localized gravitational pull on this vacuum dark matter, given that the gravitational attraction exerted on negative vacuum-dark-matter energy by voids in the positive-energy matter distribution grows along with their size. In the absence of baryonic negative-energy matter\index{baryonic negative-energy matter!overdensities} overdensities, it would appear that only such large voids can exert a gravitational attraction large enough to allow negative vacuum-dark-matter energy\index{negative vacuum-dark-matter energy!overdensities} overdensities to grow to such proportions that they may actually influence the process of structure formation in the positive-energy matter distribution. But once triggered, this process would self-amplify, as a result of the gravitational attraction exerted by negative vacuum-dark-matter energy on itself.

The presence of such very-large-scale overdensities in the negative-energy matter distribution\index{negative-energy matter distribution!very-large-scale overdensities} is not only allowed by current observational data, it is actually required in order to explain the size of those voids. In the absence of large negative-energy matter overdensities located within the largest voids in the positive-energy matter distribution, the unexpectedly large size of those voids would need to be explained through biasing\index{biasing}, an approach which amounts to assume, without justification, that galaxies have a tendency to form preferably in those regions where the density of the cold dark matter\index{cold dark matter} is already larger, at the epoch of recombination\index{epoch of recombination}. But the presence of large and growing overdensities of negative vacuum-dark-matter energy inside those voids in the positive-energy matter distribution would allow the voids themselves to grow larger than expected at an earlier time, due to the gravitational repulsion\index{gravitational repulsion} they would exert on the surrounding positive-energy galaxies.

Given that the gravitational repulsion attributable to a large negative-energy matter overdensity would merely contribute to further accelerate the local rate of expansion\index{rate of expansion!local acceleration} of the expanding positive-energy matter distribution which is already being accelerated by the presence of the void in the positive-energy matter distribution\index{positive-energy matter distribution!void} in which it would normally be located, then it may not be immediately apparent that the phenomenon is sometimes actually attributable in part to the presence of gravitationally-repulsive material. This is certainly a positive development, given that it has been known for some time that certain voids, apparent on a very large scale in the positive-energy matter distribution, do produce a larger than expected acceleration of the local rate of expansion of galaxies located on their periphery, a phenomenon which had remained unexplained until now.

When large negative-energy matter overdensities are present inside the largest voids in the visible matter distribution\index{visible matter distribution!largest voids}, it is also easier to reconcile our theory of structure formation\index{structure formation!theory} with those observations which show that there is a much smaller number of galaxies in the Local Void\index{Local Void} than is predicted by computer simulations\index{computer simulations!galaxy distribution}, because any galaxy that would form in the void would rapidly be expelled to the periphery by the repulsive gravitational forces\index{repulsive gravitational force} attributable to the presence of this negative-energy matter. Also, given that the density of negative vacuum-dark-matter energy\index{vacuum-dark-matter energy!negative} in the Local Void would not be as low as it would in our galactic neighborhood\index{galactic neighborhood}, it follows that the missing-mass effects\index{missing-mass effect} attributable to negative-energy matter underdensities would be more localized around those galaxies located closer to the center of the void and this must have accelerated their formation. That may explain why a larger than expected number of very large galaxies in the Local Sheet\index{Local Sheet} are located on the periphery of the Local Void instead of in the more crowded areas, where most of the visible matter is concentrated.

\subsection{The flatness problem\label{sec:6.3}}

I mentioned above that, inflation\index{inflation!theory} theory aside, we have no idea what the constraint is that would require the present average density\index{present average energy density!critical value} of positive matter and vacuum energy to be fixed to its critical value, given that a very precise adjustment of parameters would seem to be required in the early stages of the Big Bang\index{Big Bang!adjustment of parameters} in order to produce the present outcome. The truth, therefore, is that while the Big Bang model\index{Big Bang model!incompleteness} appears to be mathematically consistent, it is nevertheless incomplete, given that the initial conditions\index{initial conditions!Big Bang}, it would seem, cannot be determined by the theory.

Thus, while it can be predicted that the average density of positive matter, radiation, and vacuum energy\index{average energy density!critical value} $\rho(t)$ was closer to its critical value $\rho_c(t)$ (associated with a density parameter\index{density parameter} $\Omega\equiv\rho/\rho_c=1$) in the past, it remains to explain why it is that the current density of positive energy $\rho_0$ itself is fixed to its critical value $\rho_{c,0}$ to such a high degree of precision. Relativity theory\index{relativity theory} enables a positive-energy observer to predict what the rate of expansion\index{rate of expansion} of the universe was at different times in the past, given the current densities of positive matter, radiation, and vacuum energy, but this is only true in as much as the rate of expansion at the present time is empirically determined through a measurement of the Hubble constant\index{Hubble constant} $H_0$. The model, however, remains well-defined for any value of $\rho_0$ and $H_0$.

Here I will assume that there exists an upper limit to the positive and negative measures of matter energy density, which is provided by the natural vacuum-stress-energy tensors\index{natural vacuum-stress-energy tensors} entering the generalized gravitational field equations\index{generalized gravitational field equations} introduced in section \ref{sec:5}. From a quantum-gravitational viewpoint\index{quantum-gravitational viewpoint}, it must be assumed that those are the positive and negative densities of matter energy which existed in the state that emerged from the initial singularity\index{initial singularity}. This means that space cannot continue to contract, in the past direction of time, beyond the point at which a maximum positive or negative amount of matter and radiation energy is contained in every elementary unit of surface\index{elementary unit of surface}, under which conditions current quantum gravitation\index{quantum gravitation!theories} theories predict that a quantum bounce\index{quantum bounce} must take place.

What needs to be explained, therefore, is merely why it is that the rate of expansion\index{rate of expansion!critical value} of space did not begin to differ from its critical value immediately after the universe emerged from this state of maximum positive and negative matter energy densities\index{maximum matter energy densities} that is uniquely determined by the natural scale of quantum-gravitational phenomena\index{quantum-gravitational phenomena!natural scale}. From a conventional viewpoint, it would appear that the early variation of the rate of expansion\index{rate of expansion!early variation} that gives rise to a flat space\index{flat space} at the present time is merely one alternative among an enormous range of possibilities. But while the current value of gravitational potential energy\index{gravitational potential energy} for the universe as a whole (which is fixed by the present average density\index{present average energy density!universe} of positive matter and vacuum energy $\rho_0$) and the currently observed kinetic energy of expansion\index{kinetic energy of expansion} (which is determined by $H_0$) may appear to constitute free parameters of the standard model of cosmology\index{standard model of cosmology!free parameters}, I will show that they are not really independent variables in the context where energy must be null for the universe as a whole.

If such a condition must be imposed for the total energy of the universe\index{total energy of the universe} it is because the requirement of relational definition of physical attributes\index{requirement of relational definition of physical attributes} implies that, even if the Big Bang is not considered to constitute a creation event at which any conserved physical quantity must be created out of nothing, from the viewpoint of an observer of any energy sign the universe would still need to have a vanishing total energy\index{total energy of the universe!vanishing}, because if it was possible to measure a non-zero value of matter energy\index{matter energy!universe as a whole} for the universe as a whole, then this value would need to be either positive or negative and this would allow the particular direction of time\index{direction of time} relative to which this positive or negative energy would propagate to be singled out as an absolutely (non-relationally) defined direction\index{absolutely defined direction!time}, as would necessarily happen in the absence of negative-energy matter if there was no compensation of the positive energy of matter by a negative energy of the gravitational field\index{gravitational field energy!universe}.

What most people already recognize concerning the energy content of the universe is that, for a universe with a flat geometry\index{flat geometry!universe} and a zero cosmological constant\index{cosmological constant!zero}, the negative gravitational potential energy\index{gravitational potential energy!negative} of positive-energy matter and radiation is balanced by the positive kinetic energy of expansion\index{kinetic energy of expansion!positive} of this matter. When that is not the case, then an additional amount of energy is present that is attributable to the gravitational field\index{gravitational field!energy} itself (or the curvature of space\index{curvature of space!energy}) and this energy tends to become dominant very rapidly (regardless of whether it is positive or negative) as space expands, because, while the gravitational potential energy of matter decreases in inverse proportion to the volume, the energy associated with the curvature of space decreases as the inverse of the surface enclosing that volume.

The condition that energy be conserved for the universe as a whole, as it expands or contracts, is expressed by the the following initial value equation\index{initial value equation} for a homogeneous and isotropic universe\index{homogeneous and isotropic universe}, which is derived from the general-relativistic gravitational field equations\index{gravitational field equations} for a positive-energy observer.
\begin{equation}\label{eq:33}
E=K+V(a)=\left(\frac{1}{a}\frac{da}{dt}\right)^2+\left(\frac{-8\pi\rho}{3}-\frac{\Lambda}{3}+\frac{k}{a^2}\right)=0
\end{equation}
Here $E$ is the gravitational energy of the universe\index{gravitational energy of the universe}, $K$ is the kinetic energy of expansion, and $V(a)$ is the Friedmann potential\index{Friedmann potential} as a function of the scale factor\index{scale factor} $a(t)$, in the presence of a cosmological constant\index{cosmological constant} $\Lambda$, for a universe with an average matter density\index{average matter density!universe} $\rho$. In this equation, the spatial curvature parameter\index{spatial curvature parameter}, which I redefine as $-k/a^2$ and which is always precisely equal to zero for a flat universe, appears as just one particular (reversed) contribution to the Friedmann potential.

But when it is possible to assume that the magnitude of the cosmological constant was negligible initially, this equation can be rewritten as
\begin{equation}\label{eq:34}
E_g=K+U(a)=\left(\frac{1}{a}\frac{da}{dt}\right)^2-\left(\frac{8\pi\rho}{3}\right)=\frac{-k}{a^2}
\end{equation}
which clearly shows that the spatial curvature parameter is the outcome of the imperfect cancellation of the gravitational potential energy\index{gravitational potential energy!of matter} of matter by the kinetic energy of expansion\index{kinetic energy of expansion}. It must be clear, though, that while the kinetic energy of expansion is usually considered to be an attribute of the expanding matter, which may, therefore, be either positive or negative, depending on the sign of energy of the matter that produces the gravitational field that influences the rate of expansion\index{rate of expansion} of space (which for positive-energy observers would be positive), it is, nevertheless, entirely determined by this rate of expansion, which means that it is actually an energy of the gravitational field\index{gravitational field!energy} itself (as, in principle, it could exist even in a universe\index{universe!absence of matter} devoid of matter).

Thus, whenever the gravitational potential energy of matter $U(a)$ is not matched by a kinetic energy of expansion $K$ that's exactly its opposite, the energy $E_g$ associated with the gravitational field or the curvature of space\index{curvature of space!energy} itself, which is given by $-k/a^2$, is not zero and contributes to alter the expansion rate. If $k$ is positive this excess of gravitational energy\index{gravitational energy excess} is negative, which means that the negative gravitational potential energy of matter contributes predominantly to determine the gravitational field, as must be the case when the source of this gravitational field has positive energy, while when $k$ is negative there is a positive excess of gravitational energy, which means that the positive kinetic energy of expansion contributes predominantly to determine the gravitational field\index{gravitational field!universe} of the universe.

The gravitational energy $E_g$ associated with the present value of the spatial curvature parameter\index{spatial curvature parameter} $-k/a^2$ must therefore be considered to consist of a residual measure of energy, which could in principle assume any positive, negative, or null value depending on the current value of the scale factor\index{scale factor} and on whether $k$ is equal to $-1$, $+1$, or $0$. There is no \textit{a priori} reason, however, to assume that the measure of energy associated with the curvature of space\index{curvature of space!energy} on the cosmological scale should be the same for positive- and negative-energy observers at the same epoch, because the kinetic energy of expansion\index{kinetic energy of expansion} varies as a function of the rate of expansion\index{rate of expansion!observer-dependent quantity}, which is an observer-dependent quantity.

It must be clear, however, that what the original form (\ref{eq:33}) of the initial value equation\index{initial value equation} really means is that when an additional term, which is provided by the negative of the spatial curvature parameter $-k/a^2$, is added to the equation that would otherwise express the nullity of the gravitational energy of the universe\index{gravitational energy of the universe!nullity}, then this energy can be conserved even in those cases where it would not really be null initially, but it does not really amount to require that the universe comes into existence with zero gravitational energy. What the redefinition of gravitational energy provided by equation (\ref{eq:34}) means is that, once it is assumed that the cosmological constant\index{cosmological constant} $\Lambda$ is negligible initially, then it is only when the free parameter $-k/a^2$ associated with the curvature of space is zero that the positive kinetic energy of expansion $K$ can balance the negative gravitational potential energy\index{gravitational potential energy!negative} $U(a)$ attributable to the presence of positive-energy matter. The true measure of gravitational energy for the universe as whole, therefore, is really the energy $E_g$ which is associated with the curvature of space and it is only when this energy is null that the gravitational field does not contribute energy on the cosmological scale.

I believe that what allows the value of gravitational energy $E_g$ associated with the spatial curvature parameter\index{spatial curvature parameter!gravitational energy} to be null for an expanding zero-energy universe\index{zero-energy universe} is the fact that, even when negative-energy matter is present in the very first instants of the Big Bang and the total energy of matter\index{total energy of matter!null} itself is null, the negative gravitational potential energy\index{gravitational potential energy!of matter} of matter experienced by a positive-energy observer can be arbitrarily large (given that only the positive portion of the uniform distribution of matter energy exerts a gravitational force on positive-energy matter) so that it is allowed to compensate a positive kinetic energy of expansion\index{kinetic energy of expansion} of comparable magnitude. It would therefore appear that the alternative concept of negative-energy matter as consisting of missing positive vacuum energy\index{missing positive vacuum energy!negative-energy matter}, which was developed in the preceding sections of this report, is essential to the conclusion that a zero-energy universe can have a null value of gravitational energy.

If we were to assume instead that negative-energy matter was absent in the primordial universe, then the kinetic energy of expansion itself would actually need to be arbitrarily small and the negative gravitational energy of curvature\index{gravitational energy of curvature} $E_g$ arbitrarily large, because only then could the negative energy contained in the gravitational field compensate the large positive energy of matter\index{energy of matter!positive}. Such a universe, however, would be closed\index{closed universe} and highly curved and would collapse back to a spacetime singularity\index{spacetime singularity} in an arbitrarily short time. In the context where the existence of negative-energy matter would be considered unavoidable, the solution to this problem which is usually assumed to be provided by inflation\index{inflation!theory} theory would no longer be appropriate, if we still require the universe to have zero energy\index{zero-energy universe}, unless we recognize the validity of the concept of negative-energy matter developed in this report (which will be shown to make inflation unnecessary), because otherwise the gravitational potential energy of matter could be null, which means that the kinetic energy of expansion would still need to be null following inflation.

But we must still explain why the total, average density of matter energy\index{average matter energy densities!zero} was initially so close to zero that the energy of the gravitational field (associated with the global curvature of space\index{global curvature of space}) was itself required to be perfectly null. The problem is that, in principle, it is possible for the magnitude of the average density of positive matter energy\index{average density of positive matter energy} to be larger or smaller than that of the average density of negative matter energy\index{average density of negative matter energy}, even in a universe with zero energy, as long as the sum of the positive and negative contributions to the energy of matter is compensated, from the viewpoint of a given observer, by the energy of the gravitational field\index{gravitational field energy!universe} of the universe, which is determined by the rate of expansion\index{rate of expansion} measured by that observer.

What's significant here is that, when the energy of the universe\index{energy of the universe!null} is required to be null, it follows that if space\index{positively curved and closed space} was positively curved and closed from the viewpoint of a positive-energy observer, it would need to be negatively curved and open\index{negatively curved and open space} from the viewpoint of a negative-energy observer. This is due to the fact that the gravitational field\index{gravitational field!universe} of a universe that would be positively curved, from the viewpoint of a positive-energy observer, would have a negative energy and could, therefore, only compensate an excess of positive matter energy (through a reduction of the positive kinetic energy of expansion\index{kinetic energy of expansion!positive}). But while it is true that, even from the viewpoint of a negative-energy observer, an excess of positive matter energy would require the contribution of a gravitational field with negative energy, such a gravitational field would be associated not with a smaller positive kinetic energy of expansion, but with a larger negative kinetic energy of expansion\index{kinetic energy of expansion!negative} and a higher than critical expansion rate\index{rate of expansion!higher than critical}, which would actually give rise to an open universe\index{open universe}, while the opposite would be true if the total energy of matter was instead negative initially (before the early annihilation of matter and antimatter\index{early matter-antimatter annihilation}).

While those two mutually exclusive configurations may appear to merely consist of two additional possibilities, no different from the case where the average density of matter energy\index{average matter energy densities!null} happens to be null initially, just like the energy of the gravitational field\index{gravitational field energy!universe} of the universe, there is actually a very important distinction between the case of a universe with flat geometry\index{flat geometry!universe} and that of the curved space configurations. This essential difference has to do with the fact that it is only in the case where space is flat that the universe can be open from both the viewpoint of a positive-energy observer and that of a negative-energy observer. I believe that if the average density of positive matter energy must exactly compensate the average density of negative matter energy in the initial Big Bang state, it is precisely because, in the absence of any other contribution to the energy budget, if matter energy was not null, then space would need to be open from the viewpoint of a certain observer and closed from the viewpoint an observer with opposite energy sign. But given that the difference between the volume of a closed universe\index{closed universe} and that of an open universe would in principle be infinite, it follows that such a configuration would be characterized by an arbitrarily large, positive or negative value of average vacuum energy density\index{average vacuum energy density!arbitrarily large value}.

It should be clear that it cannot be assumed that the energy of matter in a universe with a non-zero curvature of space\index{curvature of space!universe} is compensated by an opposite energy that would be contained in the vacuum as a result of this curvature, because in a universe\index{universe!negative gravitational energy} with negative gravitational energy and positive space curvature\index{positive space curvature!universe} the average density of vacuum energy\index{average vacuum energy density!positive} would actually be positive and would add to the positive energy of matter, thereby requiring an even larger negative energy for the gravitational field, that would make the positive density of vacuum energy even larger.

The problem, therefore, is that the maximum positive or negative cosmological constant\index{cosmological constant!maximum positive or negative value} which would be associated with \textit{any} non-zero gravitational energy of curvature\index{gravitational energy of curvature!non-zero} would appear to forbid the emergence of an observer\index{observer emergence}, because even if the gravitational force exerted by the cosmological constant on the specific expansion rates\index{specific expansion rates} generally contributes to reduce the magnitude of the average density of vacuum energy\index{average vacuum energy density}, if this magnitude had been maximum when the expansion process began, then it could never have been reduced to a level favorable to the emergence of an observer, at least not before the average matter density itself would have become too low to allow for the development of structures (which we may assume to be essential for the emergence of an observer). Only a universe with precisely balanced initial contributions to the energy of matter and therefore, also, to the energy of the gravitational field\index{gravitational field energy!universe}, is allowed to be experienced as a long lasting process by a physical observer that is part of that universe, when it is appropriately required that the universe itself has null energy\index{energy of the universe!null}.

Thus, the `extra' principle which would allow to fix the rate of expansion\index{rate of expansion!critical value} of space on the global scale to its critical value is nothing else but the requirement of relational definition of physical attributes\index{requirement of relational definition of physical attributes}, which requires the sum of all energies\index{sum of all energies!null} to be null, for the universe as a whole, from the viewpoint of both positive- and negative-energy observers. When it is recognized that negative-energy matter must also contribute to the universe's initial energy budget, this very basic principle makes it possible to explain not only why there is expansion, but why it is that the rate of this expansion is still critical, even long after the Big Bang. Space is flat\index{flat space} and the rate of expansion remains critical, because the universe\index{universe!open space} must have an open space from both the viewpoint of positive energy observers and that of negative energy observers and the precision with which the initial rate of expansion\index{rate of expansion!initial} was adjusted to its critical value is merely a reflection of the exactness of this requirement. The flatness of space, therefore, is not a mere possibility that emerged as a byproduct of an uncertain process of inflationary expansion\index{inflationary expansion}, but rather constitutes a basic consistency requirement that must be satisfied by any viable cosmological model.

But the same constraint allows one to expect that there is no difference, initially and therefore also at later times, between the average states of motion\index{average states of motion} of positive- and negative-energy matter on the largest scale that could have given rise to a non-zero momentum for the universe\index{momentum of universe!non-zero} as a whole, because such a momentum for matter would also need to be compensated by an opposite momentum of the gravitational field\index{gravitational field!momentum}, and if the gravitational field (or the appropriate component of the curvature of space\index{curvature of space}) had a non-zero momentum on a global scale, it would also need to have a non-zero energy and this is not possible for a universe with flat geometry\index{flat geometry!universe}.

However, in the context where the initial, average density of negative-energy matter can become smaller than that of positive-energy matter following the annihilation of baryonic matter and antimatter\index{early matter-antimatter annihilation} that takes place early on, during the Big Bang, it is possible for the average value of vacuum energy density\index{average vacuum energy density}, or the cosmological constant\index{cosmological constant!initial zero value}, to grow from its initial zero value toward a larger, positive value during the matter-dominated era\index{matter-dominated era}, as I previously explained. But it is not to be expected that this divergence could develop to an arbitrarily large magnitude, because the weak anthropic principle\index{weak anthropic principle} also forbids the cosmological constant from becoming so large, as a result of this divergence, that it would no longer be compatible with the presence of a (positive-energy) observer at the present time.

It is, therefore, possible to assume that it is the weak anthropic principle and the condition of null energy\index{condition of null energy} that must be imposed on the universe, which explain that the divergence of the scale factors\index{scale factors!divergence} experienced by opposite-energy observers was as small as it appears to have been in the early universe (as a result of the relatively small violation of time reversal symmetry\index{time reversal symmetry!violation} that is responsible for the existence of all baryonic positive-energy matter\index{baryonic positive-energy matter}), because this is what allowed the current value of the cosmological constant\index{cosmological constant!current value} to be as small as it is observed to be. I believe that the fact that such a relatively simple and efficient solution to what has been called `the mother of all physics problem'\index{mother of all physics problem} had never been seriously considered before is simply a consequence of the preconceived opinion that negative-energy matter cannot exist, which is a consequence of both irrational prejudice and what always appeared to be the insurmountable difficulties preventing a consistent description of gravitationally-repulsive matter\index{gravitationally-repulsive matter}.

\subsection{The arrow of time\label{sec:6.4}}

Before I can discuss the relevance of the generalized gravitation theory\index{generalized gravitation theory} which was developed in the preceding sections of this report for a solution to the problem of the origin of the thermodynamic arrow of time\index{thermodynamic arrow of time!origin problem} it is necessary to explain what really constitute the many facets of this long-standing problem. This will allow me to properly identify the nature of the deep contradiction that dwells at the heart of theoretical physics\index{theoretical physics!contradictions}, as a result of the apparent incompatibility between the time-symmetric laws\index{time-symmetric laws} of classical mechanics\index{classical mechanics} and particle physics\index{particle physics} and the unidirectional laws\index{unidirectional laws} of thermodynamics\index{thermodynamics} and statistical mechanics\index{statistical mechanics}.

First of all, it must be clear that time irreversibility\index{time irreversibility!non-subjective} is not a subjective notion that would arise merely as a consequence of adopting a particular coarse-graining\index{coarse-graining}, that is to say, it is not due merely to the choice we make regarding what details of the microscopic state\index{microscopic state} of a system are to be ignored, because the changes which are taking place in that portion of entropy\index{entropy!of gravitational field} which is attributable to the gravitational field can be characterized in a fully objective way, given that the measure of entropy\index{entropy!black-hole event horizon} associated with black-hole event horizons does not depend on any \textit{arbitrary} definition regarding what parameters characterize the macroscopic state\index{macroscopic state} of such a system and what information\index{information!unavailability} remains unavailable. For an observer outside a black hole, information is only available about the macroscopic parameters of total mass\index{black hole!mass} $M$, angular momentum\index{black hole!angular momentum} $J$, and charge\index{black hole!charge} $Q$, while the entropy itself is dependent merely on the area of the event horizon of the object, which is entirely determined by those parameters.

It is usually understood, however, that while we are allowed to consider entropy\index{entropy!missing information} as missing information, an objective characterization of temporal irreversibility\index{temporal irreversibility!objective} does not require assuming that information is actually vanishing from reality when entropy is rising. It is simply the fact that, as entropy grows, \textit{macroscopic} parameters\index{macroscopic parameters} become increasingly less efficient at providing a full description of the structure contained in the exact microscopic state\index{microscopic state!of universe} of our universe, that explains that information\index{information!loss} appears to be lost when the number of microscopic states which are compatible with those macroscopic conditions is growing with time.

But if entropy is indeed increasing in the future, from the viewpoint of an objectively determined choice of coarse-graining\index{coarse-graining!objective choice}, then it means that entropy was definitely smaller in the past. What is deduced from observations, in fact, is that entropy\index{entropy!decrease} is continuously decreasing in the past, in every place we look and as far back in time as we can probe. This is a condition that is far more constraining than simply assuming that the universe is not in a state of thermal equilibrium\index{thermal equilibrium!state} at the present time, which would not only allow entropy\index{entropy!growth} to grow larger in the future, but also in the past.

Due to the time-symmetric nature of the fundamental physical laws\index{fundamental physical laws!time-symmetric} that describe the evolution of elementary particles, it would appear that when a macroscopic physical system with many independent microscopic degrees of freedom\index{microscopic degrees of freedom!independent} evolves in the past direction of time, starting from a present non-equilibrium state\index{non-equilibrium state} of relatively low entropy, this entropy should grow (regardless of the details of the microscopic state\index{microscopic state} of the system) for the exact same reason that we expect it to grow in the future, when the evolution that takes place on a microscopic level occurs in a random way. But even if entropy was continuously increasing in the past, from its present non-maximum value, we may still have a problem, because from the forward-in-time viewpoint\index{forward-in-time viewpoint} the evolution that would have taken place in the past would have occurred with diminishing entropy in the future and this would also remain unexplained, unless we are dealing with a \textit{momentary} fluctuation.

One of the oldest attempts at solving the problem of the origin of the thermodynamic arrow of time\index{thermodynamic arrow of time!origin problem}, which is usually recognized as inadequate, was originally proposed by Ludwig Boltzmann\index{Boltzmann, Ludwig}, the originator of the kinetic theory of gases\index{kinetic theory of gases}. It was based on the recognition that there always occur fluctuations to lower entropy\index{entropy!fluctuation} states for randomly-evolving, isolated systems\index{isolated system} which are in a state of thermal equilibrium\index{thermal equilibrium!state}. It was suggested that over a very long time-scale, it should sometimes happen that such fluctuations would be so significant as to bring even a universe which was in a state of thermal equilibrium into a state with an entropy so low that any subsequent evolution would likely be characterized by a continuous increase of entropy\index{entropy!continuous increase}.

It should be clear, however, that in such a context, the only reason we would have to expect to observe the universe in a phase of continuously growing entropy, instead of finding it in one of the much, much more common phases of unchanging maximum entropy\index{entropy!maximum} would be that this entropy growth\index{entropy!growth} is necessary for the presence of an observer which can witness such an evolution. The problem, however, is that if such a requirement was to be satisfied merely as a consequence of the occurrence of a fluctuation in an otherwise unchanging maximum entropy state, then we should not expect to observe entropy to be so low in all parts of the universe and as far back in time as the epoch of the Big Bang. A much more localized and ephemeral fluctuation, that would provide the observer with no records of a pervasive and long-lasting, time-asymmetric history\index{time-asymmetric history!pervasive and long-lasting}, would do just as well for allowing this kind of phenomenon to occur at the present time and given that such a fluctuation would be more likely to occur than a long-lived fluctuation involving the entire universe, then based on this kind of argument what we should experience is a short-lived fluctuation.

The question, therefore, remains: Why is the universe evolving irreversibly in one single direction of time in all locations and throughout its entire lifetime? It must be clear, first of all, that even though the random nature of elementary physical processes\index{elementary physical processes!random nature} and the sensibility to initial conditions\index{initial conditions!sensibility to} allow one to reject the possibility that it could be a precise adjustment of present microscopic conditions\index{present microscopic conditions!precise adjustment} that would explain the diminution of entropy\index{entropy!diminution} that is observed to take place in the past direction of time, this doesn't necessarily mean that irreversibility\index{time irreversibility!fundamental and irreducible} must be occurring at a fundamental and irreducible level in our description of physical processes. It would certainly not be appropriate to abdicate the requirement of time reversal symmetry\index{time reversal symmetry}, simply to provide an explanation for the observed unidirectionality of thermodynamic processes\index{thermodynamic processes!unidirectionality}, in the context where our most valuable physical theories are all time-symmetric at the most elementary level of description. The challenge consists in actually explaining irreversibility, not in decreeing that it is the foundation of reality, when this would require abandoning most of everything else we have learned.

\bigskip

\noindent It is well-known that, aside from the volume of space itself, it is the measure of gravitational entropy\index{gravitational entropy} that constitutes the essential distinction between the state that emerged from the past Big Bang singularity\index{past Big Bang singularity} and the present state of our universe, or that into which it can be expected to evolve in the future. Thus, there is no paradox associated with the fact that the universe has always evolved irreversibly, even though the initial state at the Big Bang\index{initial Big Bang state!thermal equilibrium} was already one of near perfect thermal equilibrium, because, as Roger Penrose\index{Penrose, Roger} first pointed out \cite{Penrose-1}, under such conditions it is only the portion of entropy\index{entropy!of gravitational field} that excludes the contribution of gravitational fields (and ultimately that of black holes) that is maximum.

Thus, it appears that it is precisely the smoothness of the initial matter distribution\index{initial matter distribution!smoothness} (which is reflected in the uniformity of the temperature of the cosmic microwave background\index{cosmic microwave background!temperature uniformity}) that is responsible for having allowed the universe to evolve irreversibly at later times, because under such conditions it is the growth of matter inhomogeneities\index{growth of matter inhomogeneities} which must have provided the dominant contribution to irreversible entropy\index{entropy!growth} growth in our universe, since the epoch of matter-antimatter annihilation\index{matter-antimatter annihilation!epoch}, given that gravitational entropy is higher for more inhomogeneous matter distributions. What needs to be explained, therefore, is why it is that matter was so homogeneously distributed initially that gravitational entropy\index{gravitational entropy!null} was almost perfectly null, even if this would appear to be a highly unlikely configuration to begin with, in the context where a much larger number of possibilities exist, for the microscopic state\index{microscopic state!matter and gravitational field} of matter and its gravitational field, which would not give rise to such a uniform matter distribution\index{uniform matter distribution}.

In order to explain those facts, it is necessary to identify the nature of the constraint imposed by the fundamental, time-symmetric physical laws\index{time-symmetric physical laws} on the boundary conditions\index{boundary conditions!Big Bang} at the Big Bang that is responsible for the minimum gravitational entropy\index{gravitational entropy!minimum} that characterizes this initial state. The problem is that a much larger number of initial microscopic states\index{initial microscopic states} would be characterized by the presence of an abundance of primordial black holes\index{primordial black holes} and other matter density fluctuations, while those initial conditions would not have had as much potential for allowing subsequent evolution to take place irreversibly. However, given that the presence of primordial black holes would have disturbed the process of structure formation\index{structure formation!process} in the initial matter distribution in ways which would have had observable consequences at the present epoch, then it seems necessary to assume that the initial Big Bang state was virtually free of macroscopic black holes\index{macroscopic black holes} (with a mass larger than the Planck mass\index{Planck mass}).

When I discussed the flatness problem\index{flatness problem} in section \ref{sec:6.3} I explained why it is that we can expect the universe to be expanding. But the fact that we are not instead observing it to be contracting at the present moment can only be explained as being the consequence of another fact, which is that the magnitude of gravitational fields\index{magnitude of gravitational fields!continuous decrease} is decreasing continuously in this direction of time relative to which space is contracting. If we perceive the universe to be expanding, it is simply because, our memory formation process\index{memory formation!thermodynamic process}, as a thermodynamic process, only occurs in the direction of time in which the inhomogeneity of the matter distribution\index{inhomogeneity of matter distribution!growth} and the associated gravitational entropy\index{gravitational entropy!growth} are growing, which means that if the inhomogeneity of the matter distribution was instead growing in the direction of time relative to which the universe is contracting, we would then necessarily perceive space to be contracting.

It should be clear, however, that the simple fact that, if an observer is to be present to witness an absence of inhomogeneities, the space surrounding her must not have collapsed into a black hole\index{black hole!singularity} singularity, does not provide strong enough a constraint to explain that the initial matter distribution\index{initial matter distribution!smoothness} was as smooth as it is observed to be. Matter could be much more inhomogeneously distributed than it currently is and the conditions necessary for the emergence of an observer\index{observer emergence} would still be satisfied in most locations, even if a large number of macroscopic black holes\index{macroscopic black holes} had been present initially. It is merely the fact that the inhomogeneity is not as pronounced as it \textit{could} have been that is unexplained.

In the context where one must acknowledge the presence of negative-energy matter, it follows that, even in a universe with an arbitrarily large volume of space, a uniform matter distribution\index{uniform matter distribution} doesn't become more likely as a randomly chosen configuration\index{randomly chosen configuration}, because even a diluted matter distribution\index{diluted matter distribution} could still contain inhomogeneities and produce arbitrarily strong gravitational fields on a very large scale\index{very large scale!arbitrarily strong gravitational fields}, as a result of the fact that negative-energy matter can be concentrated in regions of space distinct from those occupied by positive-energy matter. The homogeneity of the matter distribution itself is not apparent merely in the low magnitude of local variations in the density of positive matter energy, but also in the near absence of large-scale disparities between the positive- and negative-energy matter distributions.

In a universe that contains both positive- and negative-energy matter the state with the highest gravitational entropy\index{gravitational entropy} would necessarily be one for which the distribution of positive- and negative-energy matter would be completely polarized, in such a way that all the matter would be contained in opposite-energy black holes with arbitrarily large masses, whose magnitude would be limited solely by the amount of matter in the universe. But there is no \textit{a priori} motive for assuming that a high level of polarization of the positive- and negative-energy matter distributions could not also apply to the initial Big Bang state (regardless of the fact that the matter density is then maximum) if such a configuration is, in effect, favored from a statistical viewpoint, because a universe that would evolve without constraint, as space is contracting in the past direction of time, would have more chances to reach such a configuration.

There are strong motives, however, for believing that, even in the presence of negative-energy matter, if the initial matter distribution\index{initial matter distribution!homogeneous} is sufficiently homogeneous, it is still appropriate to consider that there should arise a state in the past which, from a classical viewpoint, would consist in a spacetime singularity\index{spacetime singularity!past}. In section \ref{sec:3} I have explained, in effect, that a globally homogeneous distribution of negative-energy matter would exert no influence on the rate of expansion\index{rate of expansion} of positive-energy matter and would not diminish the strength of the gravitational field attributable to this positive-energy matter, despite the fact that negative-energy matter overdensities do exert a repulsive gravitational force\index{repulsive gravitational force} on positive-energy matter. Thus, if the initial matter distribution is sufficiently homogeneous on the largest scale, then nothing can prevent the formation of the trapped surface\index{trapped surface} which would give rise to a past singularity\index{past singularity}, even if negative-energy matter is present in the universe.

It would, therefore, appear that the very uniformity of the matter distribution which is responsible for giving rise to the existence of a thermodynamic arrow of time\index{thermodynamic arrow of time} is actually required in order that the existence of a past singularity be considered unavoidable. I must emphasize, however, that what I have in mind when I'm referring to an initial singularity\index{initial singularity} is not a state where the laws of physics would actually break down, but simply a state where the average, positive and negative densities of matter energy\index{average matter energy densities!maximum values} have reached their maximum theoretical values, determined by the natural vacuum-stress-energy tensors\index{natural vacuum-stress-energy tensors} which enter the generalized gravitational field equations\index{generalized gravitational field equations} introduced in section \ref{sec:5}. This initial singularity, however, was of such a nature that it could not constitute the outcome of a gravitational collapse\index{gravitational collapse} of the kind that would occur in a universe in which gravitational entropy\index{gravitational entropy} is growing.

What's significant, as well, is that the presence of past singularities\index{past singularity} appears to be restricted to the one known initial singularity from which the Big Bang emerged, even if there does exist solutions of the gravitational field equations\index{gravitational field equations!solutions} that would appear to describe processes which would be the time-reverses of a black-hole gravitational collapse\index{black-hole gravitational collapse!time-reverse}. There appears to be something that constrain evolution in the past direction of time to take place with continuously decreasing gravitational entropy\index{gravitational entropy!continuous decrease}, despite the apparent improbability of this evolution, and this is precisely what remains unexplained. If such white hole\index{white hole} phenomena are never observed, therefore, it is simply because this would require a decrease of gravitational entropy in the future (which is unlikely), or equivalently, a continuous increase of gravitational entropy\index{gravitational entropy!continuous increase} in the past (which for some reason appears to be forbidden).

It was once suggested that the smoothness of the initial matter distribution\index{initial matter distribution!smoothness} might only be apparent and that a state of higher inhomogeneity might have existed initially, that was later made uniform through various smoothing processes\index{smoothing processes}. What constitutes the most significant difficulty for the smoothing hypothesis is the fact that the existence of cosmological horizons\index{cosmological horizon} would have forbidden any such process from ironing out inhomogeneities above the scale determined by the size of the horizon, at the time when the cosmic microwave background\index{cosmic microwave background} was released, and therefore we should not observe uniformity on the largest scale if the homogeneity of the distribution of matter energy is attributable to smoothing processes obeying the principle of local causality\index{principle of local causality}. As a consequence of the clear inadequacy of conventional smoothing processes and in the absence of a better alternative, it is still widely believed that inflationary expansion\index{inflationary expansion} may be the cause of the very high homogeneity of the universe's initial matter distribution which is reflected in the small amplitude of cosmic microwave background\index{cosmic microwave background!temperature fluctuations} temperature fluctuations.

However, I think that this hypothetical process of exponentially accelerated expansion\index{exponentially accelerated expansion} would not be of much help in explaining the observed time asymmetry\index{time asymmetry!cosmic evolution} that characterizes cosmic evolution, because there is no reason to expect that a contracting universe\index{contracting universe} would evolve toward a more homogeneous configuration during the epoch that would precede a hypothetical phase of exponentially accelerated \textit{contraction}\index{exponentially accelerated contraction}, which would then take the universe back to a more likely state of maximum inhomogeneity\index{maximum inhomogeneity state}. If inflation\index{inflation!theory} theory could perhaps explain why the universe evolves in an otherwise unnatural way (from the viewpoint of the growth of gravitational entropy\index{gravitational entropy}), between the moment when matter emerges from the initial singularity\index{initial singularity} and the instant at which inflationary expansion ceases, it could not explain why it evolves toward greater homogeneity from far in the future and back toward the time at which the universe would presumably begin to contract at an exponentially accelerated rate into the initial singularity, now with naturally growing inhomogeneity.

Even if inflation\index{inflation} may give rise to a homogeneous universe\index{homogeneous universe} forward in time, a Big Crunch\index{Big Crunch} would not be expected to occur with decreasing inhomogeneity forward in time, unless the state immediately preceding the exponentially accelerated contraction into the final singularity\index{final singularity} would be required to be as smooth as the state which was produced in the past following ordinary inflation. But assuming that this would occur would amount to require that causality\index{causality} operates backward in time from the final singularity, instead of forward in time from the initial singularity, because, from the viewpoint where causality operates from the past toward the future, a Big Crunch would be more likely to occur with increasing inhomogeneity in the future, right up to the moment when a process of inflationary expansion\index{inflationary expansion} would perhaps take place in reverse and merely increase the inhomogeneity that would already exist even further and produce an inhomogeneous final state\index{inhomogeneous final state}.

Assuming that this is not what occurs would amount to postulate without motive that classical (unidirectional) causality\index{causality!classical or unidirectional}\footnote{
In the previously cited report \cite{Lindner-4} on the problem of time directionality\index{time directionality|nn}, which was published after the first version of the present one, I have explained that, in a quantum mechanical context\index{quantum mechanical context|nn}, a certain concept of \textit{bidirectional} causality\index{bidirectional causality|nn}, more general than the classical concept associated with the growth of entropy\index{entropy!growth|nn} that is characteristic of thermodynamic phenomena, must be assumed to operate at the elementary particle level\index{elementary particle level|nn}, where effects are not constrained to propagate only in the future direction of time.}
 must rather operate backward, from the instant at which matter emerges from the future Big Crunch singularity\index{future Big Crunch singularity} and until the moment when the universe\index{universe!maximum volume} would begin recollapsing, after having reached its maximum volume, so that the period of inflationary expansion\index{inflationary expansion} that would occur backward in time from the instant at which matter emerges from the final singularity would give rise to a homogeneous state \textit{after} inflation, in the past direction of time. But there is no \textit{a priori} reason not to assume, instead, that it is a highly inhomogeneous final state existing \textit{before} the phase of exponentially accelerated contraction\index{exponentially accelerated contraction} that gives rise to the inhomogeneous state that would develop in the future direction of time as a result of this exponentially accelerated contraction, as we may expect based on the hypothesis that causality still operates forward in time.

The problem is that the hypothesis that classical causality operates forward in time from the past singularity\index{past singularity} is necessary for the conclusion that inflationary expansion\index{inflationary expansion} would necessarily produce a homogeneous state, because if it was assumed that it is the events in the future which can influence what occurs backward in time until the moment when matter would start contracting at an exponentially accelerated rate back into the initial singularity\index{initial singularity}, then the state we would expect to obtain following the initial phase of inflationary expansion, from the forward-in-time viewpoint\index{forward-in-time viewpoint}, would still be a state of maximum inhomogeneity\index{maximum inhomogeneity state}, while this does not correspond to reality. But classical causality, or the rule that past events always have an influence on future events and not the opposite, is simply a consequence of thermodynamic time asymmetry\index{thermodynamic time asymmetry} or irreversibility, and if this property is assumed to characterize our universe without question, then it cannot be used to explain time irreversibility\index{time irreversibility} itself.

One can only begin to understand the cause of the homogeneity of the matter distribution that emerged out of the past singularity\index{past singularity} when one acknowledges that what is significant with our current description of the physics of the early universe\index{early universe!physics} is the explicit assumption that the cosmological horizon\index{cosmological horizon} (sometimes called the particle horizon\index{particle horizon}) begins to grow at the exact moment when the magnitude of the average densities of positive and negative matter energy\index{average matter energy densities!maximum magnitude} is maximum. But why should causality\index{causality} have anything to do with the density of matter energy?

Is it even appropriate to assume that the universe could have come into existence as a set of independent entities, not causally related to one another, due to the presence of non-overlapping cosmological horizons\index{cosmological horizon!non-overlapping horizons} in the primordial state\index{primordial state}? The conventional idea is that this is not a problem as long as causal relationships\index{causal relationships} can be established at later times in the future through the propagation of effects\index{propagation of effects} at relativistic velocities. This is what would allow the causally unrelated parts\index{causally unrelated parts!Big Bang} which are assumed to exist at the Big Bang to form one single universe.

What is even problematic, however, is the assumption that the cosmological horizon\index{cosmological horizon} must begin to grow at the precise moment when the density of matter and radiation is maximum (or infinite, as one would assume from a classical viewpoint). This idea appears to be motivated by the hypothesis that time begins with the past singularity\index{past singularity!beginning of time}. But when one considers that the size of the cosmological horizon increases with time and encompasses an increasingly larger portion of space, as the universe itself expands, one is actually presuming the validity of the classical principle of causality\index{classical principle of causality}, that is, of the idea that causes always precede their effects. However, it is always \textit{past} causes that produce \textit{future} effects. It is never assumed that a future cause could produce an effect in the past.

In other words, we are assuming the existence of a preferred direction in time (the future) and a preferred instant (that of the past singularity\index{past singularity}) at which effects begin to propagate. This is usually appropriate, as we experience time in a unidirectional way as a consequence of the fact that the thermodynamic arrow of time\index{thermodynamic arrow of time} always operates from past to future and never in the opposite direction. But there is no \textit{a priori} reason why classical causality\index{causality!classical or unidirectional} could not instead operate toward the past, from an arbitrarily remote time in the future, in which case the size of the cosmological horizon\index{cosmological horizon} would already encompass all of space, or at least a very large portion of it, at the Big Bang. In fact, it would appear more natural to assume that causality begins to operate at the present time, so that the cosmological horizon would spread from there in both the future and the past direction of time.

Therefore, if what we are seeking to explain is the existence of a preferred direction in time\index{preferred direction in time}, then we cannot simply assume the validity of the hypothesis that the cosmological horizon began expanding in the future direction of time at the instant in the past when the magnitude of the average densities\index{average matter energy densities!maximum magnitude} of positive and negative matter energy was maximum. We cannot claim that there is a problem with the homogeneity of the large-scale matter distribution\index{large-scale matter distribution!homogeneity}, if this problem arises as a consequence of assumptions concerning the size of the cosmological horizon which are only meaningful in the context where there is a preferred direction to causal signals\index{causal signals} which originates from this very same homogeneity. What we must provide is a consistent justification for the very validity of this particular choice of a horizon concept. We must explain why the matter distribution in the maximum density state was configured in such a way that it allowed classical (unidirectional) causality\index{causality!classical or unidirectional} to be a meaningful concept that came into effect at the exact moment when space began to expand out of the past singularity.

I believe that what makes the cosmological horizon\index{cosmological horizon} concept acceptable is simply the fact that, as we consider increasingly earlier times, the size of the horizon would eventually reach the limit imposed by quantum theory\index{quantum theory} on the classical definiteness\index{classical definiteness!spatial distance} of any measure of spatial distance. When the size of the cosmological horizon passes below the limit at which the uncertainty that is intrinsic to quantum phenomena\index{quantum phenomena!uncertainty} would apply to spacetime relationships themselves, it is certainly no longer appropriate to assume that the limited velocity of signal propagation\index{signal propagation!limited velocity} forbids the existence of causal relationships\index{causal relationships} between regions of space separated by distances larger than the size of the cosmological horizon, but smaller than this characteristic scale of quantum-gravitational phenomena\index{quantum-gravitational phenomena!characteristic scale}, as there are no classically well-defined relationships of distance and duration below that scale.

In such a context, it would be suitable to assume that there may, after all, exist causal relationships\index{causal relationships} between all physical elements of the universe which were in contact with one another to within an elementary quantum-gravitational unit of area\index{quantum-gravitational unit of area}, equal to four Planck units of area\index{Planck unit of area}\footnote{
%%BOOK VERSION:
%%In section 3.10 of the extended version of the report \cite{Lindner-4} on which is based this book I have explained that on the scale characteristic of quantum-gravitational phenomena\index{quantum-gravitational phenomena!characteristic scale|nn} there appears to exist a minimum significant unit of area\index{minimum significant unit of area|nn}, which always encodes exactly one binary unit of information\index{binary unit of information}, and which is equal to four times a Planck unit of area\index{Planck unit of area|nn}.},
In section 3.10 of the extended version of this report \cite{Lindner-4} I have explained that on the scale characteristic of quantum-gravitational phenomena\index{quantum-gravitational phenomena!characteristic scale|nn} there appears to exist a minimum significant unit of area\index{minimum significant unit of area|nn}, which always encodes exactly one binary unit of information\index{binary unit of information}, and which is equal to four times a Planck unit of area\index{Planck unit of area|nn}.},
 at the Planck time\index{Planck time}, if we also have good reasons to expect that the area delimited by the cosmological horizon\index{cosmological horizon!elementary unit of area} was then equal to this elementary unit of area, within which the gravitational field and the metric properties of spacetime\index{metric properties of spacetime} where submitted to quantum indefiniteness\index{quantum indefiniteness}. In fact, from a quantum-gravitational viewpoint, it may be preferable to simply recognize that there is nothing smaller than the elementary unit of surface\index{elementary unit of surface} associated with this particular scale of distance.

Therefore, when the size of the cosmological horizon\index{cosmological horizon} reaches the natural limit imposed by quantum gravitation\index{quantum gravitation}, as it contracts in the past, if the most elementary particles (whose characteristic size is that of quantum-gravitational phenomena\index{quantum-gravitational phenomena!characteristic scale}) are allowed to be in contact with one another to within such an elementary unit of area\index{elementary unit of area}, then no smaller components would remain causally unrelated in the initial Big Bang state, which is probably sufficient a condition to impose, regarding the necessity for the universe\index{universe!ensemble of causally interrelated elements} to form a global ensemble whose elements (the elementary particles\index{elementary particles}) are allowed to remain causally interrelated at all times, as a result of having been in direct contact with one another when the size of the cosmological horizon\index{cosmological horizon!minimum size} was minimal.

Thus, I would suggest that all the elementary particles originally present in our universe at the Big Bang be required to have been in contact with at least one other particle to within an elementary quantum-gravitational unit of area\index{quantum-gravitational unit of area} at the Planck time\index{Planck time}. More specifically, I propose that the following condition must apply.
\begin{quote}
\textbf{Global entanglement constraint\index{global entanglement constraint}}: There must exist an entire space-like hypersurface\index{space-like hypersurface}, at one particular moment of cosmic time\index{cosmic time}, over which all elementary particles, regardless of their energy sign, are in contact with at least one neighboring elementary particle of either positive or negative energy sign to within an elementary quantum-gravitational unit of area\index{quantum-gravitational unit of area}, in a state of maximum positive and negative matter energy densities\index{maximum matter energy densities}.
\end{quote}
If this condition is fulfilled, then any elementary particle that is present in the universe today would have been in direct contact with another elementary particle that was in contact with another such particle and so on, when the size of the cosmological horizon\index{cosmological horizon!size of most elementary particle} was equal that of a most elementary particle, which means that no component of the universe\index{universe!component} could then exist that would be causally unrelated to the other components of the same universe.

Now, if only positive-energy matter was present in our universe, the global entanglement constraint could be satisfied in the initial state regardless of the measure of gravitational entropy\index{gravitational entropy}, because even if strong gravitational fields and macroscopic event horizons\index{macroscopic event horizon} existed in the instants immediately preceding the formation of the past singularity\index{past singularity}, as we approach the initial state\index{initial state!maximum matter energy densities} of maximum matter energy densities in the past direction of time, all elementary particles\index{elementary particles} would nevertheless be allowed to come into contact with their neighbors in the past singularity, because those are attractive gravitational fields\index{attractive gravitational field}.

When negative-energy matter is present, however, macroscopic event horizons may constitute potential barriers\index{potential barrier} which are impossible to overcome. If the constraint of global entanglement\index{global entanglement constraint} imposes contact between all neighboring elementary particles\index{neighboring elementary particles} at the Planck time\index{Planck time}, regardless of their energy sign, then given that gravitational repulsion\index{gravitational repulsion}, unlike gravitational attraction, may forbid local contacts between particles located within black holes\index{black holes!opposite energy signs} of opposite energy signs, it follows that event horizons\index{event horizon} can be expected to be absent initially on all but the smallest scale, even if macroscopic black holes\index{macroscopic black holes} are allowed to form at later times. If this was not the case, then certain particles could exist in our universe that would not be causally related to the rest of it, which I believe would involve a contradiction.

In the absence of a condition requiring the entanglement of all matter particles, the most likely initial state, from a purely statistical viewpoint, would be one for which all the matter in the universe would be concentrated in the smallest possible number of opposite-energy black holes\index{black holes!arbitrarily large masses} with arbitrarily large masses, which would already be in a state of maximum gravitational entropy\index{gravitational entropy!maximum}. But this was not allowed to constitute our boundary conditions\index{boundary conditions!Big Bang} at the Big Bang simply because, under such conditions, the spacetime singularities\index{spacetime singularity} at the center of the objects could never come into contact with one another in the initial state of maximum positive and negative matter energy densities\index{initial state!maximum matter energy densities} or at any other time (because gravitational entropy\index{gravitational entropy} can only grow as time passes), while this is required by the global entanglement constraint.

In the presence of negative-energy matter, global entanglement actually constitutes a very constraining requirement, because any sufficiently large fluctuation in the initial density of positive or negative matter energy would give rise to the presence of a macroscopic event horizon\index{macroscopic event horizon} that would forbid the condition from applying. The mass of any black hole that is now present in the universe must, therefore, diminish continuously in the past direction of time, as we approach the initial singularity\index{initial singularity}, so as to allow the condition of homogeneity\index{condition of homogeneity!initial matter distribution} that is imposed on the initial matter distribution to be satisfied, despite the fact that it is actually the past condition that gives rise to the future configuration, in the context where the condition that applies on the initial Big Bang state\index{initial Big Bang state} is, in effect, one of minimum gravitational entropy\index{gravitational entropy!minimum}, from which the classical (unidirectional) principle of causality\index{classical principle of causality} itself can be expected to emerge.

What had long remained unexplained is the fact that an ensemble of systems started in the same macroscopic state\index{macroscopic state} evolves to occupy all available microscopic states\index{microscopic state} in the future, while a similar ensemble, started in the same macroscopic state, usually evolves only to past states characterized by a lower entropy and more particularly, a lower gravitational entropy\index{gravitational entropy!lower}. But it is now possible to understand that the unnatural evolution that takes place in the past direction of time is a direct consequence of the limitation imposed on the initial state by the constraint of global entanglement\index{global entanglement constraint} in the presence of negative-energy matter and that this asymmetry would not merely characterize a small portion of all possible universes, but really all universes governed by the known, fundamental principles of physics\index{fundamental principles of physics}, in which negative-energy matter is present with an initial density whose average magnitude equals that of positive-energy matter. Remarkably enough, this unrecognized, but necessary condition allows to explain why it is that only the gravitational component of entropy\index{entropy!gravitational component} was not maximum at the Big Bang, while the entropy\index{entropy!of matter and radiation} of matter and radiation was allowed to be arbitrarily large, as required if the universe was to already be in a thermal equilibrium state\index{thermal equilibrium!state}.

It must be clear that the constraint of global entanglement merely imposes that the positive- and negative-energy matter particles which were present in the initial state of maximum matter energy densities\index{initial state!maximum matter energy densities} be as homogeneously distributed as necessary for an absence of \textit{macroscopic} event horizons\index{macroscopic event horizon} larger than one elementary quantum-gravitational unit of area\index{quantum-gravitational unit of area}\footnote{
%%BOOK VERSION:
%%In section 3.10 of the extended version of the report \cite{Lindner-4} on which this book is based I have explained that the smallest black holes which are present on the quantum-gravitational scale\index{quantum-gravitational scale!black holes|nn}, as a result of quantum fluctuations in the curvature of spacetime\index{curvature of spacetime!quantum fluctuations}, must contain only one matter or radiation particle and must have a surface area that is equal to one elementary unit of area\index{elementary unit of area|nn} (itself equal to a small multiple of the Planck unit of area\index{Planck unit of area|nn}).},
In section 3.10 of the extended version of this report \cite{Lindner-4} I have explained that the smallest black holes which are present on the quantum-gravitational scale\index{quantum-gravitational scale!black holes|nn}, as a result of quantum fluctuations in the curvature of spacetime\index{curvature of spacetime!quantum fluctuations}, must contain only one matter or radiation particle and must have a surface area that is equal to one elementary unit of area\index{elementary unit of area|nn} (itself equal to a small multiple of the Planck unit of area\index{Planck unit of area|nn}).},
 because it is only under such conditions that the most-elementary particles\index{most-elementary particles}, with the highest possible positive and negative energies, can all be in direct contact with one another, regardless of their energy signs, as required if they are to be part of the same universe when the area delimited by the cosmological horizon\index{cosmological horizon} is no larger than the smallest physically significant unit of area\index{area!smallest physically significant unit}.

It is appropriate to impose such a condition, even in the context where one recognizes that there are no direct interactions between positive- and negative-energy particles\index{opposite-action particles!absence of direct interactions}, while there nevertheless exists an indirect repulsive gravitational force\index{indirect repulsive gravitational force} between opposite-energy objects which is strong enough to prevent macroscopic opposite-energy black holes\index{macroscopic opposite-energy black holes!absence of contact} from coming into contact with one another, because what is involved here is not an interaction propagated across space and time, but a minimum unit of area\index{minimum unit of area!causal relationships} within which there need not even be an exchange of energy (mediated by an interaction boson\index{interaction bosons}) for causal relationships to exist. The very meaningfulness of the constraint of global entanglement\index{global entanglement constraint} is in fact dependent on the hypothesis that there exists a minimum, physically significant unit of distance\index{minimum physically significant unit of distance}, to which corresponds a minimum unit of area\index{minimum unit of area}, below which no causal signal\index{causal signals!propagation} needs to propagate for entanglement to be established.

It is merely the fact that the elementary particles inside a macroscopic, positive-energy black hole cannot come into contact with those inside a macroscopic, negative-energy black hole, that constitutes a limitation to the existence of causal relationships\index{causal relationships!particles inside opposite-energy black holes} between those particles. If all the matter in the universe was initially concentrated in two macroscopic black holes\index{macroscopic black holes!opposite energy signs} with opposite energy signs, the elementary particles inside one of the objects would necessarily remain causally independent from those inside the other black hole, due to the insurmountable gravitational repulsion\index{gravitational repulsion!opposite-energy black holes} that exists between the two objects and this is what explains that it is not possible for gravitational entropy to be maximum in the initial Big Bang state\index{initial Big Bang state!gravitational entropy}.

What's interesting is that, contrarily to the situation we would have if inflationary expansion\index{inflationary expansion} was assumed to be responsible for the smoothness of the initial matter distribution\index{initial matter distribution!smoothness}, it is now possible to explain why it is that the constraint that gives rise to a homogeneous initial state is necessarily effective in only one direction of time. Thus, gravitational entropy\index{gravitational entropy} can be expected to decrease continuously in the past direction of time from its current intermediary value, regardless of whether space is expanding or contracting, as long as we are, in effect, approaching the instant at which is formed the unique past singularity\index{past singularity!unique} on which the condition of global entanglement is imposed. It is simply the fact that the condition that applies to the initial singularity\index{initial singularity} is precisely one of minimum gravitational entropy\index{gravitational entropy!minimum}, from which can emerge a phenomenon of classical (unidirectional) causality\index{causality!classical or unidirectional} that operates toward the future from that particular instant of time, that requires the evolution that takes place at all later times to be such that it allows an initial state obeying this condition to be reached in the past direction of time, because under such circumstances causality must, in effect, operate in one unique direction of time, thereby allowing past causes to exert irreversible effects in the future.

It is important to emphasize that, in the context of this explanation of temporal irreversibility\index{temporal irreversibility}, all physical systems, regardless of how isolated they may have become at the present time, must evolve with continuously decreasing gravitational entropy\index{gravitational entropy!continuous decrease} in the same past direction of time. If the temporal parallelism\index{temporal parallelism} of the thermodynamic evolution of isolated branch systems\index{isolated branch systems} is not unexpected, from this particular viewpoint, it is because any system that is part of a given universe, regardless of how isolated it might have become, must have been entangled with the rest of the matter in this universe at the Big Bang in order that causal relationships\index{causal relationships} be established between all components of the universe and this implies that even those systems which have become isolated from one another must follow the same kind of gravitational entropy decreasing evolution that is necessary for achieving this global entanglement at some point in the past.

But if a state of maximum positive and negative matter energy densities\index{maximum matter energy densities} must necessarily occur at one time or another for the global entanglement\index{global entanglement} of all elementary particles to be achieved in the presence of negative-energy matter, then given that such a state would not likely be characterized by an absence of macroscopic event horizons\index{macroscopic event horizon} unless it constitutes the mandatory unique event at which global entanglement is enforced on the universe, then one must conclude that our Big Bang really is this unique event. In such a context, the presence of an initial singularity\index{initial singularity} would no longer be a mere fortuitous consequence of the fact that space is expanding, but would be an essential requirement for the existence of any universe\index{universe!existence requirement} obeying the known fundamental principles of physics\index{fundamental principles of physics}.

If gravitational entropy\index{gravitational entropy} does indeed rise in only one particular direction of time, it is because only evolution away from the initial singularity, either in the future or in the past, can be expected to be left unaffected by the constraint of global entanglement\index{global entanglement constraint}. It is, therefore, possible to understand why it is that the cosmological horizon\index{cosmological horizon} only begins to spread outward at the Big Bang. It is the fact that the constraint of global entanglement would only be required to apply once, even if the universe was to return to a state of maximum matter density\index{maximum matter density state} at some point in the future, that explains that the evolution that takes place from the moment at which this condition is enforced is not symmetric in time.

Thus, it is incorrect to argue that in order not to assume the very outcome we are seeking to derive (the temporal irreversibility\index{temporal irreversibility}), it is required that any condition that applies to some initial state should also apply to a final state of the universe's history\index{history of universe!initial and final states}. Once it is understood that there need only be one state of maximum matter density and minimum gravitational entropy\index{gravitational entropy!minimum} in any given universe, then the kind of evolution which can be expected to take place in the direction of time toward that unique state, either in the past or in the future direction of time, would necessarily be different from that occurring in the opposite direction and this allows to explain time asymmetry\index{time asymmetry} without assuming it in the first place.

But if time does extend past the `initial' singularity\index{initial singularity}, following a quantum bounce\index{quantum bounce} (as predicted by the most promising and least speculative of current quantum gravitation\index{quantum gravitation!theories} theories), then we can expect space to be expanding at a critical rate and the density of matter to begin decreasing from its maximum value immediately after the event (in the past direction of time), while the inhomogeneity of the matter distribution would still need to be minimum if there is to be any continuity in the evolution of the microscopic state\index{microscopic state!matter and gravitational field} of matter and its gravitational field as we pass the state of maximum matter density\index{maximum matter density state}. This means that, even for the portion of history the precedes the initial singularity, the thermodynamic arrow of time\index{thermodynamic arrow of time} would (initially at least) have the same direction as the cosmological arrow of time\index{cosmological arrow of time} associated with expansion and would actually be opposite that we observe on our side in time of the initial singularity. As a consequence, whatever occurred during the portion of history that preceded the Big Bang would remain unknowable to observers in the current portion of history.

Now, it is sometimes argued that if there was a history prior to the Big Bang, then the final singularity\index{final singularity} which would be produced in the future direction of time (which would constitute our initial singularity) would likely be in a high gravitational entropy\index{gravitational entropy} state (as any future state reached after a long period of random evolution), which would require the state following it (our initial state) to have a similar configuration. But in fact, it is exactly the opposite which is true and the state preceding the initial singularity must actually be very homogeneous, because the constraint of global entanglement\index{global entanglement constraint} applies to the singularity itself, while it is the evolution away from it, in \textit{any} direction of time, which is unconstrained. Continuity merely imposes that the configuration be similar on both sides of the initial singularity, but it does not allow one to determine what this configuration actually is.

It is only in the absence of an appropriate constraint to be imposed on the initial singularity\index{initial singularity} that gravitational entropy would have to be maximum in both the immediate past and the immediate future of the initial state and indeed at all times. Not recognizing this would, again, amount to favor one particular direction of time (that relative to which entropy would be assumed to grow before the initial singularity) without justification, instead of explaining why such a preferred direction naturally emerges, as I have done.

When time does unfold toward states of higher gravitational entropy\index{gravitational entropy} past the `initial' Big Bang singularity, then the history of the universe\index{history of universe!global time symmetry} is allowed to be globally symmetric with respect to past and future, from a global viewpoint, because gravitational entropy can be expected to grow continuously in both the future and the past of its mandatory state of maximum matter density\index{maximum matter density state} and minimum gravitational entropy\index{gravitational entropy!minimum}. This irreversible evolution\index{irreversible evolution} can be expected to continue regardless of whether space keeps expanding or eventually begins to recontract. But in the context where gravitational entropy\index{gravitational entropy!continuous increase} is continuously growing, as a consequence of the polarization of the positive- and negative-energy matter distributions, it follows that, if there is an infinite amount of matter in the universe, then there may never arise a state of maximum stability\index{maximum stability state}, equivalent to thermal equilibrium\index{thermal equilibrium}, where gravitational entropy\index{gravitational entropy!maximum} would become maximum and would no longer rise.

\bigskip

\noindent Given that the distribution of matter energy was so uniform at the time when the cosmic microwave background\index{cosmic microwave background} was released it may seems that what remains unexplained is really that the temperature was not perfectly smooth and free of any fluctuations initially. But I believe that this smoothness problem\index{smoothness problem} is a mere consequence of the fact that we do not properly understand what gives rise to the high level of uniformity of the initial matter distribution\index{initial matter distribution!smoothness}, above the cosmological horizon\index{cosmological horizon} scale. In the present context, it is merely the upper bound of fluctuations in the initial density of matter energy\index{initial density of matter energy!upper bound of fluctuations} which is constrained and therefore it is to be expected that certain local variations in matter energy density would necessarily be present, as the absence of macroscopic event horizons\index{macroscopic event horizon} can be satisfied even when some fluctuations are present.

What is truly remarkable is that the spectrum of fluctuations in the initial density of matter energy\index{initial density of matter energy!spectrum of fluctuations} which is deduced from observations of cosmic microwave background\index{cosmic microwave background!temperature fluctuations} temperature fluctuations is a (nearly) scale-independent spectrum\index{cosmic microwave background!scale-independent spectrum} of the Harrison-Zel'dovich\index{Harrison-Zel'dovich spectrum} type (for fluctuations larger than the scale of the cosmological horizon at the epoch of the decoupling of matter from radiation\index{decoupling of matter from radiation}) while this is the only spectrum which, according to specialists, does not allow the magnitude of early fluctuations in matter energy density to diverge on either large or small scales and which, therefore, does not give rise to the creation of a large number of primordial black holes\index{primordial black holes}, while those are precisely the conditions which are required by the theoretical constraint of global entanglement\index{global entanglement constraint}. It would therefore appear that the requirement that the spectrum of temperature fluctuations in the cosmic microwave background\index{cosmic microwave background!scale-invariant spectrum of fluctuations} be scale invariant is not a unique property of inflationary cosmology\index{inflationary cosmology}.

Also, the idea that only inflation\index{inflation!initial perturbations in matter energy density} allows the initial perturbations in the density of matter energy to begin oscillating (between compressions and rarefactions) at the same epoch of cosmic time\index{cosmic time}, as required to explain the existence of harmonics in the spectrum of cosmic microwave background temperature fluctuations, may not be justified, because any scale-invariant spectrum of fluctuations in the density of matter and radiation energy which would be present initially on scales larger than the cosmological horizon\index{cosmological horizon}, would have the same consequences. But this is precisely what we can expect to occur as a consequence of the constraint of global entanglement, in the presence of negative-energy matter. Thus, even in the absence of inflationary expansion\index{inflationary expansion}, the required fluctuations in matter density would already exist in the `initial' state and those perturbations would all begin oscillating as soon as they are encompassed by the cosmological horizon, which means that, on a given angular scale\index{angular scale}, all maximum compressions and rarefactions would be reached at the same time.

The fact that the power spectrum of density fluctuations in the temperature of the cosmic microwave background\index{cosmic microwave background!scale-invariant spectrum of fluctuations} appears to only be \textit{nearly} scale invariant, on the other hand, may not be a consequence of inflation either. It is usually assumed, in effect, that it is because the gravitational repulsion\index{gravitational repulsion} driving inflationary expansion becomes weaker as the process is occurring, that the smaller-scale fluctuations, which are produced later by the inflation\index{inflation!process} process, have a smaller amplitude. But what may be happening, instead, is that, given that there remained only a negligible portion of baryonic negative-energy matter\index{baryonic negative-energy matter} following the early annihilation of matter with antimatter\index{early matter-antimatter annihilation}, then one can expect that inhomogeneities in the negative-energy matter distribution were less developed on smaller scales when the CMB was released, because, even though there existed as much negative vacuum-dark-matter energy\index{negative vacuum-dark-matter energy} before and after the annihilation process, no significant small-scale fluctuations could develop in this matter distribution in the absence of baryonic negative-energy matter (as I have explained in section \ref{sec:6.2}).

Thus, while the distribution of negative vacuum-dark-matter energy would be inhomogeneous on larger scales, at the epoch of last scattering, it would be more homogeneous at the same epoch on smaller scales. But given that negative-energy matter inhomogeneities do contribute to produce temperature fluctuations in the cosmic microwave background\index{cosmic microwave background!temperature fluctuations} at the epoch of last scattering\index{epoch of last scattering} (as a result of the gravitational forces they exert on positive-energy matter and radiation), then there naturally arises a deficit of fluctuations on smaller scales, even when inflation\index{inflation} is not assumed to be responsible for producing the inhomogeneities which were present on both smaller and larger scales at the epoch of last scattering. I believe that this is the true reason why the spectrum of CMB\index{cosmic microwave background!slightly tilted spectrum of fluctuations} temperature fluctuations is slightly tilted (with a power-law index\index{power-law index!smaller than one} smaller than one) compared to a pure Harrison-Zel'dovich spectrum\index{Harrison-Zel'dovich spectrum}.

But if it cannot be expected that the inhomogeneities which were present on smaller scales in the early distribution of negative-energy matter were as developed as those which were present at the same epoch in the positive-energy matter distribution, it remains that, in the presence of negative vacuum-dark-matter energy\index{negative vacuum-dark-matter energy}, there should be more inhomogeneities in the matter distribution than we could attribute to positive-energy matter, at least on a very large scale, and therefore more fluctuations in the temperature of the cosmic microwave background as well. As a result, it would appear necessary to revise the magnitude of density fluctuations attributable to positive-energy matter downward.

Now, if the actual magnitude of density fluctuations in the distribution of positive-energy matter at the epoch of last scattering\index{epoch of last scattering} can be assumed to be smaller than the value currently deduced from observations of CMB temperature fluctuations\index{cosmic microwave background!temperature fluctuations}, due to the fact that some of the observed fluctuations are attributable to the presence of negative-energy matter, then it is natural to expect that the magnitude of fluctuations in the positive-energy matter distribution obtained by evolving the fluctuations which existed at the epoch of last scattering forward in time to their present state, should be smaller than that which is actually observed, even when one recognizes that the presence of negative-energy matter must have accelerated the process of structure formation\index{structure formation!process} in the early universe, on smaller scales, and at later times, on larger scales. As a result, the presence of negative-energy matter inhomogeneities would allow to ease the tension that emerged from relatively recent observations \cite{Asgari-1} which indicate that the current magnitude of fluctuations in the positive-energy matter distribution is smaller than the value deduced from CMB temperature fluctuations extrapolated to the present epoch (a problem known technically as the $S_8$ tension\index{S8 tension@$S_8$ tension}).

In any case, it is possible to deduce that the temperature of the cosmic microwave background\index{cosmic microwave background!homogeneous large-scale temperature} must have been very (but not perfectly) homogeneous, even on a scale larger than the size of the cosmological horizon\index{cosmological horizon} at the epoch of last scattering, because the absence of macroscopic event horizons\index{macroscopic event horizon} is required on all scales and this imposes very stringent conditions on the fluctuations of matter energy density that could be observed, even on the largest scale. This means that no smoothing process\index{smoothing processes} is required to make the temperature of the cosmic microwave background uniform, because the distribution of matter energy was mostly uniform on all scales right from the beginning, even if the size of the cosmological horizon decreases more rapidly than the scale factor\index{scale factor} as we approach the initial Big Bang singularity\index{initial Big Bang singularity} in the past direction of time, so that regions which are now in contact must have been causally disjoint at the epoch of decoupling\index{epoch of decoupling} (despite the existence of causal relationships\index{causal relationships} between all elementary particles\index{elementary particles} which were present in the universe at this epoch).

When one properly recognizes the limitations imposed by the global entanglement constraint on the initial state at the Big Bang, the horizon problem\index{horizon problem} simply no longer exists and no independent assumption is required to confirm the relevance of the cosmological principle\index{cosmological principle} for a description of the early universe\index{early universe}. There is no longer any mystery associated with the fact that only one parameter (the scale factor\index{scale factor}) is required to describe the state of the universe at all but the most recent epoch. In fact, it would now appear that the cosmological principle must be obeyed as accurately as we are considering increasingly larger regions of space, corresponding to times increasingly closer to the initial singularity\index{initial singularity}.

\section{Conclusion\label{sec:7}}

One of the objectives of research in theoretical physics is to find what is not well understood in current models and in particular to find the implicit assumptions which might constitute weak points of these models. I have given arguments to the effect that current ideas regarding negative-energy matter show such weaknesses. I believe that if those inconsistencies have been part of the common scientific paradigm for such a long time it is because there never was any real motivation to consider that negative-energy matter could actually be an element of physical reality. For one thing, no observations seemed to require the existence of this matter. But the strongly held view that energy must always be positive also seems to have its roots in the thermodynamic definition of the concept of energy\index{energy!thermodynamic definition} as being related to heat\index{heat} and temperature\index{temperature}. When measuring the amount of heat transferred from one reservoir to another you would never imagine that such a quantity could take on negative values. This bias has remained with us even long after energy became a property of quantum particles and the source of spacetime curvature\index{spacetime curvature}.

In this report I have explained what would really be the properties of negative-action particles, based on the most unavoidable theoretical requirements and empirical constraints. I also introduced a quantitative framework based on Einstein's general theory of relativity\index{general relativity theory}, that enables those requirements to be fulfilled. Now, ten basic postulates\index{basic postulates} (and one additional cosmological hypothesis\index{cosmological hypothesis}) may seem like a lot, but as I'm seeking to provide as much precision as possible on what really constitutes my viewpoint, they all appeared to be absolutely necessary. However, it is also true that the only fundamental assumptions are contained in the first, fourth, and fifth principles and that all the other postulates are simple consistency requirements following more or less directly from those three basic hypotheses. What's more, I provided arguments to the effect that even my first postulate is not really an assumption as much as it is itself a basic consistency requirement\index{basic consistency requirement} founded on the deepest precepts inherited from the long and fruitful tradition of scientific analysis which gave rise to the standard models\index{standard models!particle physics and cosmology} of particle physics and cosmology. If I have been successful in conveying the rationale behind those principles, then it should be clear that this is the only approach toward a solution to the problem of negative energy states\index{negative energy states!problem of} in modern physical theory that is both theoretically and observationally viable.

Looking forward, it may be interesting to examine the consequences of the classical model proposed here for a quantum mechanical description of the gravitational interaction. Theories have already been developed which are consistent at once with quantum theory\index{quantum theory} and general relativity. The approaches which are generally considered most promising are formulated in a background-independent\index{background-independent approaches} manner and rely on spin networks\index{spin networks}. One interesting outcome of this program is that its success seems to depend on the introduction of the same kind of restrictions toward advanced and retarded propagation\index{advanced and retarded propagation} of positive and negative frequencies\index{positive and negative frequencies} which are found in traditional quantum field theory\index{quantum field theory}. This treatment was shown to be necessary for implementing causality\index{causality} into spin-foam models of quantum gravitation\index{quantum gravitation!spin-foam models} \cite{Livine-1}. This was actually the first confirmation we had that the distinction between forward and backward propagation in time, which led to the introduction of negative energy states in quantum field theory, also applies in the case of gravitation.

At this point I'm not claiming that the simple reformulation of the existing classical theory of gravitation I have proposed would alone allow to solve the remaining problems that still stand in the way of a fully satisfactory integration of quantum principles to gravitation theory, but it does appear that the more consistent perspective that it brings, particularly to our conception of the vacuum, has the potential to be a key element in creating a viable framework where, at least, calculations of transition probabilities\index{transition probabilities!gravitation} involving gravitation provide finite and useful results.

\section*{Acknowledgements}
I would like to thank my colleagues Terrence A. Simpson and Fran\c{c}ois E. Racicot from Universit\'{e} de Montr\'{e}al, as well as Emma Roux from UQAM for helpful discussions concerning the epistemological foundations of this work. I must also acknowledge the support of all members of my family during the rather slow process of intellectual development that finally gave rise to this concrete result.

%\newpage
\addcontentsline{toc}{section}{\protect{Bibliography}}
\bibliographystyle{unsrtnat}
\bibliography{References}

%\newpage
\addcontentsline{toc}{section}{\protect{Index}}
\small
\printindex
\normalsize

\end{document}